%% file: main.tex
\documentclass[nofootinbib,preprint,superscriptaddress,prC]{revtex4-2}

\input{preamble}

\begin{document}
\title{Cold Neutron-Deuteron Capture and Wigner-SU(4) Symmetry}

\author{Xincheng Lin\,\orcidlink{0000-0001-9068-6787}}
\email{xincheng.lin@duke.edu}
\affiliation{Department of Physics, Box 90305, Duke University, Durham, North Carolina 27708, USA}

\author{Hersh Singh\,\orcidlink{0000-0002-2002-6959}}
\email{hershsg@uw.edu}
\affiliation{InQubator for Quantum Simulation (IQuS), Department of Physics, University of Washington, Seattle, Washington 98195-1550, USA}
\affiliation{Institute for Nuclear Theory, University of Washington, Seattle, Washington 98195-1550, USA}

\author{Roxanne P. Springer\,\orcidlink{0000-0002-5023-0205}}
\email{rps@phy.duke.edu}
\affiliation{Department of Physics, Box 90305, Duke University, Durham, North Carolina 27708, USA}
\author{Jared Vanasse\,\orcidlink{0000-0001-5593-6971}}
\email{jvanass3@fitchburgstate.edu}
\affiliation{Fitchburg State University, 160 Pearl St., Fitchburg, MA 01420-2697}

\preprint{INT-PUB-22-029}

\begin{abstract}

We calculate the cold neutron-deuteron ($nd$) capture cross section, $\sigma_{nd}$, to next-to-next-to leading order (\NNLO) using the model-independent approach of pionless effective field theory (\EFT). At leading order we find $\sigma_{nd} = 0.314 \pm 0.217$ mb, while the experimental result is 0.508(15)~mb~\cite{Jurney:1982zz} for a laboratory neutron velocity of 2200 m/s. At next-to-leading-order (NLO), we show that $\sigma_{nd}$ is sensitive to the low energy constant (LEC), $L_1^{(0)}$, of the two-nucleon isovector  current appearing at NLO. A fit of $L_1^{(0)}$ at NLO to the triton magnetic moment yields a NLO prediction of $\sigma_{nd}=0.393 \pm 0.164$ mb, where the error comes from propagating the error from the $L_1^{(0)}$ fit. 
At next-to-next-to-leading-order (\NNLO), we find that a new three-nucleon magnetic moment counterterm is required for renormalization group invariance of both $\sigma_{nd}$ and the triton magnetic moment. Fitting the \NNLO correction to $L_1^{(0)}$ (denoted $L_1^{(1)}$) to cold neutron-proton capture ($\sigma_{np}$) yields a \NNLO prediction of $\sigma_{nd}=0.447 \pm 0.130$ mb, where the error comes from propagating the error from the $L_1^{(1)}$ fit. We also study different fittings of $L_1^{(0)}$ and $L_1^{(1)}$ to $\sigma_{np}$, $\sigma_{nd}$, and/or the triton magnetic moment. For example, fitting $L_1^{(0)}$ simultaneously to $\sigma_{np}$, $\sigma_{nd}$, and the triton magnetic moment at NLO, and fitting $L_1^{(1)}$ simultaneously to $\sigma_{np}$ and $\sigma_{nd}$ at \NNLO, yields $\sigma_{nd} = 0.480 \pm 0.114$ mb and $0.511 \pm 0.042$ mb, respectively, where errors are naively estimated from \eftnopi power counting. In addition, we discuss how Wigner-SU(4) symmetry may alter the naive \eftnopi expansion of $\sigma_{nd}$.
\end{abstract}

\maketitle
\newpage

\section{Introduction}

Cold neutron-deuteron capture into a triton  and a photon ($nd\to t\gamma$) is one of the simplest reactions involving three nucleons and an external current.
The study of $nd\to t\gamma$ is a precursor for understanding its isospin mirror process,  the proton-deuteron capture into Helium-3 and a photon ($pd\to \jjvHe\gamma$),  which is more complicated than $nd\to t\gamma$ because of the Coulomb interaction. $pd\to \jjvHe \gamma$ is important for precisely determining deuterium abundance from Big Bang nucleosynthesis (BBN) and stellar nucleosynthesis as it leads to the loss of deuterium~\cite{Bertulani:2016eru}. Understanding the  $nd\to t\gamma$ reaction also yields  insights into the electromagnetic properties of the three-nucleon bound states of the triton ($\jjvH$) and $\jjvHe$ because they depend upon some of the same two- and three-nucleon currents.  Moreover, understanding  $nd\to t\gamma$ is essential for additional calculations of few-nucleon systems with external currents, such as  two- and three-body photo-disintegration~\cite{Faul:1981zz}, polarization~\cite{Skopik:1981zz,Konijnenberg:1988gq,Konijnenberg:1990zha,Schmid:1995zz,Schmid:1996zz} and parity-violating ~\cite{Alberi:1988fd,Avenier:1984is} asymmetries, and  beyond-the-standard-model (BSM) physics such as dark-matter-nucleon interactions.

The process $nd\to t\gamma$ was measured by Jurney et al. \cite{Jurney:1982zz}, who found a capture cross section of  0.508$\pm$0.015 mb at a neutron laboratory velocity of 2200~m/s.
References~\cite{Viviani:1996ui} and \cite{Marcucci:2005zc} used potential models, including AV18/UIX to calculate  $nd\to t\gamma$ using one-, two- (Ref.~\cite{Viviani:1996ui}), and three- (Ref.~\cite{Marcucci:2005zc}) nucleon currents that preserve gauge symmetry.
Reference~\cite{Marcucci:2005zc} found the following for the $nd\to t\gamma$ cross section: at low energies it is dominated by a magnetic dipole transition, there is a small electric quadrupole contribution, including only one-nucleon currents underpredicts the cross section by about a factor of two, after including two-nucleon currents their prediction is 0.523 mb, and including three-nucleon currents increases the predicted cross section to 0.556 mb.
The
$nd\to t\gamma$ process has also been studied using chiral effective field theory~\cite{Girlanda:2010vm} ($\chi$EFT), where plots in agreement with experiment were shown, as well as by using heavy baryon chiral perturbation theory~\cite{Song:2008zf}, where a value of  $0.490\pm0.008$~mb was found.

In order to understand the role of  external currents and their renormalization group (RG) behavior, in this paper we use pionless effective field theory (\eftnopi) (See Refs.~\cite{vanKolck:1999mw,Beane:2000fx,Bedaque:2002mn} for reviews) as a simple model-independent approach for calculating few-body low-energy nuclear observables. At momenta well below the pion mass all mesons can be integrated out, leaving a theory consisting solely of nucleons and possible external currents. \eftnopi has a systematic expansion (power counting) in powers of $Q \sim p/\lambdanopi$, where $\lambdanopi\approx m_{\pi}$ is the cutoff of \eftnopi and $p< \lambdanopi$ is a typical momentum scale in the process.  Power counting provides a naive a priori (parametric) theoretical error estimate at each order of a  calculation. \EFT  has advantages over its higher energy counterpart, $\chi$EFT, since it has a more straight-forward power counting~\cite{Valderrama:2016koj} that gives RG invariant results as well as analytical calculations in the two-nucleon sector~\cite{Chen:1999tn,Chen:1999vd,Rupak:1999rk}.  \EFT has been successful in calculating three-nucleon processes such as nucleon-deuteron ($N\!d$) scattering~\cite{Bedaque:1998mb,Bedaque:1999ve,Gabbiani:1999yv,Bedaque:2002yg,Griesshammer:2004pe,Vanasse:2013sda,Rupak:2001ci,Koenig:2011lmm,Konig:2013cia,Vanasse:2014kxa,Konig:2014ufa,Konig:2016iny},  charge radii~\cite{Vanasse:2015fph,Kirscher:2017fqc,Konig:2019xxk},  magnetic moments~\cite{Vanasse:2017kgh,Kirscher:2017fqc,De-Leon:2020glu}, triton beta decay~\cite{DeLeon:2016wyu}, and  polarizabilities~\cite{Kirscher:2017fqc}.  It was also used to investigate $nd\to t\gamma$~\cite{Sadeghi:2006fc,Arani:2014qsa}. However, Refs.~\cite{Sadeghi:2006fc,Arani:2014qsa} used the partial resummation technique \cite{Gabbiani:1999yv, Bedaque:2002yg}, which includes an infinite subset of higher order diagrams in addition to the contributions at the desired order, and is therefore not strictly perturbative.  
Further, Refs.~\cite{Sadeghi:2006fc,Arani:2014qsa} lack an additional three-nucleon current counterterm  that we find necessary for RG invariance in the strictly perturbative approach taken in this work. 

Here we describe the relative importance of multi-nucleon currents by finding the order at which they appear in \eftnopi, as well as provide the first fully perturbative calculation of $nd\to t\gamma$ in \EFT using the methods  outlined in Refs.~\cite{Vanasse:2014sva,Vanasse:2015fph}. At next-to-next-to-leading order (\NNLO) we identify a new three-nucleon magnetic moment counterterm necessary for the RG invariance of both $nd\to t\gamma$ and the triton magnetic moment.

In the Wigner-SU(4)~\cite{Wigner:1936dx} limit the nucleon-nucleon (\NN) scattering lengths (as well as other parameters in the effective range expansion) in the ${}^3S_1$ and ${}^1S_0$ channels are equal.\footnote{ An observable is Wigner-SU(4)-symmetric (-antisymmetric) if upon interchanging spin and isospin degrees of freedom the quantity remains unchanged (changes sign).}  Although the physical scattering lengths, $1/a^{{}^3S_1}\approx 45$~MeV and $1/a^{{}^1S_0}\approx -8.3$~MeV, are far  from the Wigner-SU(4) limit, the Wigner-SU(4) breaking is parameterized by~\cite{Mehen:1999qs} 
\begin{equation}
    \delta=\frac{1}{2}\left(1/a^{{}^3S_1}-1/a^{{}^1S_0}\right)\approx 27\mathrm{\,MeV} \ \ ,
    \label{eq:deltaintro}
\end{equation}
which is much smaller than a momentum scale associated with three-body binding, $\kappa^*$.  Therefore, the ratio $\delta/\kappa^*$ serves as an appropriate expansion parameter for the three-nucleon bound state portion of the $nd\to t\gamma$ process. Vanasse and Phillips showed \cite{Vanasse:2016umz} that the triton and $\jjvHe$ charge radii were reproduced well in this expansion.  In addition, they demonstrated that the Wigner-SU(4)-antisymmetric contribution to the three-nucleon vertex function is $\approx$10\% of the leading Wigner-SU(4)-symmetric contribution to the three-nucleon vertex function.   
In this paper, we investigate the impact of Wigner-SU(4) symmetry on $nd\to t\gamma $ and use it to understand the relatively large (for some parameter fits) NNLO correction to  the cross section as compared to the LO contribution for this process in \eftnopi.
  The triton wavefunction is nearly an eigenstate of the one-nucleon magnetic current operator, and orthogonal to the $nd$ scattering state in the absence of energy-dependent three-body forces.  Therefore,  the matrix element of the one-nucleon magnetic current operator between the triton wavefunction and the $nd$ scattering state is nearly zero, leading to a small contribution from the one-nucleon current to cold $nd$ capture. In this work we explain the size of this matrix element in the context of Wigner-SU(4) symmetry and its breaking. 

This paper is organized as follows. Section~\ref{sec:two} provides the Lagrangians used,  introduces relevant notation, and demonstrates how the two-nucleon propagators and three-nucleon vertex functions are constructed.  In Sec.~\ref{sec:cap} we show the integral equations that describe the $nd$ capture amplitude. Section~\ref{sec:mag} presents the triton magnetic moment calculation to \NNLO and Sec.~\ref{sec:zero} gives the integral equation for the $nd$ capture amplitude in the zero-recoil limit.  The fitting of the $L_1$ and $L_2$ LECs is discussed in Sec.~\ref{sec:L1L2} and consequences of Wigner-SU(4) symmetry are shown in Sec.~\ref{sec:Wigner}.  Observables of interest are derived in Sec.~\ref{sec:obs}, results are discussed in Sec.~\ref{sec:results}, and a summary is given in Sec.~\ref{sec:conclusion}.  The appendices contain details of the zero-recoil limit calculation, the impact of Wigner-SU(4) symmetry on one-nucleon currents, the matching between the nucleon and dibaryon auxiliary field formalisms,  the \eftnopi-Wigner-SU(4) dual expansion~\cite{Vanasse:2016umz} for $nd$ capture and the three-body magnetic moments, and error analysis. 

\section{\label{sec:two} Pionless EFT in the Auxiliary field formalism}

The Lagrangian for \EFT in the two-nucleon sector up to and including \NNLO corrections is given by
\begin{align}
 \label{eq:two}   \mathcal{L}_2=&\hat{N}^{\dagger}\left(i\partial_0+\frac{\nabla^2}{2M_N}\right)\hat{N}+\hat{t}_i^{\dagger}\left[\Delta_t-\sum_{n=0}^{1}c_{0t}^{(n)}\left(i\partial_0+\frac{\nabla^2}{4M_N}+\frac{\gamma_t^2}{M_N}\right)\right]\hat{t}_i\\\nonumber
    &+\hat{s}_a^{\dagger}\left[\Delta_s-\sum_{n=0}^{1}c_{0s}^{(n)}\left(i\partial_0+\frac{\nabla^2}{4M_N}+\frac{\gamma_s^2}{M_N}\right)\right]\hat{s}_a\\\nonumber
    &+y\left[\hat{t}_i^{\dagger}\hat{N}^TP_i\hat{N}+\hat{s}_a^{\dagger}\hat{N}^T\bar{P}_a\hat{N}+\mathrm{H.c.}\right],
\end{align}
where $\hat{N}$ is a nucleon field, $\hat{t}_i$ is a deuteron field, and $\hat{s}_a$ is a dibaryon field representing two nucleons in the $^1S_0$ channel. $P_i=\frac{1}{\sqrt{8}}\sigma_2\sigma_i\tau_2$ ($\bar{P}_a=\frac{1}{\sqrt{8}}\sigma_2\tau_2\tau_a$) projects out nucleons in the $^3S_1$ ($^1S_0$) state.  We use the $Z$ parametrization~\cite{Phillips:1999hh,Griesshammer:2004pe}, which fits the parameters in Eq.~\eqref{eq:two} to the $^3S_1$ bound state pole, the $^1S_0$ virtual bound state pole, and the residues of each pole,  yielding~\cite{Griesshammer:2004pe}
\begin{align}
    y^2=\frac{4\pi}{M_N}\quad,\quad&\Delta_t=\gamma_t-\mu\quad,\quad c_{0t}^{(n)}=(-1)^n\frac{M_N}{2\gamma_t}(Z_t-1)^{n+1}\\\nonumber
    &\Delta_s=\gamma_s-\mu\quad,\quad c_{0s}^{(n)}=(-1)^n\frac{M_N}{2\gamma_s}(Z_s-1)^{n+1},
\end{align}
where $\gamma_t=45.7025$~MeV ($\gamma_s=-7.890$~MeV) is the deuteron ($^1S_0$ virtual state) binding momentum. $Z_t=1.6908$ ($Z_s=0.9015$) is the residue about the deuteron ($^1S_0$ virtual state) pole.  $\mu$ is a scale introduced by using dimensional regularization with the power divergence subtraction scheme ~\cite{Kaplan:1998tg}. All physical observables must be independent of $\mu$.

At LO in \EFT a three-body force is necessary~\cite{Bedaque:1998kg,Bedaque:1998km,Bedaque:1999ve}, and it receives corrections at higher orders~\cite{Hammer:2001gh,Bedaque:2002yg}.
 An energy-dependent three-body force, $H_2(\Lambda)$, is necessary at \NNLO~\cite{Bedaque:2002yg,Ji:2012nj}.  The LO three-body force and corrections up to and including \NNLO can be written as
\begin{align}
    \label{eq:3BF}
    &\mathcal{L}_3=\frac{\pi}{3}(H_{\mathrm{LO}}(\Lambda)+H_{\mathrm{NLO}}(\Lambda)+H_{\mathrm{NNLO}}(\Lambda))\\\nonumber
    &\hspace{4cm}\times\left[\hat{N}^\dagger(\vec{t}\cdot\vec{\boldsymbol{\sigma}})^\dagger-\hat{N}^\dagger(\vec{s}\cdot\vec{\boldsymbol{\tau}})^\dagger\right]\left[(\vec{t}\cdot\vec{\boldsymbol{\sigma}})\hat{N}-(\vec{s}\cdot\vec{\boldsymbol{\tau}})\hat{N}\right]\\\nonumber
    &+\frac{\pi}{3} M_N H_{2}(\Lambda){\frac{4}{3}}\left[\hat{N}^\dagger(\vec{t}\cdot\vec{\boldsymbol{\sigma}})^\dagger-\hat{N}^\dagger(\vec{s}\cdot\vec{\boldsymbol{\tau}})^\dagger\right]\left(i\vec{\partial}_0-E_B\right)\left[(\vec{t}\cdot\vec{\boldsymbol{\sigma}})\hat{N}-(\vec{s}\cdot\vec{\boldsymbol{\tau}})\hat{N}\right],
\end{align}
where the LO three-body force $H_{\mathrm{LO}}(\Lambda)$, its NLO correction $H_{\mathrm{NLO}}(\Lambda)$, its \NNLO correction $H_{\mathrm{NNLO}}(\Lambda)$, and the \NNLO energy-dependent counterterm $H_2(\Lambda)$, are functions of $\Lambda$, the cutoff used in sharp cutoff regularization. The cutoff dependence of $H_{\mathrm{LO}}(\Lambda)$, $H_{\mathrm{NLO}}(\Lambda)$, and $H_{\mathrm{\NNLO}}(\Lambda)$ is chosen so that three-nucleon observables converge as $\Lambda\to\infty$. This can be accomplished by either fitting the LO three-body force to the triton binding energy, $E_B=-8.48$~MeV, or to the $nd$ scattering length in the doublet $S$-wave channel, $a_{nd}=0.65\pm0.04$~fm~\cite{Dilg:1971gqb}. Here we fit $H_{\mathrm{LO}}(\Lambda)$ and its higher order corrections to the triton binding energy to avoid perturbative corrections to the triton binding energy at each order. Including perturbative corrections to the triton binding energy makes calculating cold $nd$ capture more difficult.  We fit $H_2(\Lambda)$ to $a_{nd}$. While dimensional regularization is used in the two-nucleon sector, a sharp cutoff regularization is used in the three-nucleon sector.  This is equivalent~\cite{Bedaque:1999ve} to using two different sharp cutoffs $\Lambda'$ and $\Lambda$ in the two- and three-nucleon sectors, respectively, and taking $\Lambda'\to\infty$ before taking $\Lambda\to\infty$.  The difference between the treatment used here and setting $\Lambda'=\Lambda$ is higher order and absorbed in the three-body force order-by-order.

Equation~\eqref{eq:3BF} is a useful parametrization of the three-body force for $nd$ scattering, whereas an equivalent but more useful  parametrization for three-nucleon bound states involves the introduction of a three-nucleon auxiliary field and is given by the Lagrangian 
\begin{equation}
    \label{eq:3BFb}
    \mathcal{L}_3=\hat{\psi}^{\dagger}\left[\Omega-h_2(\Lambda)\left(i\partial_0+\frac{\nabla^2}{6M_N}-E_B\right)\right]\hat{\psi}+\sum_{n=0}^2\omega_0^{(n)}\left[\hat{\psi}^{\dagger}\sigma_i\hat{N}\hat{t}_i-\hat{\psi}^{\dagger}\tau_a\hat{N}\hat{s}_a+\mathrm{H.c.}\right],
\end{equation}
where $\hat{\psi}$ is a three-nucleon isodoublet of the triton and $\jjvHe$.  Parameters in Lagrangians~\eqref{eq:3BF} and~\eqref{eq:3BFb} can be matched~\cite{Vanasse:2015fph,Bedaque:2002yg}.

The LO magnetic interaction between photons and nuclear states is
\begin{equation}
\mathcal{L}^{mag}_{1,0} = 
\frac{e}{2M_N} \hat{N}^\dagger\left(\kappa_0 +  \kappa_1 \tau_3 \right) \vec{\boldsymbol{\sigma}} \cdot \vect{\mathrm{B}} \  \hat{N},
\label{eq:maglag}
\end{equation}
where $ \kappa_0= 0.43990$ ($\kappa_1=2.35295$) is the dimensionless isoscalar (isovector) nucleon magnetic moment. At NLO in \EFT there are two-nucleon-one-magnetic-photon contact interactions given by
\begin{equation}
    \mathcal{L}_{2,1}^{mag}=\left(e\frac{L_1^{(0)}}{2}\hat{t}^{j\dagger}\hat{s}_3\mathbf{B}_j+\mathrm{H.c.}\right)-e\frac{L_2^{(0)}}{2}i\epsilon^{ijk}\hat{t}_i^{\dagger}\hat{t}_j\mathbf{B}_k.
    \label{eq:MEClagNLO}
\end{equation}
At \NNLO there are corrections to Eq.~\eqref{eq:MEClagNLO}  given by
\begin{equation}
    \mathcal{L}_{2,2}^{mag}=\left(e\frac{L_1^{(1)}}{2}\hat{t}^{j\dagger}\hat{s}_3\mathbf{B}_j+\mathrm{H.c.}\right)-e\frac{L_2^{(1)}}{2}i\epsilon^{ijk}\hat{t}_i^{\dagger}\hat{t}_j\mathbf{B}_k,
    \label{eq:MEClagNNLO}
\end{equation}
where the first subscript in each $\mathcal{L}^{mag}$ gives the number of nucleons involved and the second refers to the order at which the terms occur in \eftnopi. Finally, to maintain RG invariance at \NNLO requires a three-nucleon magnetic moment term
\begin{equation}
\mathcal{L}^{mag}_{3,2} = 
\frac{e}{2M_N} \hat{\psi}^\dagger\left(\widetilde{\kappa}_0(\Lambda) +  \widetilde{\kappa}_1(\Lambda) \tau_3 \right) \vec{\boldsymbol{\sigma}} \cdot \vect{\mathrm{B}} \  \hat{\psi}. 
\label{eq:maglag3}
\end{equation}
We fit the combination $\widetilde{\kappa}_0(\Lambda)-\widetilde{\kappa}_1(\Lambda)$ to the triton magnetic moment.
\subsection{\label{sec:3B}Two-Body System}

The $^3S_1$ ($^1S_0$) two-nucleon channel possesses a shallow (shallow virtual) bound state relative to the cutoff of \eftnopi.  To reproduce the shallow state poles an infinite number of diagrams must be summed at LO in \EFT, which can be carried out analytically via a geometric series (see, for example, Ref.~\cite{Kaplan:1998tg}). Higher-order range corrections are then included perturbatively.  This procedure gives the following dibaryon propagators up to and including \NNLO corrections in the $Z$ parametrization
\begin{align}
    \label{eq:dibprop}
    iD_{\{t,s\}}(E,p)=&\frac{i}{\gamma_{\{t,s\}}-\sqrt{\frac{1}{4}p^2-M_NE-i\epsilon}}\\\nonumber
    &\left[\underbrace{1\vphantom{\frac{Z_{\{t,s\}}-1}{2\gamma_{\{t,s\}}}\left(\gamma_{\{t,s\}}+\sqrt{\frac{1}{4}p^2-M_NE-i\epsilon}\right)}}_{\mathrm{LO}}+\underbrace{\frac{Z_{\{t,s\}}-1}{2\gamma_{\{t,s\}}}\left(\gamma_{\{t,s\}}+\sqrt{\frac{1}{4}p^2-M_NE-i\epsilon}\,\right)}_{\mathrm{NLO}}\right.\\\nonumber
    &\left.+\underbrace{\left(\frac{Z_{\{t,s\}}-1}{2\gamma_{\{t,s\}}}\right)^2\left(\frac{1}{4}p^2-M_NE-\gamma^2_{\{t,s\}}\right)}_{\mathrm{NNLO}}+\cdots\right],
\end{align}
 with the $t$ ($s$) subscript labeling the spin-triplet (spin-singlet) dibaryon propagator. The deuteron wavefunction renormalization is given by the residue about the deuteron pole, yielding
\begin{equation}\label{d.wave.rn}
    Z_{d}=\frac{2\gamma_t}{M_N}\left[\underbrace{1\vphantom{(Z_t-1)}}_{\mathrm{LO}}+\underbrace{(Z_t-1)}_{\mathrm{NLO}}\right].
\end{equation}
In the $Z$ parametrization $Z_{d}$ is exact at NLO  by construction. Taking the square root of $Z_d$ and expanding it perturbatively gives.
\begin{equation}\label{eq:drenorm}
    \sqrt{Z_{d}}=\sqrt{\frac{2\gamma_t}{M_N}}\left[\underbrace{1\vphantom{\frac{1}{2}(Z_t-1)}}_{\mathrm{LO}}+\underbrace{\frac{1}{2}(Z_t-1)}_{\mathrm{NLO}}-\underbrace{\frac{1}{8}(Z_t-1)^2}_{\mathrm{NNLO}}+\cdots\right].
\end{equation}

\subsection{Three-Body System}

To determine properties of the three-nucleon system the three-nucleon wavefunction, or equivalently the three-nucleon  vertex function, is required.  At LO in \EFT the three-nucleon vertex function is given by an infinite sum of diagrams, which can be solved via the integral equation 
\begin{equation}
    \label{eq:LOvertex}
    \Gb_0(E,p)=\oneb+\Kb(q,p,E)\otimes_q\Gb_0(E,q),
\end{equation}
where  the inhomogeneous term $\oneb$ is a vector in cluster configuration (c.c.) space~\cite{Griesshammer:2004pe} given by
\begin{equation}
    \label{eq:oneb}
    \oneb=\left(\!\!\begin{array}{r}
    1\\
    -1\\
    \end{array}\!\!\right).
\end{equation}
The kernel of the integral equation is given by
\begin{equation}
    \Kb(q,p,E)=\mathbf{R}_0(q,p,E)\Db\left(E-\frac{q^2}{2M_N},\vectn{q}\right),
\end{equation}
where 
\begin{equation}
    \mathbf{R}_0(q,p,E)=-\frac{2\pi}{qp}Q_0\left(\frac{q^2+p^2-M_NE-i\epsilon}{qp}\right)\left(\!\!\begin{array}{rr}
    1 & -3 \\
    -3 & 1 
    \end{array}\!\right),
\end{equation}
and
\begin{equation}
    \Db(E,\vectn{q})=\left(\!\!\begin{array}{cc}
    D_t(E,q) & 0 \\
    0 & D_s(E,q) 
    \end{array}\!\!\right)
\end{equation}
are both matrices in c.c.~space. $\Db(E,\vectn{q})$ is constructed from the LO spin-triplet and spin-singlet dibaryon propagators in Eq.~\eqref{eq:dibprop}.  $Q_L(a)$ is a Legendre function of the second kind defined by\footnote{This differs from the conventional definition of Legendre functions of the second kind by a phase of $(-1)^L$.}
\begin{equation}
    Q_L(a)=\frac{1}{2}\int_{-1}^1dx\frac{P_L(x)}{x+a},
\end{equation}
where $P_L(x)$ are the Legendre polynomials.  The symbol $\otimes_q$ defines the operation
\begin{equation}
    A(q)\otimes_qB(q)=\frac{1}{2\pi^2}\int_0^{\Lambda}dqq^2A(q)B(q).
\end{equation}

The NLO correction to the three-nucleon vertex function is given by the integral equation~\cite{Vanasse:2015fph}
\begin{equation}
    \label{eq:NLOvertex}
    \Gb_1(E,p)=\mathbf{R}_1\left(E-\frac{p^2}{2M_N},\vectn{p}\right)\Gb_0(E,p)+\Kb(q,p,E)\otimes_q\Gb_1(E,q),
\end{equation}
where 
\begin{equation}
    \mathbf{R}_1(p_0,\vectn{p})=\left(\begin{array}{cc}
    \frac{Z_t-1}{2\gamma_t}\left(\gamma_t+\sqrt{\frac{1}{4}p^2-M_Np_0-i\epsilon}\,\right) & 0\\
    0 & \frac{Z_s-1}{2\gamma_s}\left(\gamma_s+\sqrt{\frac{1}{4}p^2-M_Np_0-i\epsilon}\,\right)
    \end{array}\right)
\end{equation}
is a matrix in c.c.~space.  At NNLO the correction to the three-nucleon vertex function is~\cite{Vanasse:2015fph}
\begin{align}
    \label{eq:NNLOvertex}
    &\Gb_2(E,p)=\mathbf{R}_1\left(E-\frac{p^2}{2M_N},\vectn{p}\right)\Gb_1(E,p)\\\nonumber
    &\hspace{2cm}+\mathbf{R}_2\left(E-\frac{p^2}{2M_N},\vectn{p}\right)\Gb_0(E,p)+\Kb(q,p,E)\otimes_q\Gb_2(E,q),
\end{align}
where $\mathbf{R}_2(p_0,\vectn{p})$ is a matrix in c.c.~space given by
\begin{equation}
    \mathbf{R}_2(p_0,\vectn{p})=-\left(\!\!\begin{array}{cc}
    Z_t-1 & 0\\
    0 & Z_s-1
    \end{array}\!\!\right)\mathbf{R}_1(p_0,\vectn{p}).
\end{equation}
From the three-nucleon vertex function the three-nucleon propagator in the three-nucleon rest frame up to and including NNLO corrections is~\cite{Vanasse:2015fph}
\begin{align}
    \label{eq:D3}
    &\Delta_3(E)=\frac{1}{\Omega}\frac{1}{1-H_{\mathrm{LO}}\Sigma_0(E)}\left[\underbrace{\vphantom{\frac{H_{\mathrm{LO}}\Sigma_1(E)+H_{\mathrm{NLO}}\Sigma_0(E)}{1-H_{\mathrm{LO}}\Sigma_0(E)}}1}_{\mathrm{LO}}+\underbrace{\frac{H_{\mathrm{LO}}\Sigma_1(E)+H_{\mathrm{NLO}}\Sigma_0(E)}{1-H_{\mathrm{LO}}\Sigma_0(E)}}_{\mathrm{NLO}}\right.\\\nonumber
    &+\underbrace{\left(\frac{H_{\mathrm{LO}}\Sigma_1(E)+H_{\mathrm{NLO}}\Sigma_0(E)}{1-H_{\mathrm{LO}}\Sigma_0(E)}\right)^2}_{\mathrm{\NNLO}}\\\nonumber
    &\left.+\underbrace{\frac{H_{\mathrm{LO}}\Sigma_2(E)+H_{\mathrm{NLO}}\Sigma_1(E)+H_{\mathrm{NNLO}}\Sigma_0(E)+\frac{4}{3}M_N(E-E_B)H_2/H_{\mathrm{LO}}}{1-H_{\mathrm{LO}}\Sigma_0(E)}}_{\mathrm{\NNLO}}\right] , 
\end{align}
where the functions $\Sigma_n(E)$ are defined by 
\begin{equation}
    \label{eq:Sigmadef}
    \Sigma_n(E)=-\pi\Db\left(E-\frac{q^2}{2M_N},\vectn{q}\right)\left(\begin{array}{rr}
    1 & 0\\
    0 & -1
    \end{array}\right)\otimes_q\Gb_{n}(E,q).
\end{equation}
The LO three-body force $H_{\mathrm{LO}}$ is fit by ensuring the three-nucleon propagator has a pole at the triton binding energy, yielding the condition
\begin{equation}
    \label{eq:HLOdef}
    H_\mathrm{LO}=\frac{1}{\Sigma_0(E_B)}.
\end{equation}
$H_{\mathrm{NLO}}$ and $H_{\mathrm{\NNLO}}$ are fit by ensuring that the double pole occurring in Eq.~\eqref{eq:D3} vanishes at each order,
yielding 
\begin{equation}
    \label{eq:HNLOdef}
    H_{\mathrm{NLO}}=-\frac{\Sigma_1(E_B)}{\left(\Sigma_0(E_B)\right)^2}\quad,\quad H_{\mathrm{NNLO}}=\frac{\left(\Sigma_1(E_B)\right)^2-\Sigma_2(E_B)\Sigma_0(E_B)}{\left(\Sigma_0(E_B)\right)^3}.
\end{equation}
The three-nucleon wavefunction renormalization is given by the square root of the residue of the three-nucleon propagator about the triton pole, yielding~\cite{Vanasse:2015fph} 

\begin{align}
    &\sqrt{Z_\psi}=\sqrt{-\frac{1}{\Omega}\frac{1}{H_{\mathrm{LO}}\Sigma_0'}}\left[\underbrace{\vphantom{\frac{1}{2}\left(\frac{\Sigma_1'}{\Sigma_0'}-\frac{\Sigma_1}{\Sigma_0}\right)}1}_{\mathrm{LO}}-\underbrace{\frac{1}{2}\left(\frac{\Sigma_1'}{\Sigma_0'}-\frac{\Sigma_1}{\Sigma_0}\right)}_{\mathrm{NLO}}\right.\\\nonumber
    &\left.\underbrace{-\frac{1}{2}\left[\frac{\Sigma_2'}{\Sigma_0'}+\frac{1}{2}\frac{\Sigma_1}{\Sigma_0}\frac{\Sigma_1'}{\Sigma_0'}-\frac{\Sigma_2}{\Sigma_0}+\frac{1}{4}\left(\frac{\Sigma_1}{\Sigma_0}\right)^2-\frac{3}{4}\left(\frac{\Sigma_1'}{\Sigma_0'}\right)^2+\frac{4}{3}M_NH_2\frac{\Sigma_0^2}{\Sigma_0'}\right]}_{\mathrm{\NNLO}}+\cdots\right],
\end{align}
where the functions $\Sigma_n(E)$ and $\Sigma_n'(E)$ are understood to be evaluated at $E=E_B$.  To find the wavefunction renormalization for the vertex function additional factors of $\omega_{0}^{(n)}$ must be included, giving~\cite{Vanasse:2015fph}
\begin{align}
    \label{eq:tritonrenorm}
    &\sqrt{Z_t}=\sqrt{\frac{\pi}{\Sigma_0'(E_B)}}\left[\underbrace{1\vphantom{\frac{1}{2}\frac{\Sigma_1'(E_B)}{\Sigma_0'(E_B)}}}_{\mathrm{LO}}-\underbrace{\frac{1}{2}\frac{\Sigma_1'(E_B)}{\Sigma_0'(E_B)}}_{\mathrm{NLO}}\right.\\\nonumber
    &\hspace{4cm}\left.+\underbrace{\frac{3}{8}\left(\frac{\Sigma_1'(E_B)}{\Sigma_0'(E_B)}\right)^2-\frac{1}{2}\frac{\Sigma_2'(E_B)}{\Sigma_0'(E_B)}-\frac{2}{3}M_NH_2\frac{\Sigma_0(E_B)^2}{\Sigma_0'(E_B)^2}}_{\mathrm{NNLO}}+\cdots\right],
\end{align}
where we have used the matching conditions between Eqs.~\eqref{eq:3BF} and \eqref{eq:3BFb} for $\omega^{(n)}_0$ and then used Eq.~\eqref{eq:HNLOdef}.
  Further details on this procedure and the fitting of $H_2$ are explained in Ref.~\cite{Vanasse:2015fph}.

\section{\label{sec:cap}Capture reaction}

\begin{figure}[hbt]
    \centering
    \includegraphics[width=90mm]{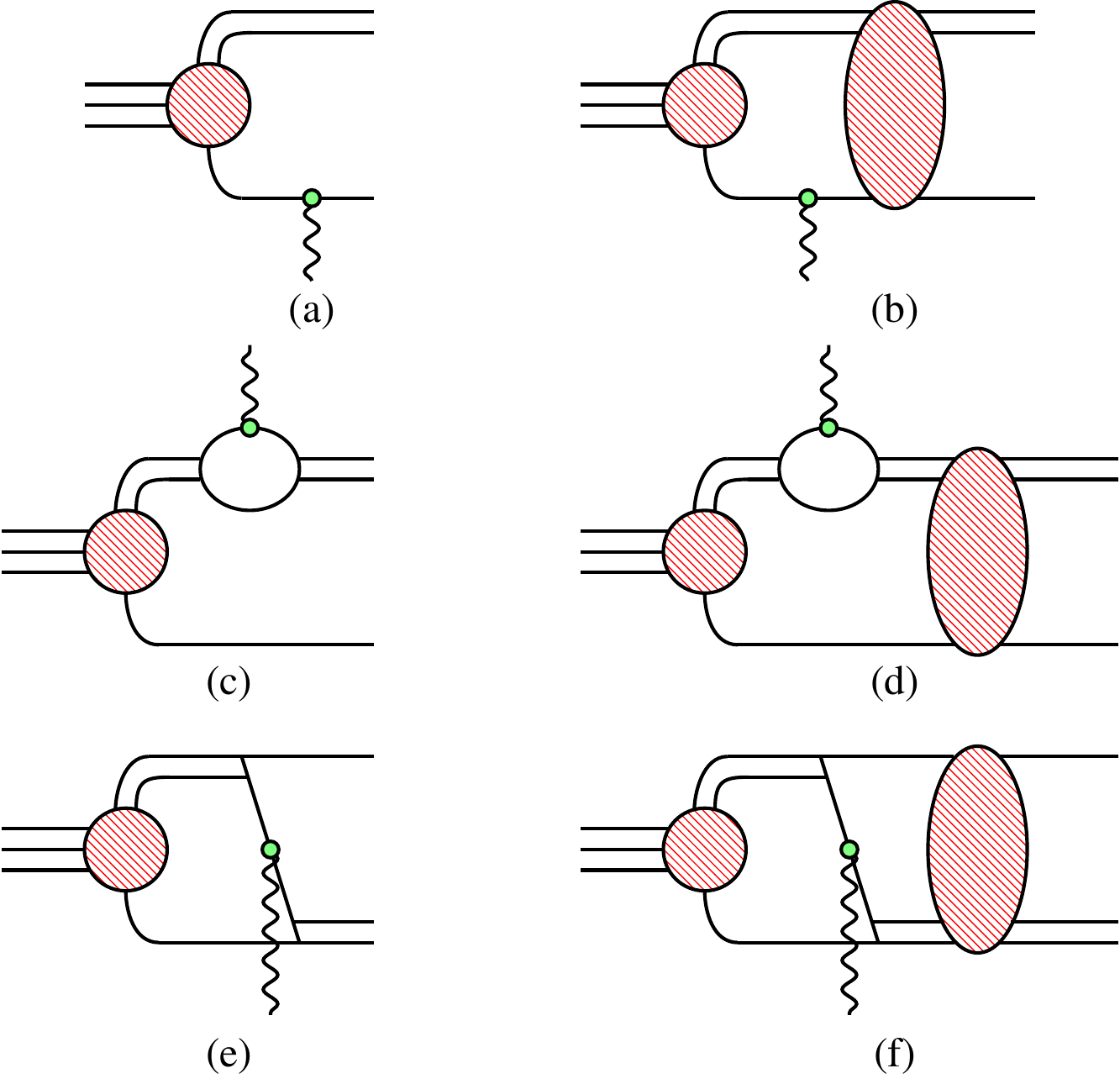}
    \caption{Diagrams for the LO two-body triton photo-disintegration amplitude.  Single lines are nucleon propagators,  double lines are dibaryon propagators (either ${}^3S_1$ or $^1S_0$ channel), the shaded circle is the LO vertex functon, and the shaded oval is the LO $nd$ scattering amplitude.  Wavy lines are photons and the small green dot is the magnetic photon interaction vertex given in Eq.~\eqref{eq:maglag}.
    }
    \label{fig:photodis}
\end{figure}

By time-reversal symmetry the amplitude for $nd$ capture is equivalent to the amplitude for two-body triton photo-disintegration.  The two-body triton photo-disintegration amplitude at LO is given by the sum of diagrams in Fig.~\ref{fig:photodis}, where the shaded oval on the right represents an insertion of the $nd$ scattering amplitude, the shaded circle an insertion of the triton vertex function, and the photons are magnetically coupled via Eq.~\eqref{eq:maglag}.\footnote{For cold $nd$ capture we do not consider the contribution from the electric dipole (E1) transition by electrically coupled photons as they come with powers of nucleon momentum and are thus highly suppressed at low energies.  However, the electric quadrupole (E2) moment does have a contribution at \NNLO from the $SD$ mixing term.  Based on previous potential model calculations~\cite{Marcucci:2005zc, Viviani:1996ui} the contribution from E2 is expected to be at most 2\%, less than our theoretical error at \NNLO, which is $\approx 4\%$. }

\begin{figure}[hbt]
    \centering
    \includegraphics[width=100mm]{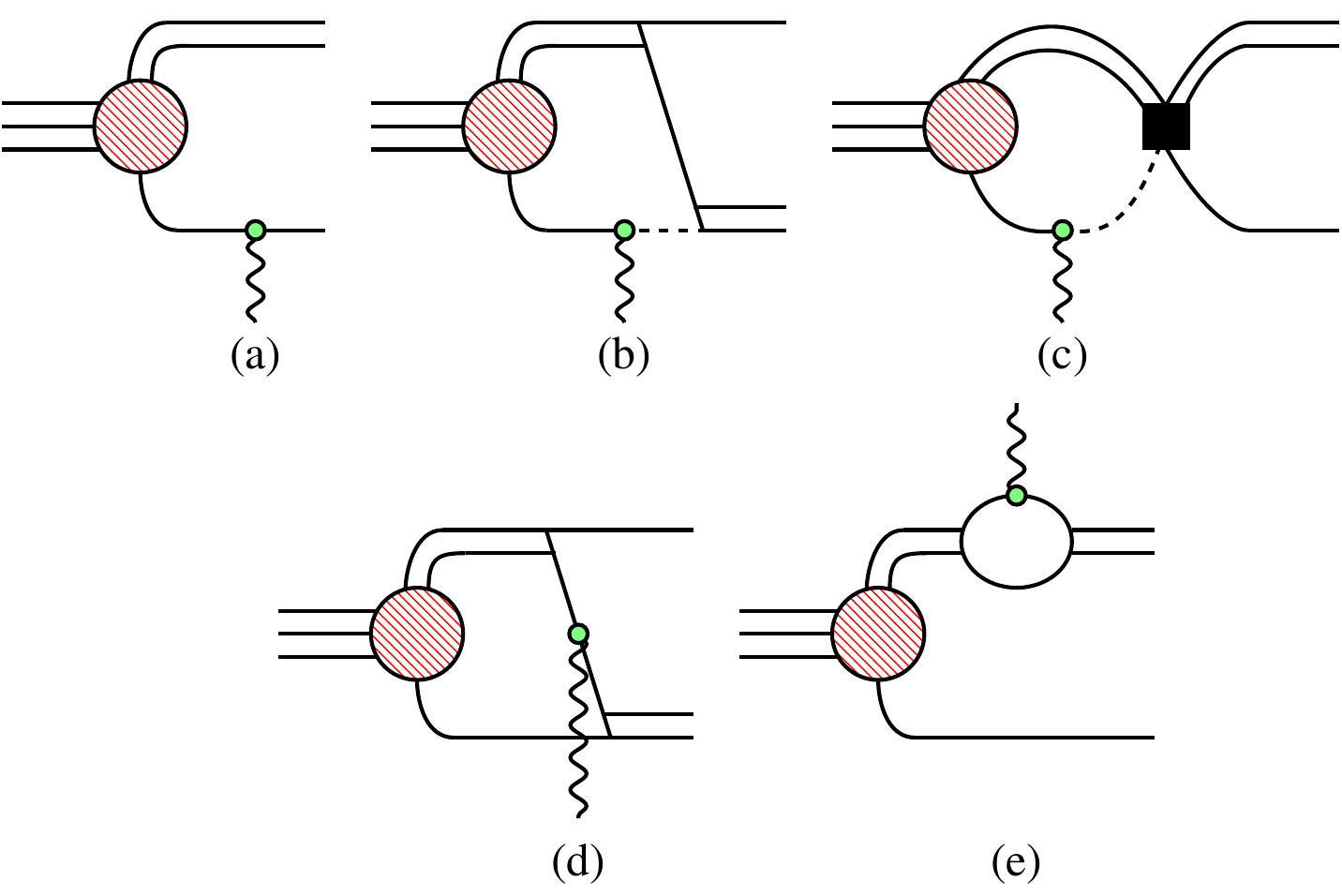}
    \caption{Diagrams for the LO inhomogeneous term of Eq.~\eqref{eq:TampLO}. Dashed single lines are nucleon propagators whose poles are not picked up in the energy loop integral.  The solid black box is the LO three-body force $H_{\mathrm{LO}}$ and all other symbols are defined in Fig~\ref{fig:photodis}. }
    \label{fig:inhomLO}
\end{figure}

  In principle the two-body triton photo-disintegration amplitude can be calculated from the Feynman diagrams in Fig.~\ref{fig:photodis}, where the $nd$ scattering amplitude and the triton vertex function, both calculated from integral equations, are used.
  However, we take a different approach, as in Ref.~\cite{Vanasse:2013sda}, in which the final state $nd$ scattering amplitude is included directly through an integral equation. Without wavefunction renormalization factors, the contribution to the LO two-body triton photo-disintegration amplitude is given by the coupled set of integral equations in c.c.~space 
\begin{align}
    \label{eq:TampLO}
    {\Tb^{[0]}}_{L'S',LS}^{J,J'}(p,k)=&\frac{e}{2M_N}{\Bb^{[0]}}_{L'S',LS}^{J,J'}(p,k)\\\nonumber
    &+\sum_{L'',S''}\Kb^{J'}_{L'S',L''S''}(q,p,E)\mathbf{D}\left(E-\frac{q^2}{2M_N},\vectn{q}\right)\otimes_q {\Tb^{[0]}}_{L''S'',LS}^{J,J'}(q,k),
\end{align}
where $k_0$ and $k$ are the photon energy and magnitude of the photon momentum, respectively, in the triton rest frame, and $p$ is the magnitude of the neutron momentum in the center of mass (c.m.) frame. $E = 3p^2/(4M_N) -\gamma_t^2/M_N$ is the total energy in the c.m.~frame. 
$L$ and $S$ ($L'$ and $S'$) are the quantum numbers for the total orbital and spin angular momentum of the incoming (outgoing) nuclear state, respectively, while $J$ ($J'$) is the quantum number for the total angular momentum of the incoming (outgoing) nuclear state.  Since the photon injects angular momentum, $J\neq J'$. However, after the photon injects its angular momentum the total angular momentum is fixed and the final state scattering amplitude has total angular momentum $J'$.  The kernel is the same as that for $nd$ scattering (cf. Ref.~\cite{Bedaque:1998mb,Bedaque:1999ve}) and is a matrix in c.c.~space given by
\begin{align}
    \label{eq:scattkernel}
    &\Kb^J_{L'S',LS}(q,p,E)=\\\nonumber
    &\delta_{LL'}\delta_{SS'}\left\{
    \begin{array}{lcc}
          -\frac{2\pi}{qp}Q_L\left(\frac{q^2+p^2-M_NE-i\epsilon}{qp}\right)\left(\!\!\begin{array}{rr}
         1 & -3\\
         -3 & 1
         \end{array}\!\right)-\pi H_{\mathrm{LO}}\delta_{L0}\left(\!\!\begin{array}{rr}
         1 & -1\\
         -1 & 1
         \end{array}\!\right)
         & ,& S=\frac{1}{2}\\[7mm]
         \phantom{-}\frac{4\pi}{qp}Q_L\left(\frac{q^2+p^2-M_NE-i\epsilon}{qp}\right)\left(\begin{array}{rr}
         1 & 0\\
         0 & 0
         \end{array}\right)
         & ,& S=\frac{3}{2}
    \end{array}
    \right. 
\end{align}
 for both the spin-quartet and -doublet channels, but the LO three-body force $H_{\mathrm{LO}}$ only appears in the \DS channel.   The inhomogeneous term ${\Bb^{[0]}}_{L'S',LS}^{J,J'}(p,k)$ of Eq.~\eqref{eq:TampLO} is given by the sum of the diagrams in Fig.~\ref{fig:inhomLO},
\begin{equation}
    {\Bb^{[0]}}_{L'S',LS}^{J,J'}(p,k)=\sum_{x=a,b,c,d,e}{\Bb^{[0]}}_{L'S',LS}^{J,J',(x)}(p,k),
\end{equation}
and the action of the integral equation is to attach a final state scattering amplitude onto the inhomogeneous term. 
Diagram~\ref{fig:photodis}(b) picks up a pole from each of the two internal nucleon lines when integrating over the energy in the loop. Inserting diagram~\ref{fig:inhomLO}(a) into the integral equation, Eq.~\eqref{eq:TampLO}, gives the contribution of diagram~\ref{fig:photodis}(b) from the pole of the nucleon propagator to the right of the photon coupling.  In  diagram~\ref{fig:photodis}(b) the pole from the nucleon propagator to the left of the photon coupling causes the nucleon propagator to the right of the photon coupling to be off-shell. This off-shell nucleon propagates into the scattering amplitude. To compute the contribution from the scattering amplitude with its incoming nucleon off-shell we insert diagram~\ref{fig:inhomLO}(b) (as well as diagram~\ref{fig:inhomLO}(c) in the \DS channel) into the integral equation, Eq.~\eqref{eq:TampLO}.  In diagrams~\ref{fig:inhomLO}(b) and ~\ref{fig:inhomLO}(c) only the pole of the nucleon propagator to the left of the photon coupling is included when integrating over the energy in the loop.

\begin{figure}[hbt]
    \centering
    \includegraphics[width=100mm]{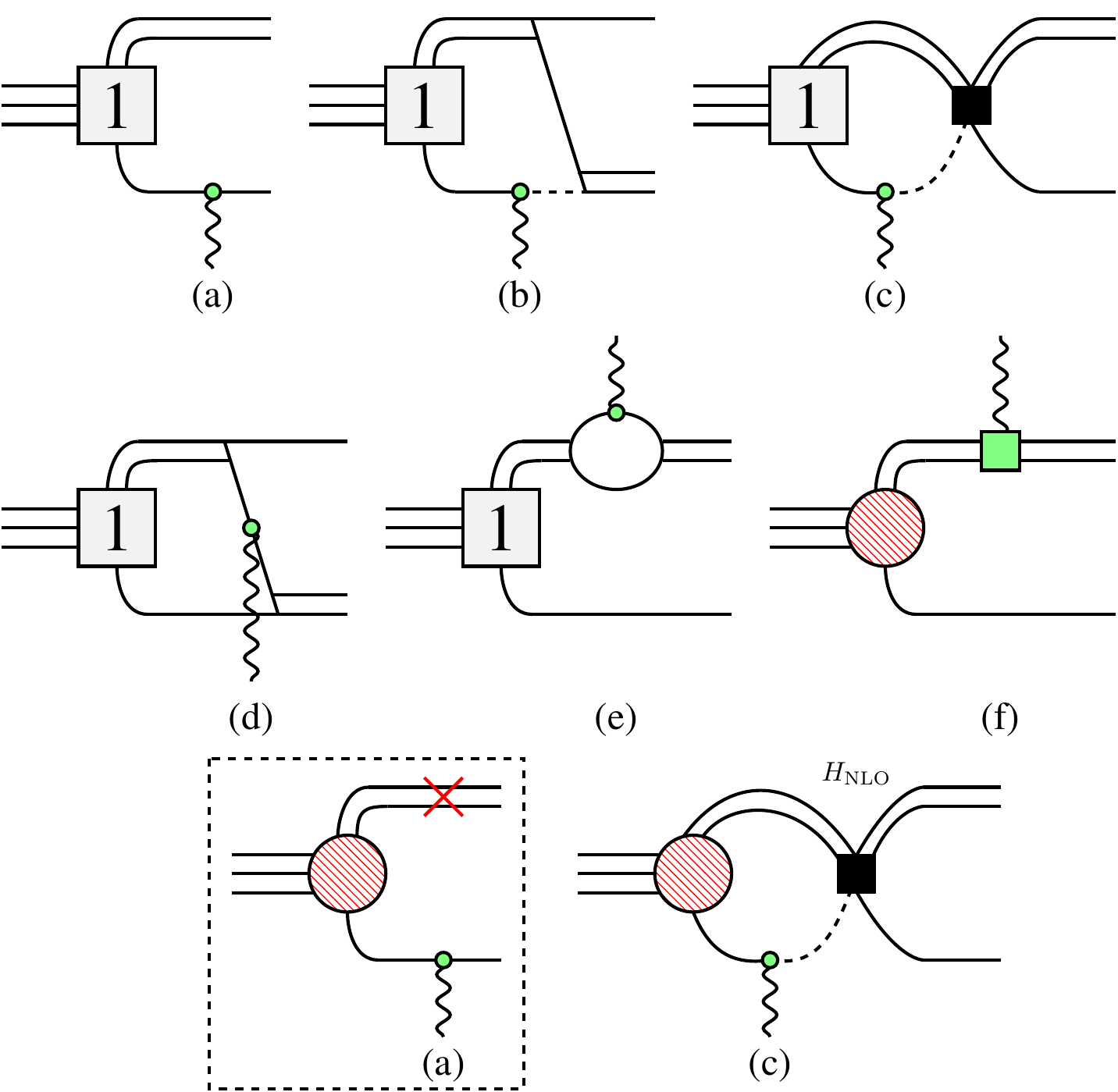}
    \caption{Diagrams for the inhomogeneous term ${\Bb^{[1]}}_{L'S',LS}^{J,J'}(p,k)$ of the NLO correction to the two-body triton photo-disintegration amplitude, Eq.~\eqref{eq:TampNLO}.  The box with a ``1'' is the NLO correction to the vertex function, the solid black box with the $H_{\mathrm{NLO}}$ label is the NLO correction to the LO three-body force, the solid green box attached to the photon is the interaction from  Eq.~\eqref{eq:MEClagNLO}, and the red  \textcolor{red}{$\times$} is a range correction.  All other symbols are the same as in Figs.~\ref{fig:photodis} and~\ref{fig:inhomLO}. The boxed diagram is subtracted to avoid double counting.  }
    \label{fig:inhomNLO}
\end{figure}

At NLO the correction to the two-body triton photo-disintegration amplitude can be found via the integral equation (complete wavefunction renormalizations are included in Eq.~\eqref{eq:T-renormed})
\begin{align}
    \label{eq:TampNLO}
    {\Tb^{[1]}}_{L'S',LS}^{J,J'}(p,k)=&\frac{e}{2M_N}{\Bb^{[1]}}_{L'S',LS}^{J,J'}(p,k)\\\nonumber
    &+\mathbf{R}_1\left(E-\frac{p^2}{2M_N},\vectn{p}\right){\Tb^{[0]}}_{L'S',LS}^{J,J'}(p,k)\\\nonumber
    &-\pi H_{\mathrm{NLO}}\delta_{L'0}\delta_{S'\frac{1}{2}}\left(\begin{array}{rr}
    1 & -1\\
    -1 & 1
    \end{array}\right)\Db\left(E-\frac{q^2}{2M_N},\vectn{q}\right)\otimes_q{\Tb^{[0]}}_{L'S',LS}^{J,J'}(q,k)
    \\\nonumber
    &+\sum_{L'',S''}\Kb^{J'}_{L'S',L''S''}(q,p,E)\mathbf{D}\left(E-\frac{q^2}{2M_N},\vectn{q}\right)\otimes_q {\Tb^{[1]}}_{L''S'',LS}^{J,J'}(q,k),
\end{align}
where the inhomogeneous term ${\Bb^{[1]}}_{L'S',LS}^{J,J'}(p,k)$ is given by the sum of diagrams in Fig.~\ref{fig:inhomNLO} 
\begin{equation}
    {\Bb^{[1]}}_{L'S',LS}^{J,J'}(p,k)=\sum_{x=a,b,c,d,e,f}{\Bb^{[1]}}_{L'S',LS}^{J,J',(x)}(p,k),
\end{equation}
and  the inhomogeneous term $\mathbf{R}_1{\Tb^{[0]}}_{L'S',LS}^{J,J'}(p,k)$ by diagram (\rom{1}) in Fig.~\ref{fig:inhomothers}.  
The third and last diagrams of Fig.~\ref{fig:inhomNLO} are both labeled with (c) since they both have three-body force contributions.  To avoid double counting, diagram (a) in the dashed box is subtracted since it is included both in the first diagram (a) of Fig.~\ref{fig:inhomNLO} and in diagram (\rom{1}) of Fig.~\ref{fig:inhomothers}. Diagram (f) comes from the $L_1^{(0)}$ and $L_2^{(0)}$ interactions in Eq.~\eqref{eq:MEClagNLO}.  

\begin{figure}[hbt]
    \centering
    \includegraphics[width=100mm]{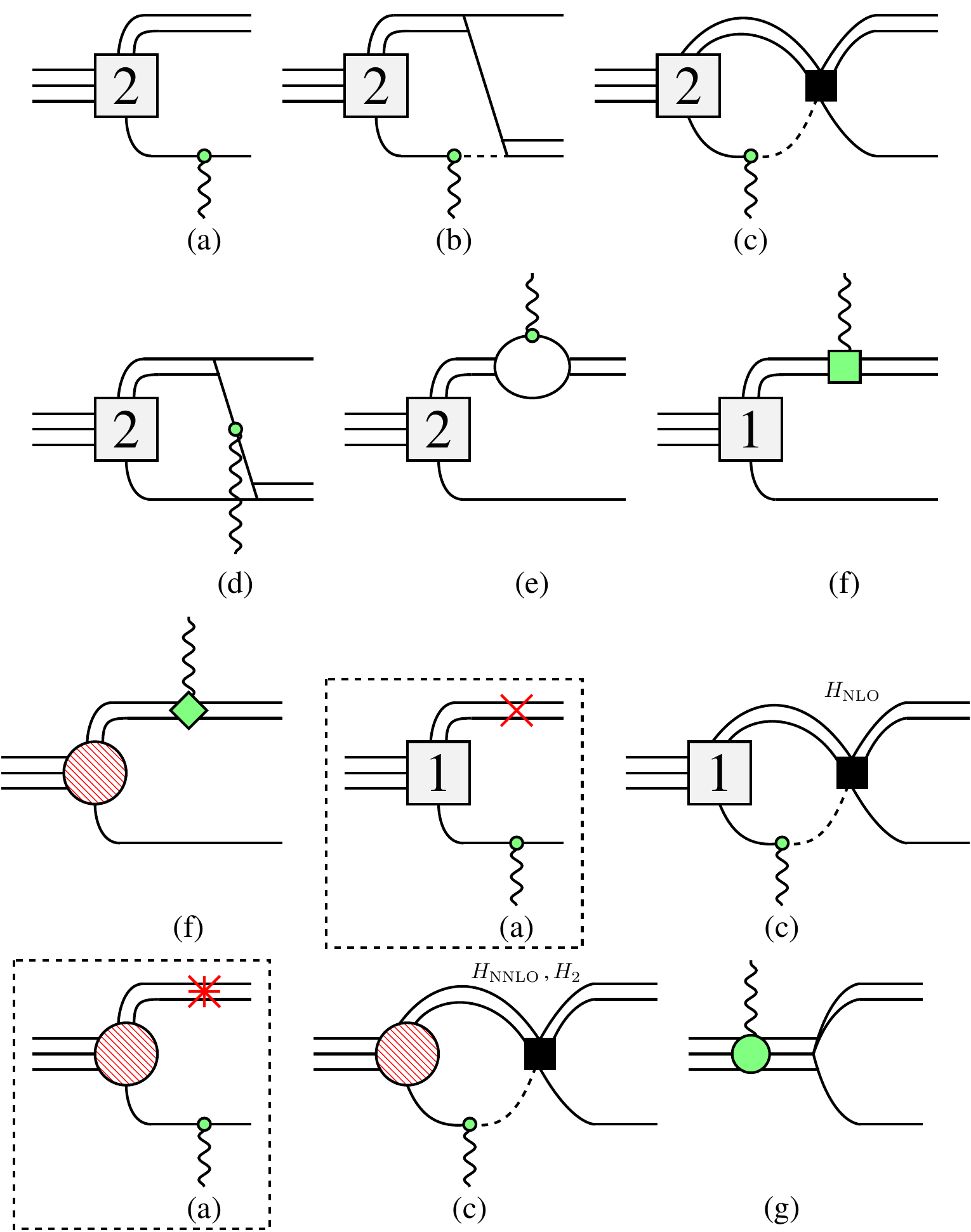}
    \caption{Diagrams of the inhomogeneous term ${\Bb^{[2]}}_{L'S',LS}^{J,J'}(p,k)$ for the integral equation of the \NNLO correction to the two-body triton photo-disintegration amplitude, Eq.~\eqref{eq:TampNNLO}.  The box with a ``2" is the \NNLO correction to the vertex function, the green diamond is the interaction from Eq.~\eqref{eq:MEClagNNLO}, and the large circle on the trimer attached to the photon is the \NNLO three-nucleon magnetic moment correction in Eq.~\eqref{eq:maglag3}. There are two diagrams living in the third instance of the diagram labeled (c) (in the last row); one with the three-body counterterm $H_{\mathrm{\NNLO}}$ and one with the three-body counterterm $H_2$.  Finally, the red symbol  \mbox{\textcolor{red}{\Large$\times$\hspace{-4.02mm}\raisebox{0.45mm}{\normalsize$+$}}}
    \normalsize  represents the contribution from either $c_{0t}^{(1)}$ or $c_{0s}^{(1)}$. All other symbols are the same as in Figs.~\ref{fig:photodis}, \ref{fig:inhomLO}, and \ref{fig:inhomNLO}. Boxed diagrams are subtracted to avoid double counting.} 
    \label{fig:inhomNNLO}
\end{figure}

\begin{figure}[hbt]
    \centering
    \includegraphics[width=110mm]{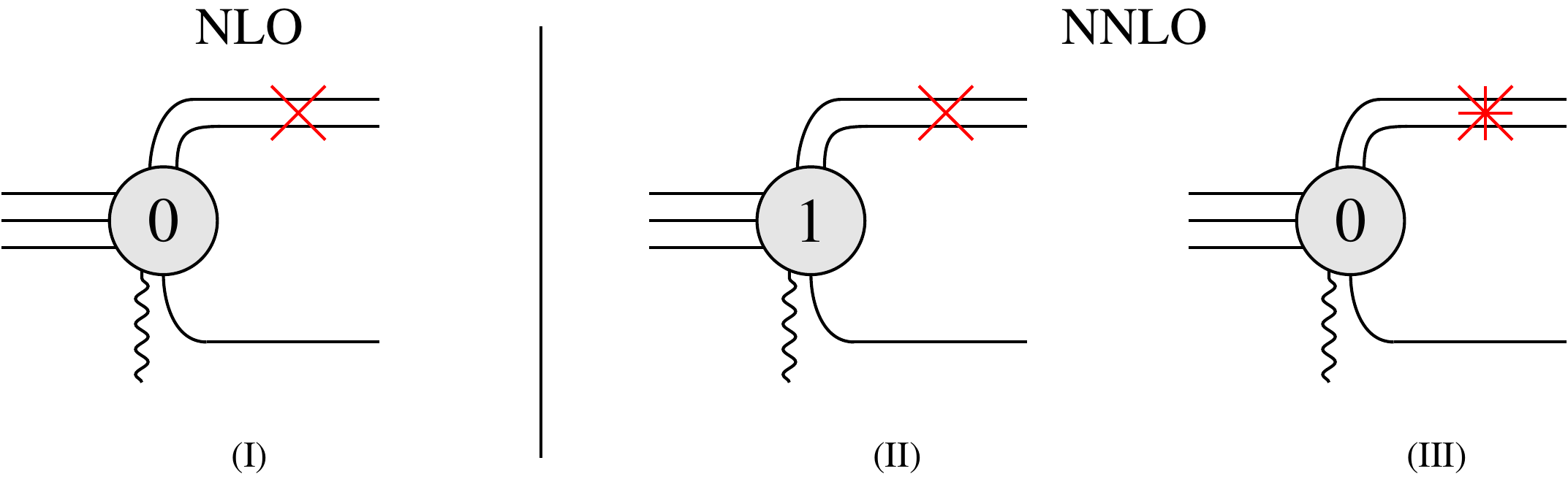}
    \caption{Diagrams for the inhomogeneous term  $\mathbf{R}_1{\Tb^{[0]}}_{L'S',LS}^{J,J'}(p,k)$ (I) of the NLO correction to the two-body triton photo-disintegration amplitude, Eq.~\eqref{eq:TampNLO}, and $\mathbf{R}_1{\Tb^{[1]}}_{L'S',LS}^{J,J'}(p,k)$ (II) and $\mathbf{R}_2{\Tb^{[0]}}_{L'S',LS}^{J,J'}(p,k)$ (III) of the NNLO correction to the two-body triton photo-disintegration amplitude, Eq.~\eqref{eq:TampNNLO}.  The gray circle with a ``0" (``1") and a photon attached to it is the LO (NLO correction to the) two-body triton photo-disintegration amplitude. All other symbols are the same as in Figs.~\ref{fig:photodis}, \ref{fig:inhomNLO}, and \ref{fig:inhomNNLO}.}
    \label{fig:inhomothers}
\end{figure}

The \NNLO correction to the two-body triton photo-disintegration amplitude is given by the integral equation
\begin{align}
    \label{eq:TampNNLO}
    &{\Tb^{[2]}}_{L'S',LS}^{J,J'}(p,k)=\frac{e}{2M_N}{\Bb^{[2]}}_{L'S',LS}^{J,J'}(p,k)\\\nonumber
    &+\mathbf{R}_1\left(E-\frac{p^2}{2M_N},\vectn{p}\right){\Tb^{[1]}}_{L'S',LS}^{J,J'}(p,k)+\mathbf{R}_2\left(E-\frac{p^2}{2M_N},\vectn{p}\right){\Tb^{[0]}}_{L'S',LS}^{J,J'}(p,k)\\\nonumber
    &-\pi H_{\mathrm{NLO}}\delta_{L'0}\delta_{S'\frac{1}{2}}\left(\begin{array}{rr}
    1 & -1\\
    -1 & 1
    \end{array}\right)\Db\left(E-\frac{q^2}{2M_N},\vectn{q}\right)\otimes_q{\Tb^{[1]}}_{L'S',LS}^{J,J'}(q,k)
    \\\nonumber
    &-\pi \left(H_{\mathrm{NNLO}}+\frac{4}{3}M_Nk_0H_2\right)\delta_{L'0}\delta_{S'\frac{1}{2}}\left(\begin{array}{rr}
    1 & -1\\
    -1 & 1
    \end{array}\right)\Db\left(E-\frac{q^2}{2M_N},\vectn{q}\right)\otimes_q{\Tb^{[0]}}_{L'S',LS}^{J,J'}(q,k)
    \\\nonumber
    &+\sum_{L'',S''}\Kb^{J'}_{L'S',L''S''}(q,p,E)\mathbf{D}\left(E-\frac{q^2}{2M_N},\vectn{q}\right)\otimes_q {\Tb^{[2]}}_{L''S'',LS}^{J,J'}(q,k),
\end{align}
where the inhomogeneous term ${\Bb^{[2]}}_{L'S',LS}^{J,J'}(p,k)$ is given by the sum of the diagrams in Fig.~\ref{fig:inhomNNLO} 
\begin{equation}
    {\Bb^{[2]}}_{L'S',LS}^{J,J'}(p,k)=\sum_{x=a,b,c,d,e,f,g}{\Bb^{[2]}}_{L'S',LS}^{J,J',(x)}(p,k),
\end{equation}
and the inhomogenous terms $\mathbf{R}_1\mathbf{\Tb^{[1]}}_{L'S',LS}^{J,J'}(p,k)$ and $\mathbf{R}_2\mathbf{\Tb^{[0]}}_{L'S',LS}^{J,J'}(p,k)$ by diagrams (\rom{2}) and (\rom{3}) in Fig.~\ref{fig:inhomothers}, respectively.
Diagrams in dashed boxes are again subtracted to avoid double counting from diagram (\rom{2}) and (\rom{3}) in Fig.~\ref{fig:inhomothers}.  The last diagram in Fig.~\ref{fig:inhomNNLO} contains the three-nucleon magnetic moment term in Eq.~\eqref{eq:maglag3}.  This term is necessary to remove divergences at \NNLO and we determine it by fitting to the triton magnetic moment at \NNLO. 
Since magnetic photons do not couple to nucleon momenta they do not change the orbital angular momentum of nuclear states.  However, their coupling to nucleon spin allows transitions in the total nuclear spin and total nuclear angular momentum.  The relevant transition amplitudes for magnetic photons take the \DS state of the triton to either a \DS or \QS final $nd$ scattering state.  To project out the amplitudes into a partial-wave basis we use the projectors of Ref.~\cite{Griesshammer:2004pe}.  The contribution to the inhomogeneous term from type-(a) diagrams at LO, NLO, and \NNLO ($n=$ 0, 1, and 2, respectively) for the outgoing $nd$ state in the \DS channel is 
\begin{align}
&{\Bb^{[n]}}_{0\frac{1}{2},0\frac{1}{2}}^{\frac{1}{2},\frac{1}{2},(a)}(p,k)=\frac{1}{2\sqrt{3}}\left(\begin{array}{cc}
-(\kappa_0+\tau_3\kappa_1) & 0\\
0 & 3\kappa_0-\tau_3\kappa_1
\end{array}\right)\\\nonumber
&\hspace{1cm}\int_{-1}^{1}\!\!dx\frac{1}{k_0+\frac{pkx}{M_N}+\frac{k^2}{6M_N}}\left\{\Gbt_n\left(\sqrt{p^2+\frac{4}{3}pkx+\frac{4}{9}k^2},E_B+k_0+\frac{pkx}{M_N}+\frac{k^2}{6M_N}\right)\right.\\\nonumber
&\hspace{1cm}\left.-\sum_{m=1}^{n}\mathbf{R}_m\left(E-\frac{p^2}{2M_N},\vectn{p}\right)\Gbt_{n-m}\left(\sqrt{p^2+\frac{4}{3}pkx+\frac{4}{9}k^2},E_B+k_0+\frac{pkx}{M_N}+\frac{k^2}{6M_N}\right)\right\},
\end{align}
and in the outgoing \QS channel is 
\begin{align}
&{\Bb^{[n]}}_{0\frac{3}{2},0\frac{1}{2}}^{\frac{1}{2},\frac{3}{2},(a)}(p,k)=\frac{1}{2\sqrt{3}}\left(\begin{array}{cc}
2(\kappa_0+\tau_3\kappa_1) & 0\\
0 & 0
\end{array}\right)\\\nonumber
&\hspace{1cm}\int_{-1}^{1}\!\!dx\frac{1}{k_0+\frac{pkx}{M_N}+\frac{k^2}{6M_N}}\left\{\Gbt_n\left(\sqrt{p^2+\frac{4}{3}pkx+\frac{4}{9}k^2},E_B+k_0+\frac{pkx}{M_N}+\frac{k^2}{6M_N}\right)\right.\\\nonumber
&\hspace{1cm}\left.-\sum_{m=1}^n\mathbf{R}_m\left(E-\frac{p^2}{2M_N},\vectn{p}\right)\Gbt_{n-m}\left(\sqrt{p^2+\frac{4}{3}pkx+\frac{4}{9}k^2},E_B+k_0+\frac{pkx}{M_N}+\frac{k^2}{6M_N}\right)\right\}.
\end{align}
These expressions come from subtracting the boxed diagrams (a) from the unboxed diagrams (a) of Figs.~\ref{fig:inhomNLO} and \ref{fig:inhomNNLO}.   $\Gbt_0(p,E')$, $\Gbt_1(p,E')$, and $\Gbt_2(p,E')$  are the LO, NLO , and \NNLO off-shell three-nucleon vertex functions in a boosted reference frame, respectively, defined by
\begin{equation}
    \Gbt_0(p,E')=\oneb+\mathbf{R}_0(q,p,E')\Db\left(E_B-\frac{q^2}{2M_N},\vectn{q}\right)\otimes_q\Gb_0(E_B,q),
\end{equation}
\begin{align}
    \Gbt_1(p,E')=&\mathbf{R}_1\left(E'-\frac{p^2}{2M_N},\vectn{p}\right)\Gbt_0(p,E')\\\nonumber
    &\hspace{2cm}+\mathbf{R}_0(q,p,E')\Db\left(E_B-\frac{q^2}{2M_N},\vectn{q}\right)\otimes_q\Gb_1(E_B,q),
\end{align}
and 
\begin{align}
    &\Gbt_2(p,E')=\mathbf{R}_1\left(E'-\frac{p^2}{2M_N},\vectn{p}\right)\Gbt_1(p,E')+\mathbf{R}_2\left(E'-\frac{p^2}{2M_N},\vectn{p}\right)\Gbt_0(p,E')\\\nonumber
    &\hspace{2cm}+\mathbf{R}_0(q,p,E')\Db\left(E_B-\frac{q^2}{2M_N},\vectn{q}\right)\otimes_q\Gb_2(E_B,q).
\end{align}
The ordering of momentum and energy arguments in the c.m.~vertex function and boosted vertex function are different to further distinguish them in addition to the presence of the tilde symbol. The LO, NLO, and NNLO diagrams (b) for the outgoing \DS channel give the contribution
\begin{align}
    &{\Bb^{[n]}}_{0\frac{1}{2},0\frac{1}{2}}^{\frac{1}{2},\frac{1}{2},(b)}(p,k)=-\frac{2\pi}{\sqrt{3}}\left(\begin{array}{cc}
    \kappa_0+\tau_3\kappa_1 & 3(3\kappa_0-\tau_3\kappa_1)\\
    -3(\kappa_0+\tau_3\kappa_1) & -(3\kappa_0-\tau_3\kappa_1)
    \end{array}\right)\\\nonumber
    &\hspace{.5cm}\int_{-1}^1\!\!dx\int_{-1}^1\!\!dy\int_0^{2\pi}\!\!\!d\phi\frac{1}{k_0+\frac{qkx}{M_N}-\frac{k^2}{2M_N}}\Db\left(E-k_0+\frac{k^2}{6M_N}-\frac{q^2}{2M_N},\vectn{q}\right)\otimes_q\Gb_n(E_B,q)\\\nonumber
    &\hspace{.5cm}\frac{1}{q^2+qp(xy+\sqrt{1-x^2}\sqrt{1-y^2}\cos\phi)+p^2-M_N(E-k_0)-\frac{1}{3}qkx-\frac{2}{3}kpy-\frac{1}{18}k^2-i\epsilon},
\end{align}
and for an outgoing \QS channel the contribution
\begin{align}
    &{\Bb^{[n]}}_{0\frac{3}{2},0\frac{1}{2}}^{\frac{1}{2},\frac{3}{2},(b)}(p,k)=-\frac{2\pi}{\sqrt{3}}\left(\begin{array}{cc}
    4(\kappa_0+\tau_3\kappa_1) & 0\\
    0 & 0
    \end{array}\right)\\\nonumber
    &\hspace{.5cm}\int_{-1}^1\!\!dx\int_{-1}^1\!\!dy\int_0^{2\pi}\!\!\!d\phi\frac{1}{k_0+\frac{qkx}{M_N}-\frac{k^2}{2M_N}}\Db\left(E-k_0+\frac{k^2}{6M_N}-\frac{q^2}{2M_N},\vectn{q}\right)\otimes_q\Gb_n(E_B,q)\\\nonumber
    &\hspace{.5cm}\frac{1}{q^2+qp(xy+\sqrt{1-x^2}\sqrt{1-y^2}\cos\phi)+p^2-M_N(E-k_0)-\frac{1}{3}qkx-\frac{2}{3}kpy-\frac{1}{18}k^2-i\epsilon}.
\end{align}
Since type-(c) diagrams have a three-body force they only contribute to outgoing states in the \DS channel, where their contribution at LO, NLO, and \NNLO to the inhomogeneous term is
\begin{align}
 &{\Bb^{[n]}}_{0\frac{1}{2},0\frac{1}{2}}^{\frac{1}{2},\frac{1}{2},(c)}(p,k)=\sum_{m=0}^n\frac{\pi M_N H^{(n)}(\Lambda)}{3\sqrt{3}}\left(\begin{array}{cc}
 -3(\kappa_0+\tau_3\kappa_1) & -3(3\kappa_0-\tau_3\kappa_1)\\
 3(\kappa_0+\tau_3\kappa_1) & 3(3\kappa_0-\tau_3\kappa_1)
 \end{array}
 \right)\\\nonumber
 &\hspace{1cm}\frac{1}{qk}Q_0\left(\frac{M_Nk_0-\frac{1}{2}k^2}{qk}\right)\Db\left(E-k_0+\frac{k^2}{6M_N}-\frac{q^2}{2M_N},\vectn{q}\right)\otimes_q\Gb_{n-m}(E_B,q),
\end{align}
and the functions $H^{(n)}(\Lambda)$ are defined by
\begin{align}
    H^{(0)}(\Lambda)=H_{\mathrm{LO}}(\Lambda)\,,\, H^{(1)}(\Lambda)=H_{\mathrm{NLO}}(\Lambda)\,,\, H^{(2)}(\Lambda)=H_{\mathrm{NNLO}}(\Lambda)+\frac{4}{3}M_Nk_0H_2(\Lambda).
\end{align}
The LO, NLO, and \NNLO contribution from diagrams (d) to the inhomogenous term for an outgoing \DS channel is
\begin{align}
    &{\Bb^{[n]}}_{0\frac{1}{2},0\frac{1}{2}}^{\frac{1}{2},\frac{1}{2},(d)}(p,k)=\frac{2\pi M_N}{\sqrt{3}}\left(\begin{array}{cc}
    -5(\kappa_0-\tau_3\kappa_1) & (3\kappa_0+\tau_3\kappa_1)\\
    (3\kappa_0+\tau_3\kappa_1) & (3\kappa_0+5\tau_3\kappa_1)
    \end{array}\right)\int_{-1}^1\!\! dx \Db\left(E_B-\frac{q^2}{2M_N},\vectn{q}\right)\\\nonumber
    &\hspace{2cm}\otimes_q\Gb_n(E_B,q)\frac{1}{\sqrt{B^2-4AC}}\left\{Q_0\left(\frac{B}{\sqrt{B^2-4AC}}\right)-Q_0\left(\frac{2A+B}{\sqrt{B^2-4AC}}\right)\right\},
\end{align}
and for an outgoing \QS channel is given by
\begin{align}
    &{\Bb^{[n]}}_{0\frac{3}{2},0\frac{1}{2}}^{\frac{1}{2},\frac{3}{2},(d)}(p,k)=\frac{2\pi M_N}{\sqrt{3}}\left(\!\!\begin{array}{cc}
    -2(\kappa_0-\tau_3\kappa_1) & -2(3\kappa_0+\tau_3\kappa_1)\\
    0 & 0
    \end{array}\right)\int_{-1}^1\!\! dx\Db\left(E_B-\frac{q^2}{2M_N},\vectn{q}\right)\\\nonumber &\hspace{2cm}\otimes_q\Gb_n(E_B,q)\frac{1}{\sqrt{B^2-4AC}}\left\{Q_0\left(\frac{B}{\sqrt{B^2-4AC}}\right)-Q_0\left(\frac{2A+B}{\sqrt{B^2-4AC}}\right)\right\}.
\end{align}
For diagrams (d) the values $A$, $B$, and $C$ are defined by
\begin{equation}
    A=\left(pkx+\frac{1}{6}k^2-M_Nk_0\right)^2-q^2k^2,
\end{equation}
\begin{equation}
    B=2\left(q^2+p^2-M_NE_B-\frac{2}{3}pkx+\frac{1}{9}k^2\right)\left(pkx+\frac{1}{6}k^2-M_Nk_0\right)-q^2\left(2pkx-\frac{2}{3}k^2\right),
\end{equation}
and
\begin{equation}
    C=\left(q^2+p^2-M_NE_B-\frac{2}{3}pkx+\frac{1}{9}k^2\right)^2-q^2\left(p^2-\frac{2}{3}pkx+\frac{1}{9}k^2\right).
\end{equation}
Diagrams (e) at LO, NLO, and \NNLO for an outgoing \DS channel give the contribution
\begin{align}
    &{\Bb^{[n]}}_{0\frac{1}{2},0\frac{1}{2}}^{\frac{1}{2},\frac{1}{2},(e)}(p,k)=-\frac{M_N}{2\sqrt{3}k}\left(\begin{array}{cc}
    8\kappa_0 & 4\tau_3\kappa_1 \\
    4\tau_3\kappa_1 & 0
    \end{array}\right)\\\nonumber
    &\hspace{1cm}\int_{-1}^{1}\!\!dx\Db\left(E-k_0+\frac{pkx}{2M_N}+\frac{k^2}{12M_N}-\frac{p^2}{2M_N},\vectn{p}\right)\Gbt_n\left(\sqrt{p^2-\frac{2}{3}pkx+\frac{1}{9}k^2},E_B\right)\\\nonumber
    &\hspace{1cm}\tan^{-1}\left(\frac{k}{2\sqrt{\frac{3}{4}p^2-\frac{1}{2}pkx-\frac{1}{12}k^2-M_N(E-k_0)}+2\sqrt{\frac{3}{4}p^2-M_NE}}\right),
\end{align}
and for an outgoing \QS channel the contribution is
\begin{align}
    &{\Bb^{[n]}}_{0\frac{3}{2},0\frac{1}{2}}^{\frac{1}{2},\frac{3}{2},(e)}(p,k)=-\frac{M_N}{2\sqrt{3}k}\left(\begin{array}{cc}
    -4\kappa_0 & 4\tau_3\kappa_1 \\
    0 & 0
    \end{array}\right)\\\nonumber
    &\hspace{1cm}\int_{-1}^{1}\!\!dx\Db\left(E-k_0+\frac{pkx}{2M_N}+\frac{k^2}{12M_N}-\frac{p^2}{2M_N},\vectn{p}\right)\Gbt_n\left(\sqrt{p^2-\frac{2}{3}pkx+\frac{1}{9}k^2},E_B\right)\\\nonumber
    &\hspace{1cm}\tan^{-1}\left(\frac{k}{2\sqrt{\frac{3}{4}p^2-\frac{1}{2}pkx-\frac{1}{12}k^2-M_N(E-k_0)}+2\sqrt{\frac{3}{4}p^2-M_NE}}\right).
\end{align}
The inverse tangent function comes from the two-nucleon sub-diagram appearing in diagrams (e).  Type-(f) diagrams arising from two-nucleon currents only contribute at NLO and \NNLO.  For an outgoing \DS channel the type-(f) diagram contribution is given by
\begin{align}
    &{\Bb^{[n]}}_{0\frac{1}{2},0\frac{1}{2}}^{\frac{1}{2},\frac{1}{2},(f)}(p,k)=-\sum_{m=0}^{n-1}\frac{M_N}{2\sqrt{3}}\left(\begin{array}{cc}
    2L_2^{(m)} & \tau_3 L_1^{(m)} \\
    \tau_3L_1^{(m)} & 0
    \end{array}\right)\\\nonumber
    &\hspace{1cm}\int_{-1}^{1}\!\!dx\Db\left(E-k_0+\frac{pkx}{2M_N}+\frac{k^2}{12M_N}-\frac{p^2}{2M_N},\vectn{p}\right)\Gbt_{n-1-m}\left(\sqrt{p^2-\frac{2}{3}pkx+\frac{1}{9}k^2},E_B\right),
\end{align}
and for an outgoing \QS channel the contribution is
\begin{align}
    &{\Bb^{[n]}}_{0\frac{3}{2},0\frac{1}{2}}^{\frac{1}{2},\frac{3}{2},(f)}(p,k)=-\sum_{m=0}^{n-1}\frac{M_N}{2\sqrt{3}}\left(\begin{array}{cc}
    -L_2^{(m)} & \tau_3 L_1^{(m)} \\
    0 & 0
    \end{array}\right)\\\nonumber
    &\hspace{1cm}\int_{-1}^{1}\!\!dx\Db\left(E-k_0+\frac{pkx}{2M_N}+\frac{k^2}{12M_N}-\frac{p^2}{2M_N},\vectn{p}\right)\Gbt_{n-1-m}\left(\sqrt{p^2-\frac{2}{3}pkx+\frac{1}{9}k^2},E_B\right).
\end{align}
Finally, diagram (g) coming from the the three-nucleon magnetic moment term in Eq.~\eqref{eq:maglag3} only contributes to the outgoing \DS channel and is given by
\begin{align}
    &{\Bb^{[n]}}_{0\frac{1}{2},0\frac{1}{2}}^{\frac{1}{2},\frac{1}{2},(g)}(p,k)=-\delta_{n2}\sqrt{3}\frac{1}{\Omega}(\widetilde{\kappa}_0(\Lambda)+\tau_3\widetilde{\kappa}_1(\Lambda))\oneb.
\end{align}
 Including complete wavefunction renormalizations and taking the component of  ${\mathbf{T}^{[n]}}{}^{J,J'}_{L'S',LS}(p,k)$ relevant for $nd$ capture yields 
\begin{align}
\label{eq:T-renormed}
    &{T^{[n]}}{}^{J,J'}_{L'S',LS}(p,k)=\left\{\sqrt{Z_t}_{0}\sqrt{Z_d}_{0}{\mathbf{T}^{[n]}}{}^{J,J'}_{L'S',LS}(p,k)\right.\\\nonumber
    &\hspace{1cm}+\left(\sqrt{Z_t}_{1}\sqrt{Z_d}_{0}-\sqrt{Z_t}_{0}\sqrt{Z_d}_{1}\right){\mathbf{T}^{[n-1]}}{}^{J,J'}_{L'S',LS}(p,k)\\\nonumber
    &\hspace{1cm}\left.+\left(\sqrt{Z_t}_{2}\sqrt{Z_d}_{0}+\sqrt{Z_t}_{1}\sqrt{Z_d}_{1}+\sqrt{Z_t}_{0}\sqrt{Z_d}_{2}\right){\mathbf{T}^{[n-2]}}{}^{J,J'}_{L'S',LS}(p,k)\right\}^T\left(\begin{array}{c}
    1\\
    0
    \end{array}\right),
\end{align}
where the superscript $T$ indicates the transpose of the vector.  $\sqrt{Z_t}_{n}$ ($\sqrt{Z_d}_{n}$) is the N$^n$LO contribution to the  wavefunction renormalization for the triton vertex function (deuteron) given in Eq.~\eqref{eq:tritonrenorm} (Eq.~\eqref{eq:drenorm}).  The negative sign on the second line comes from the fact the integral equation has a built-in wavefunction renormalization factor of $2\sqrt{Z_d}_1$ that must be adjusted~\cite{Vanasse:2015fph} in this case. The full LO two-body triton photo-disintegration amplitude and its perturbative corrections ${\Mc^{[n]}}{}^{J'M',JM\lambda}_{L'S',LS}(p,k)$ are given by
\begin{equation}
    \label{eq:amprel}
    {\Mc^{[n]}}{}^{J'M',JM\lambda}_{L'S',LS}(p,k)={T^{[n]}}{}^{J,J'}_{L'S',LS}(p,k)\epsilon_{r\ell m}k_r\epsilon_{\gamma}^\ell(\lambda)\CG{J}{1}{J'}{M}{m}{M'},
\end{equation}
where the Clebsch-Gordan coefficient comes from the projection into an angular momentum basis and  $\epsilon_{r\ell m}k_r\epsilon_{\gamma}^\ell(\lambda)$ comes from the magnetic coupling of the photon with $\epsilon_{\gamma}^\ell(\lambda)$, the photon polarization vector. $\lambda$ refers to the specific state of the photon polarization vector while the superscript $\ell$ picks a component of the polarization vector. $M$ ($M'$) is the $z$ component of the total nuclear angular momentum $J$ ($J'$).

\section{\label{sec:mag} Triton Magnetic Moment}

Without the three-nucleon magnetic moment counterterm, Eq.~\eqref{eq:maglag3}, the NNLO correction to the two-body triton photo-disintegration amplitude as well as the \NNLO correction to the triton magnetic moment diverge as $\Lambda\to\infty$.  This three-nucleon magnetic moment term renormalizes both \NNLO corrections. We fit the three nucleon magnetic moment counterterm to the triton magnetic moment.  The triton magnetic moment has been calculated previously in \EFT up to NLO~\cite{Vanasse:2017kgh,Kirscher:2017fqc,De-Leon:2020glu}.  Using Ref.~\cite{Vanasse:2017kgh}, the LO triton magnetic moment is given by (see  Ref.~\cite{Vanasse:2017kgh} for details)

\begin{align}
\label{eq:LOmagmom}
&\mu_0^{\jjvH}=(\kappa_0+\kappa_1)+2\pi M_{N}\frac{2}{3}\kappa_1\left(\widetilde{\boldsymbol{\Gamma}}_{0}(q)\right)^{T}\otimes_qM(q,\ell)\left(\!\!\begin{array}{cc}
1 & 1\\[-2mm]
1 & 1
\end{array}\!\!\right)\otimes_\ell\widetilde{\boldsymbol{\Gamma}}_{0}(\ell)\\\nonumber
\end{align}
where
\begin{equation}
    M(q,\ell)=\frac{\pi}{2}\frac{\delta(q-\ell)}{q^{2}\sqrt{\frac{3}{4}q^{2}-M_{N}E_B}}+\frac{2}{q^{2}\ell^{2}-(q^{2}+\ell^{2}-M_{N}E_B)^{2}}
\end{equation}
and
\begin{equation}
    \widetilde{\boldsymbol{\Gamma}}_n(q)=\Db\left(E_B-\frac{q^2}{2M_N},\vectn{q}\right)\sum_{m=0}^n\sqrt{Z_t}_{n-m}\Gb_m(E_B,q).
\end{equation}
This simplified expression for the LO triton magnetic moment is obtained by separating out the term proportional to the LO charge form factor at $Q^2=0$ found in Ref.~\cite{Vanasse:2017kgh}.  Similarly, rewriting the NLO correction to the magnetic moment of Ref.~\cite{Vanasse:2017kgh} by separating out terms proportional to the NLO correction to the charge form factor at $Q^2=0$ yields
\begin{align}
&\mu_1^{\jjvH}=4\pi M_{N}\frac{2}{3}\kappa_1\left(\widetilde{\boldsymbol{\Gamma}}_{1}(q)\right)^{T}\otimes_qM(q,\ell)\left(\!\!\begin{array}{cc}
1 & 1\\[-2mm]
1 & 1
\end{array}\!\!\right)\otimes_\ell\widetilde{\boldsymbol{\Gamma}}_{0}(\ell)\\\nonumber
&-4\pi M_{N}\left(\widetilde{\boldsymbol{\Gamma}}_{0}(q)\right)^{T}\otimes_q\left\{\frac{\pi}{2}\frac{\delta(q-\ell)}{q^{2}}
\left(\begin{array}{cc}
-2\frac{c_{0t}^{(0)}}{M_N}\frac{\kappa_1+2\kappa_0}{3}-\frac{2}{3}L_2^{(0)} & \frac{1}{3}L_1^{(0)} \\
\frac{1}{3}L_1^{(0)} & -\frac{2}{3}\frac{c_{0s}^{(0)}}{M_N}\kappa_1
\end{array}\right)\right\}\otimes_\ell\widetilde{\boldsymbol{\Gamma}}_{0}(\ell) , \nonumber 
\end{align}
where the symmetry of $M(q,\ell)$ under $q \leftrightarrow \ell$ has been used to combine expressions.
The \NNLO correction to the magnetic moment can also be written as a term proportional to the \NNLO correction to the charge form factor at $Q^2=0$~\cite{Vanasse:2015fph} plus everything else:  
\begin{align}
&\mu_2^{\jjvH}=4\pi M_{N}\frac{2}{3}\kappa_1\left(\widetilde{\boldsymbol{\Gamma}}_{2}(q)\right)^{T}\otimes_qM(q,\ell)\left(\!\!\begin{array}{cc}
1 & 1\\[-2mm]
1 & 1
\end{array}\!\!\right)\otimes_\ell\widetilde{\boldsymbol{\Gamma}}_{0}(\ell)\\\nonumber
&+2\pi M_{N}\frac{2}{3}\kappa_1\left(\widetilde{\boldsymbol{\Gamma}}_{1}(q)\right)^{T}\otimes_qM(q,\ell)\left(\!\!\begin{array}{cc}
1 & 1\\[-2mm]
1 & 1
\end{array}\!\!\right)\otimes_\ell\widetilde{\boldsymbol{\Gamma}}_{1}(\ell)\\\nonumber
&-8\pi M_{N}\left(\widetilde{\boldsymbol{\Gamma}}_{1}(q)\right)^{T}\otimes_q\left\{\frac{\pi}{2}\frac{\delta(q-\ell)}{q^{2}}
\left(\begin{array}{cc}
-2\frac{c_{0t}^{(0)}}{M_N}\frac{\kappa_1+2\kappa_0}{3}-\frac{2}{3}L_2^{(0)} & \frac{1}{3}L_1^{(0)} \\
\frac{1}{3}L_1^{(0)} & -\frac{2}{3}\frac{c_{0s}^{(0)}}{M_N}\kappa_1
\end{array}\right)\right\}\otimes_\ell\widetilde{\boldsymbol{\Gamma}}_{0}(\ell)\\\nonumber
&-4\pi M_{N}\left(\widetilde{\boldsymbol{\Gamma}}_{0}(q)\right)^{T}\otimes_q\left\{\frac{\pi}{2}\frac{\delta(q-\ell)}{q^{2}}
\left(\begin{array}{cc}
-2\frac{c_{0t}^{(1)}}{M_N}\frac{\kappa_1+2\kappa_0}{3}-\frac{2}{3}L_2^{(1)} & \frac{1}{3}L_1^{(1)} \\
\frac{1}{3}L_1^{(1)} & -\frac{2}{3}\frac{c_{0s}^{(1)}}{M_N}\kappa_1
\end{array}\right)\right\}\otimes_\ell\widetilde{\boldsymbol{\Gamma}}_{0}(\ell)\\\nonumber
&-\frac{4}{3}M_NH_2\frac{\Sigma_0^2(E_B)}{\Sigma_0'(E_B)}(\kappa_0+\kappa_1)-\frac{1}{\Omega H_{\mathrm{LO}}\Sigma_0'(E_B)}(\widetilde{\kappa}_0(\Lambda)-\widetilde{\kappa}_1(\Lambda)).
\end{align}
The value of $\widetilde{\kappa}_0(\Lambda) - \widetilde{\kappa}_1(\Lambda)$ is chosen to reproduce the triton magnetic moment at NNLO yielding 
\begin{equation}
    \label{eq:mufit}
    (\widetilde{\kappa}_0(\Lambda)-\widetilde{\kappa}_1(\Lambda))=-\Omega H_{\mathrm{LO}}\Sigma_0'(E_B)(\mu_t-\mu_0^{\jjvH}-\mu_1^{\jjvH}-\widetilde{\mu}_2^{\jjvH}),
\end{equation}
where $\mu_t=2.978960$ is the experimental value of the (dimensionless) triton magnetic moment and $\widetilde{\mu}_2^{\jjvH}$ is the \NNLO correction to the triton magnetic moment without the three-nucleon magnetic moment counterterm $\widetilde{\kappa}_0(\Lambda) - \widetilde{\kappa}_1(\Lambda)$. 

\begin{figure}[hbt]
    \centering
    \includegraphics{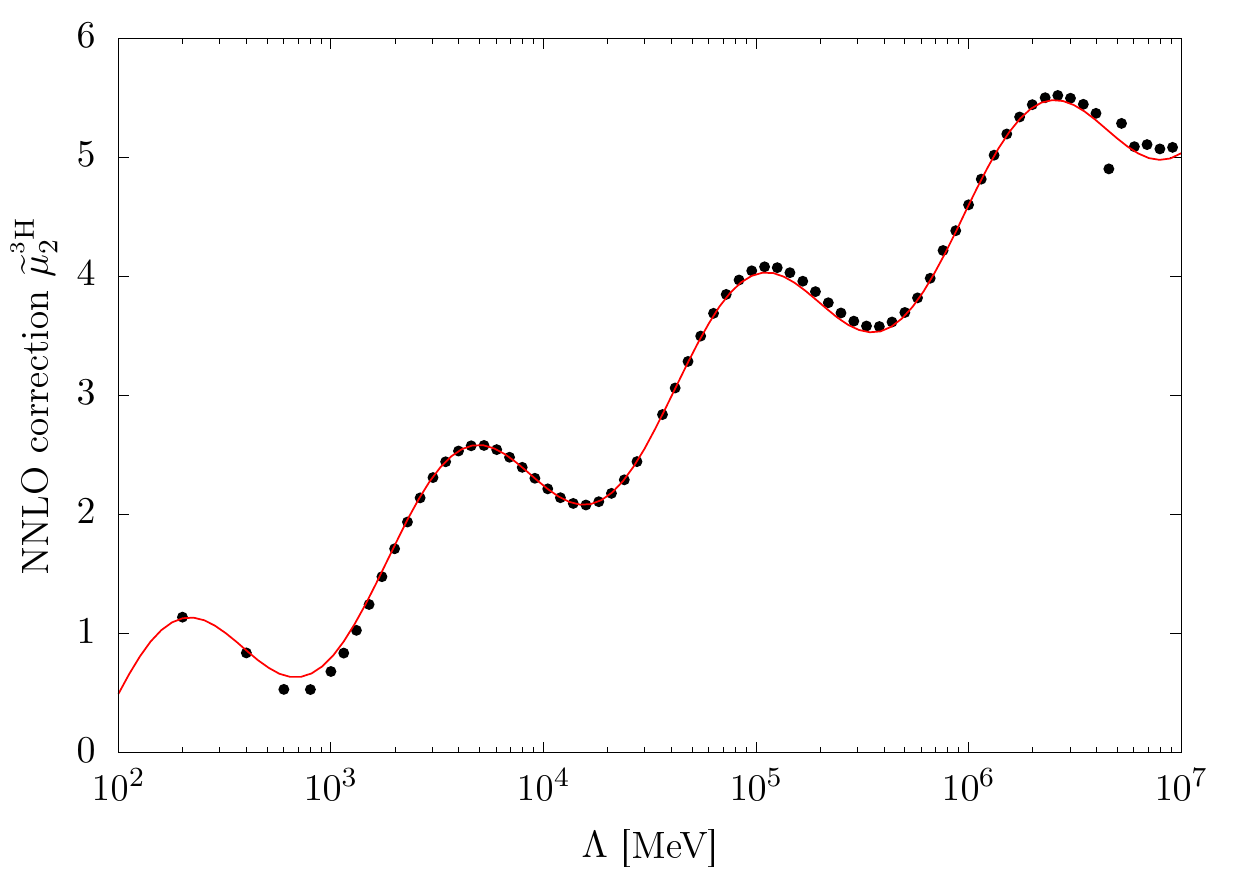}
    \caption{Cutoff dependence of the \NNLO correction to the (dimensionless) triton magnetic moment  without the three-nucleon magnetic moment counterterm $\widetilde{\kappa}_0(\Lambda)-\widetilde{\kappa}_1(\Lambda)$.  The contribution from two-nucleon currents has been removed by setting $L_1^{(m)}=L_2^{(m)}=0$.  Black dots are numerical results while the red line is a fit of Eq.~\eqref{eq:asymptotic} to the numerical data.  \label{fig:magmoment}}
\end{figure}

To demonstrate the need for the counterterm of Eq.~\eqref{eq:maglag3}, the NNLO correction $\widetilde{\mu}_2^{\jjvH}$ as a function of cutoff with $L_1^{(m)}=L_2^{(m)}=0$ is shown in Fig.~\ref{fig:magmoment}. (Different choices for $L_1^{(m)}$ and $L_2^{(m)}$ yield similar cutoff dependence as they do not remove the divergence.) 
Black dots are the numerical computations and the red line is the function
\begin{equation}
    \label{eq:asymptotic}
    f(\Lambda)=a\ln(\Lambda)+b\sin(2s_0\ln(\Lambda)+c)+d,\end{equation}
where the values $a$, $b$, $c$, and $d$ are determined by fitting to the numerical data between cutoffs $\Lambda=5,000$~MeV and $\Lambda=10,000$~MeV.  At larger cutoffs numerical instabilities lead to the numerical results having a marked deviation from the function.  It is readily apparent that the \NNLO correction to the triton magnetic moment is not converging as a function of cutoff and a new counterterm is needed to remove this divergence.  The function $f(\Lambda)$  is obtained by a rough asymptotic analysis and therefore should also disagree with the numerical data at smaller cutoffs, as is observed.  At LO the Wigner-SU(4)-symmetric part of the triton vertex function scales like $\sin(s_0\ln(q/\Lambda^*))/q$, where $s_0=1.00624\ldots$~\cite{Bedaque:1998kg,Bedaque:2002yg}, while the Wigner-SU(4)-antisymmetric NLO correction to the vertex function scales like $\sin(s_0\ln(q/\Lambda^*)+\phi)$.  The contribution to the \NNLO correction to the triton magnetic moment that comes from two NLO corrections to the vertex function in the asymptotic limit will give an integral of the form
\begin{equation}
    \int^\Lambda dq \frac{1}{q}\sin^2(s_0\ln(q/\Lambda^*)+\phi).
\end{equation}
This integral gives an expression of the form in Eq.~\eqref{eq:asymptotic}.  To predict the values of $a$, $b$, $c$, and $d$, a more detailed asymptotic analysis in which finite $\Lambda$ effects are removed will be required. This can be done by calculating the LO vertex function at a large cutoff and then using this to calculate the NLO and \NNLO corrections at smaller cutoffs (see, e.g., Ref.~\cite{Ji:2012nj}).  These finite $\Lambda$ effects may also explain the small differences observed at larger cutoffs between the numerical prediction of $\widetilde{\mu}_2^{\jjvH}$ and the fit to the functional form of Eq.~\eqref{eq:asymptotic}.

\section{\label{sec:zero} The  Zero-recoil limit}

The zero-recoil limit is a good approximation for cold $nd$ capture and  threshold two-body triton photo-disintegration  and simplifies the expressions considerably.
In the zero-recoil limit we drop terms with photon momentum $k$ (e.g., $k^2/M_N$) while keeping terms with photon energy $k_0$, giving errors of order $k/M_N \approx  1\%$ for cold $nd$ capture. By using the coupled integral equations for the vertex function at each order and redefining the integral equation for the two-body triton photo-disintegration amplitude at each order, the sum of diagrams for the inhomogeneous term of the two-body triton photo-disintegration amplitude can be greatly simplified. For details of this simplification see App.~\ref{app:zerorecoil}.  The zero-recoil limit result for the LO, NLO correction, and \NNLO correction to the two-body triton photo-disintegration amplitude Eqs.~\eqref{eq:TampLO}, \eqref{eq:TampNLO}, and \eqref{eq:TampNNLO}, respectively, is obtained by making the following substitutions
\begin{align}
    \label{eq:TampLOmod}
    {\Tb^{[n]}}{}^{J,J'}_{L'S',LS}(p,k)\to{\Tbt^{[n]}}{}^{J,J'}_{L'S',LS}(p,k)\quad,\quad {\Bb^{[n]}}{}_{L'S',LS}^{J,J'}(p,k)\to{\Bbt^{[n]}}{}_{L'S',LS}^{J,J'}(p,k).
\end{align}

The relationship between the original two-body triton photo-disintegration amplitude and the modified one for on-shell values of $p$ is given by (see App.~\ref{app:zerorecoil})
\begin{equation}
\label{eq:Tshift}
     {\Tbt^{[n]}}{}^{J,J'}_{L'S',LS}(p,k)
    ={\Tb^{[n]}}{}^{J,J'}_{L'S',LS}(p,k)+\frac{e}{2M_N}\frac{1}{\sqrt{3}k_0}(\gamma_t-\gamma_s)2\tau_3\kappa_1\left(\!\!\begin{array}{cc}
    0 & 0\\[-2mm]
    1 & 0
    \end{array}\!\!\right)\Dbb\left(E_B\right)\Gb_n,
 \end{equation}
 where we use the notation
 \begin{equation}
     \Dbb(E_B)=\Db\left(E_B-\frac{p^2}{2M_N},\vectn{p}\right)\quad,\quad \Gb_n=\Gb_n(E_B,p).
 \end{equation}
 ${\Tbt^{[n]}}{}^{J,J'}_{L'S',LS}(p,k)$ and ${\Tb^{[n]}}{}^{J,J'}_{L'S',LS}(p,k)$ are equivalent for two-body photo-disintegration because they only differ by a channel with an unphysical outgoing spin-singlet dibaryon.  However, for three-body photo-disintegration the outgoing state with a spin-singlet dibaryon  cannot be neglected.
 
 The  inhomogeneous term ${\Bbt^{[n]}}{}_{L'S',LS}^{J,J'}(p,k)$  for an outgoing \DS channel in the zero-recoil limit is given by
\begin{align}
    \label{eq:LOinhommod}
    &{\Bbt^{[n]}}{}_{0\frac{1}{2},0\frac{1}{2}}^{\frac{1}{2},\frac{1}{2}}(p,k)=\frac{2\tau_3\kappa_1}{\sqrt{3}k_0}\left(\!\!\begin{array}{rr}
    0 & 1\\
    -1 & 0
    \end{array}\!\!\right)\Dbb\left(E_B\right)\\\nonumber
    &\left\{(\gamma_t-\gamma_s)\Gb_n-\frac{1}{M_N}\sum_{m=0}^{n-1}\left[\left(c_{0t}^{(m)}-c_{0s}^{(m)}\right)(M_NE-\frac{3}{4}p^2)+c_{0t}^{(m)}\gamma_t^2-c_{0s}^{(m)}\gamma_s^2\right]\Gb_{n-1-m}\right\}\\\nonumber
 &+\frac{1}{\sqrt{3}}\sum_{m=0}^{n-1}\left(\!\!\begin{array}{cc}
    -4\kappa_0c_{0t}^{(m)}-2M_NL_2^{(m)} & -2\tau_3\kappa_1c_{0s}^{(m)}-\tau_3M_NL_1^{(m)}\\
    -2\tau_3\kappa_1c_{0t}^{(m)}-\tau_3M_NL_1^{(m)} & 0
    \end{array}\!\!\right)\Dbb\left(E_B\right)\Gb_{n-1-m}\\\nonumber
    &-\delta_{n2}\sqrt{3}\frac{4}{3}M_NH_2\Sigma_0(E_B)(\kappa_0-\tau_3\kappa_1)\oneb-\delta_{n2}\sqrt{3}\frac{1}{\Omega}(\widetilde{\kappa}_0(\Lambda)+\tau_3\widetilde{\kappa}_1(\Lambda))\oneb,
\end{align}
and for an outgoing \QS channel by
\begin{align}
&{\Bbt^{[n]}}{}_{0\frac{3}{2},0\frac{1}{2}}^{\frac{1}{2},\frac{3}{2}}(p,k)=\frac{2\tau_3\kappa_1}{\sqrt{3}k_0}\left(\begin{array}{cc}
    0 & 1 \\
    0 & 0
    \end{array}\right)\Dbb\left(E_B\right)\\\nonumber
    &\left\{(\gamma_t-\gamma_s)\Gb_n-\frac{1}{M_N}\sum_{m=0}^{n-1}\left[\left(c_{0t}^{(m)}-c_{0s}^{(m)}\right)(M_NE-\frac{3}{4}p^2)+c_{0t}^{(m)}\gamma_t^2-c_{0s}^{(m)}\gamma_s^2\right]\Gb_{n-1-m}\right\}\\\nonumber
&-\frac{1}{\sqrt{3}}\sum_{m=0}^{n-1}\left(\begin{array}{cc}
    -M_NL_2^{(m)}-2\kappa_0c_{0t}^{(m)} & \tau_3 M_NL_1^{(m)}+2\tau_3\kappa_1c_{0s}^{(m)} \\
    0 & 0
    \end{array}\right)\Dbb\left(E_B\right)\Gb_{n-1-m}.
\end{align}

\section{\label{sec:L1L2}Fitting $L_1$ and $L_2$}

Values for $L_1$ (i.e., $L_1^{(0)} $ and/or $L_1^{(1)}$) can be obtained by various fittings to combinations of the cold $np$ capture cross section, the cold $nd$ capture cross section, and the triton magnetic moment, while $L_2$ (i.e., $L_2^{(0)} $ and/or $L_2^{(1)}$) can be obtained by fitting to the deuteron magnetic moment. Here we show how $L_1$ is fit to cold $np$ capture. The $np$ capture amplitude has been calculated to NLO in \EFT at threshold~\cite{Chen:1999tn,Rupak:1999rk} and above~\cite{Vanasse:2014sva}.  The $np$ capture cross section near threshold to \NNLO in \EFT is given by~\cite{Rupak:1999rk,Vanasse:2014sva}\footnote{Near threshold the contribution from the electric dipole transition is suppressed by the momentum $p$ and is therefore neglected.}
\begin{equation}
    \sigma_{np}=\frac{2\alpha(p^2+\gamma_t^2)^3}{|\vect{v}_{rel}|M_N^3}\left[|Y_{\mathrm{LO}}+Y_{\mathrm{NLO}}|^2+2\mathrm{Re}\left[Y_{\mathrm{LO}}^*Y_{\mathrm{NNLO}}\right]+2|X_{\mathrm{NLO}}|^2\right],
\end{equation}
where $\alpha = e^2/(4\pi
)$ and $p$ is the magnitude of the c.m.~momentum of the neutron. $Y_{\mathrm{LO}}$ is the LO isovector magnetic dipole (M1) moment given by~\cite{Rupak:1999rk}
\begin{equation}
\label{eq:YLO}    Y_{\mathrm{LO}}=\frac{2\kappa_1}{M_N}\sqrt{\gamma_t\pi}\frac{1}{\gamma_t^2+p^2}\left(1-\frac{\gamma_t+ip  }{\gamma_s+ip}\right),
\end{equation}
$Y_{\mathrm{NLO}}$ is its NLO correction in the Z parametrization~\cite{Rupak:1999rk,Vanasse:2014sva}
\begin{align}
    \label{eq:YNLO}
    Y_{\mathrm{NLO}}=&\frac{2\kappa_1}{M_N}\sqrt{\gamma_t\pi}\frac{1}{\gamma_t^2+p^2}\left(\frac{1}{2}(Z_t-1)\right.\\\nonumber
    &\left.-\frac{1}{\gamma_s+ip}\left\{\frac{1}{2}(Z_t-1)+\frac{Z_s-1}{2\gamma_s}(\gamma_s-ip)\right\}(\gamma_t+ip)\right)\\\nonumber
    &-\frac{L_1^{(0)}}{M_N}\sqrt{\gamma_t\pi}\frac{1}{\gamma_s+ip},
\end{align}
and $Y_{\mathrm{NNLO}}$ is its \NNLO correction in the Z parametrization~\cite{Rupak:1999rk}
\begin{align}
    \label{eq:YNNLO}
    Y_{\mathrm{NNLO}}=&\frac{2\kappa_1}{M_N}\sqrt{\gamma_t\pi}\frac{1}{\gamma_t^2+p^2}\left(-\frac{1}{8}(Z_t-1)^2\right.\\\nonumber
    &-\frac{1}{\gamma_s+ip}\frac{1}{2}(Z_t-1)\frac{Z_s-1}{2\gamma_s}(\gamma_s-ip)(\gamma_t+ip)\\\nonumber
    &\left.-\frac{1}{\gamma_s+ip}\left\{-\frac{1}{8}(Z_t-1)^2-\left(\frac{Z_s-1}{2\gamma_s}\right)^2(p^2+\gamma_s^2)\right\}(\gamma_t+ip)\right)\\\nonumber
    &-\frac{L_1^{(0)}}{M_N}\sqrt{\gamma_t\pi}\frac{1}{\gamma_s+ip}\left\{\frac{1}{2}(Z_t-1)+\frac{Z_s-1}{2\gamma_s}(\gamma_s-ip)\right\}\\\nonumber
    &-\frac{L_1^{(1)}}{M_N}\sqrt{\gamma_t\pi}\frac{1}{\gamma_s+ip}.
\end{align}

The LO isoscalar M1 moment is exactly zero in the zero-recoil limit. $X_{\mathrm{NLO}}$, the NLO correction to the isoscalar M1 moment, is given by~\cite{Rupak:1999rk,Vanasse:2014sva}
\begin{equation}
    X_{\mathrm{NLO}}=\frac{L_2^{(0)}}{M_N}\sqrt{\gamma_t\pi}\frac{1}{\gamma_t+ip}.
    \label{eq:XNLO}
\end{equation}

The deuteron magnetic moment in the $Z$ parametrization up to and including \NNLO corrections is given by\footnote{See Refs.~\cite{Kaplan:1998sz,Chen:1999vd} for expressions in different formalisms.}
\begin{equation}
    \label{eq:mud}
    \mu_d=\left(\underbrace{2\kappa_0\vphantom{2(Z_t-1)\kappa_0+2L_2\gamma_t}}_{\mathrm{LO}}+\underbrace{2(Z_t-1)\kappa_0+2L_2^{(0)}\gamma_t}_{\mathrm{NLO}}+\underbrace{2(Z_t-1)L_2^{(0)}\gamma_t+2L_2^{(1)}\gamma_t}_{\mathrm{NNLO}}\right).
\end{equation}
Fitting $L_2^{(0)}$ and $L_2^{(1)}$ to obtain the deuteron magnetic moment $\mu_d=0.85741 \frac{e}{2M_N}$ at each order yields central values $L_2^{(0)}=-1.36$~fm and $L_2^{(1)}=0.940$~fm. For different fits of $L_2^{(m)}$ see Secs.~\ref{sec:Wigner} and \ref{sec:results}. 

One way to determine the value of $L_1^{(0)}$ is by fitting it to the cold $np$ capture cross section $\sigma_{np}=334.2(5)$~mb~\cite{COX1965497} at a neutron laboratory velocity of $|\vect{v}_{rel}|=2200$ m/s at NLO, which yields a central value $L_1^{(0)}=-6.90$~fm.   
Since $\widetilde{\kappa}_0(\Lambda)-  \widetilde{\kappa}_1(\Lambda)$ is fit to the triton magnetic moment at \NNLO, $L_1^{(1)}$ is determined by ensuring $\sigma_{np}$ is properly reproduced at \NNLO.  For $L_1^{(0)}=-6.90$~fm,  $L_1^{(1)}=3.85$~fm. For different fits of $L_1^{(m)}$ see Secs.~\ref{sec:Wigner} and \ref{sec:results}.

\section{\label{sec:Wigner}Consequences of Wigner-SU(4) symmetry}

In the Wigner-SU(4) limit $\gamma_t=\gamma_s=\gamma$ and $Z_s=Z_t=Z$. In this section we will consider a dual \eftnopi-Wigner-SU(4) expansion, where we treat deviations from the Wigner-SU(4) limit as perturbative corrections. To determine the impact of Wigner-SU(4) breaking on two- and three-nucleon observables within \eftnopi it is useful to define
\footnote{The definition of $\delta$ here is equivalent to Eq.~\eqref{eq:deltaintro} up to range corrections. Also, these definitions of $\rho$ and $\delta_r$ differ by a factor of $\frac{1}{2}$ from Ref.~\cite{Vanasse:2016umz}.  In the Wigner-SU(4) limit our value of $\rho$ is equivalent to the effective range.}
\begin{align}
    \label{eq:Wignervalues}
    &\gamma=\frac{1}{2}(\gamma_t+\gamma_s)\quad,\quad\rho=\left(\frac{Z_t-1}{2\gamma_t}+\frac{Z_s-1}{2\gamma_s}\right)\\\nonumber
    &\delta=\frac{1}{2}(\gamma_t-\gamma_s)\quad,\quad\delta_r=\left(\frac{Z_t-1}{2\gamma_t}-\frac{Z_s-1}{2\gamma_s}\right) .
\end{align}
In the Z parametrization the ratio of $\delta_r$ and $\rho$ is~\cite{Griesshammer:2010nd}
\begin{equation}
    \frac{\delta_{r}}{\rho}=0.095\sim Q^2 ,
\end{equation}
where $Q\sim \gamma/\lambdanopi\sim\frac{1}{3}$.  Therefore the limit $\delta_r=0$ can be taken, resulting in errors that are roughly next-to-next-to-next-to leading order (N$^3$LO) in \eftnopi.
In Ref.~\cite{Vanasse:2016umz} Phillips and Vanasse showed that powers of $\delta/\kappa_*$, where $\kappa_*$ is a three-body scale, give a good expansion parameter for three-nucleon bound state observables.  They argued that $\mathcal{O}(\delta)\sim\mathcal{O}(\rho^2)$, or in other words that $\mathcal{O}(\delta)$ terms are approximately the same size as NNLO corrections in \EFT. 
This suggests the use of a power counting that combines the typical \EFT power counting with an expansion in powers of the Wigner-SU(4) breaking parameter $\delta$~\cite{Vanasse:2016umz}.

\subsection{cold $np$ capture}

Taking the Wigner-SU(4) limit (where $\delta=\delta_r=0$) in Eqs.~\eqref{eq:YLO} and \eqref{eq:YNLO} the LO and NLO isovector M1 moments for $np$ capture become
\begin{align}
    Y_{\mathrm{LO}}=0\quad,\quad Y_{\mathrm{NLO}}=&-\frac{2\kappa_1}{M_N}\sqrt{\gamma\pi}\frac{1}{\gamma+ip}\frac{Z-1}{2\gamma}-\frac{L_1^{(0)}}{M_N}\sqrt{\gamma\pi}\frac{1}{\gamma+ip},
\end{align}
and Eq.~\eqref{eq:YNNLO} becomes 
\begin{align}
    Y_{\mathrm{NNLO}}=\frac{2\kappa_1}{M_N}\sqrt{\gamma\pi}\left(\frac{Z-1}{2\gamma}\right)^2\frac{ip}{\gamma+ip}+\frac{L_1^{(0)}}{M_N}\sqrt{\gamma\pi}\frac{Z-1}{2\gamma}\frac{ip-2\gamma}{\gamma+ip}-\frac{L_1^{(1)}}{M_N}\sqrt{\gamma\pi}\frac{1}{\gamma+ip}.
\end{align}
Thus at LO in \eftnopi the isovector M1 moment is zero in the Wigner-SU(4) limit.  The NLO and \NNLO isovector M1 corrections in \eftnopi to $\sigma_{np}$ are also zero in the zero-recoil limit if
\begin{equation}
\label{eq:L1wig}
    L_1^{(0)}=-2\kappa_1\frac{Z-1}{2\gamma}\quad \quad  \text{and}  \quad \quad L_1^{(1)}=2\kappa_1\frac{(Z-1)^2}{2\gamma}.
\end{equation}
These values cancel off higher order corrections from the deuteron wavefunction renormalization, yielding an $np$ capture cross section that is zero not just at LO but also at NLO in \eftnopi. Eq.~\eqref{eq:L1wig} is equivalent to setting the isovector two-nucleon current in the nucleon (as opposed to auxiliary field) formalism \cite{Chen:1999tn} of \eftnopi to zero in the Wigner-SU(4) limit. (See App.~\ref{app:dual-expansion} for the matching between the auxiliary field formalism and the nucleon formalism.) At NNLO in \eftnopi the isoscalar M1 moment, Eq.~\eqref{eq:XNLO}, gives a non-zero contribution to $\sigma_{np}$.

The limit $\delta_r = 0$ can also be taken,  independent of the value for $\delta$, with errors at the N$^3$LO level in \eftnopi, as argued above. Using $\rho$ as defined in Eq.~\eqref{eq:Wignervalues} and fitting $L_1^{(0)}$ and $L_1^{(1)}$ to $\sigma_{np}$ at each order with $\delta_r = 0$, while taking the physical values of $\rho$, $\gamma$ and $\delta$, yields central values $L_1^{(0)} = -6.08$~fm and $L_1^{(0)} = 3.17$~fm.

\subsection{deuteron magnetic moment}

The NLO and \NNLO correction to the deuteron magnetic moment, Eq.~\eqref{eq:mud}, can be made zero (either in the Wigner-SU(4) limit or not) provided that
\begin{equation}
\label{eq:L2wig}
    L_2^{(0)}=-2\kappa_0\frac{Z_t-1}{2\gamma_t}\quad \quad \text{and} \quad \quad L_2^{(1)}=2\kappa_0\frac{(Z_t-1)^2}{2\gamma_t}.
\end{equation}
Similar to the argument above, Eq.~\eqref{eq:L2wig} is equivalent to setting the isoscalar two-nucleon current in the nucleon (as opposed to auxiliary field) formalism \cite{Chen:1999tn} of \eftnopi to zero. (See App.~\ref{app:dual-expansion} again.) Using these values for $L_2^{(0)}$ and $L_2^{(1)}$ gives the deuteron magnetic moment in the so-called Schmidt limit~\cite{Schmidt1937} in which the magnetic moment is given by the sum of the magnetic moments of unpaired nucleons. This is analogous to what is shown in Ref.~\cite{Vanasse:2017kgh}: in the Wigner-SU(4) limit if the values for $L_1^{(0)}$ and $L_2^{(0)}$ in Eqs.~\eqref{eq:L1wig} and~\eqref{eq:L2wig}, respectively, are chosen then the triton and $\jjvHe$ magnetic moments (in the absence of Coulomb interactions) reproduce the Schmidt limit in \EFT up to NLO. 

Similar to cold $np$ capture, for the deuteron magnetic moment the limit $\delta_r=0$ can be taken, independent of the value of $\delta$, with errors at the N$^3$LO level in \eftnopi. Fitting $L_2^{(0)}$ and $L_2^{(1)}$ to $\mu_d=0.85741 \frac{e}{2M_N}$ at each order with $\delta_r=0$, while using the physical values of $\rho$, $\gamma$ and $\delta$, yields central values $L_2^{(0)}=-1.25$~fm and $L_2^{(1)}=0.786$~fm. 

\subsection{triton magnetic moment}

We define
\begin{equation}
    \label{eq:specL1val}
    \deltaLoz=L_1^{(0)}+\kappa_1\rho\quad,\quad \deltaLoo =L_1^{(1)} -\kappa_1\left(\gamma(\rho^2+\delta_r^2)+2\rho\delta\delta_r\right),
\end{equation}
and
\begin{equation}
    \label{eq:specL2val}
    \deltaLtm{m}= L_2^{(m)}+(-1)^m2\kappa_0\frac{(Z_t-1)^{m+1}}{2\gamma_t}.
\end{equation}
In the Wigner-SU(4) limit, when $L_1^{(0)}$ and $L_1^{(1)}$ take on the values of Eq.~\eqref{eq:L1wig}, $\deltaLoz$ and $\deltaLoo$ will be zero.  Similarly, when $L_2^{(0)}$ and $L_2^{(1)}$ take on the values of Eq.~\eqref{eq:L2wig}, $\deltaLtz$ and $\deltaLto$ will be zero. If $\deltaLoz=0$, and physical values for $\gamma$, $\delta$, $\rho$, and $\delta_r$ are used, the NLO cold $np$ capture cross section is 347.6~mb, which agrees with experiment within $\approx 4\%$. Using the values for $L_2^{(m)}$ from Eq.~\eqref{eq:L2wig}, the (dimensionless) deuteron magnetic moment at \NNLO is $0.8798$, which agrees with experiment within $\approx 3\%$. These results suggest that $\deltaLom{m}$($\deltaLtm{m}$) can be treated as perturbative corrections to $L_1^{(m)}$ $\big(L_2^{(m)}\big)$.

In the Wigner-SU(4) limit, with $\deltaLoz = \deltaLtz = 0$, the LO \eftnopi prediction of the triton magnetic moment and its NLO correction  become~\cite{Vanasse:2017kgh}
\begin{equation}
    \mu_{0,W}^{\jjvH}=\mu_p\quad,\quad \mu_{1,W}^{\jjvH}=0,
\end{equation}
respectively, where $\mu_p=\kappa_0+\kappa_1$ is the proton magnetic moment and the subscript ($n, W$) indicates it is $n$-th order in \eftnopi in the Wigner-SU(4) limit with $\deltaLoz = \deltaLtz = 0$.  The \NNLO correction in \eftnopi to the triton magnetic moment in the Wigner-SU(4) limit with $\deltaLom{m}=\deltaLtm{m}=0$ is (see App.~\ref{app:magwigner} for details)
\begin{align}
\label{eq: NNLO mu_t in the Wigner limit}
&\mu_{2,W}^{\jjvH}=-\frac{4}{3}M_NH_2\frac{\Sigma_0^2(E_B)}{\Sigma_0'(E_B)}(\kappa_0+\kappa_1)-\frac{1}{\Omega H_{\mathrm{LO}}\Sigma_0'(E_B)}(\widetilde{\kappa}_0(\Lambda)-\widetilde{\kappa}_1(\Lambda)).
\end{align}
Using these values for the triton magnetic moment with Eq.~\eqref{eq:mufit}, the LEC for the three-nucleon magnetic moment counterterm in the Wigner-SU(4) limit with $\deltaLom{m}=\deltaLtm{m}=0$ is given by
\begin{equation}
\label{3magmomW}
    (\widetilde{\kappa}_0(\Lambda)-\widetilde{\kappa}_1(\Lambda))=-\Omega H_{\mathrm{LO}}\Sigma_0'(E_B)\left(\mu_t-\mu_p+\frac{4}{3}M_NH_2\frac{\Sigma_0^2(E_B)}{\Sigma'_0(E_B)}(\kappa_0+\kappa_1)\right).
\end{equation}

Similar to fits to cold $np$ capture and the deuteron magnetic moment, the limit $\delta_r = 0$ can be taken for the triton magnetic moment, with errors at the N$^3$LO level in \eftnopi. In this limit, fitting $L_1^{(0)}$ ($L_1^{(1)}$) to $\mu_t$ ($\sigma_{np}$), while using physical values for $\rho$, $\gamma$, and $\delta$, yields central values $L_1^{(0)} = -5.51$~fm and $L_1^{(0)} = 2.45$~fm. 

\subsection{cold $nd$ capture}

As in $np$ capture, the M1 moment for $nd$ capture (and threshold two-body triton  photo-disintegration) in the zero-recoil limit is also zero in the Wigner-SU(4) limit at LO in \EFT and at NLO  if $\deltaLoz=\deltaLtz=0$. 
Using the definitions in Eqs.~\eqref{eq:Wignervalues}, \eqref{eq:specL1val}, and \eqref{eq:specL2val},  the inhomogeneous term for an outgoing \DS channel, Eq.~\eqref{eq:LOinhommod}, can be written as
\begin{align}
    \label{eq:inhomwig}
    &{\Bbt^{[n]}}{}_{0\frac{1}{2},0\frac{1}{2}}^{\frac{1}{2},\frac{1}{2}}(p,k)=\frac{\delta}{\sqrt{3}k_0}4\tau_3\kappa_1\left(\!\!\begin{array}{rr}
    0 & 1\\
    -1 & 0
    \end{array}\!\!\right)\Dbb\left(E_B\right)\Gb_{n}\\\nonumber
    &-\frac{2\tau_3\kappa_1}{\sqrt{3}k_0}\left\{\delta_r(M_NE-\frac{3}{4}p^2)+\delta_r(\gamma^2+\delta^2)+2\rho\gamma\delta\right\}\left(\!\!\begin{array}{rr}
    0   & 1 \\
    -1     & 0
    \end{array}\!\!\right)\Dbb(E_B)\Gb_{n-1}\\\nonumber
    &+\frac{2\tau_3\kappa_1}{\sqrt{3}k_0}\left\{\left(\delta(\rho^2+\delta_r^2)+2\rho\gamma\delta_r\right)(M_NE-\frac{3}{4}p^2)\right.\\[-5mm]\nonumber
    &\hspace{2cm}\left.+2\gamma\rho\delta_r(\gamma^2+3\delta^2)+\delta(3\gamma^2+\delta^2)(\rho^2+\delta_r^2)\right\}\left(\!\!\begin{array}{rr}
    0   & 1 \\
    -1     & 0
    \end{array}\!\!\right)\Dbb(E_B)\Gb_{n-2}\\\nonumber
    &+\frac{M_N}{\sqrt{3}}\left(\begin{array}{cc}
    -2\deltaLtz & -\tau_3\deltaLoz+\frac{1}{2}\kappa_1\tau_3\delta_r\\
    -\tau_3\deltaLoz-\frac{1}{2}\kappa_1\tau_3\delta_r & 0
    \end{array}\right)\Dbb\left(E_B\right)\Gb_{n-1}\\\nonumber
    &+\frac{M_N}{\sqrt{3}}\left(\begin{array}{cc}
    -2\deltaLto & -\tau_3\deltaLoo-\kappa_1\tau_3(\delta(\rho^2+\delta_r^2)+2\rho\gamma\delta_r)\\
    -\tau_3\deltaLoo+\kappa_1\tau_3(\delta(\rho^2+\delta_r^2)+2\rho\gamma\delta_r) & 0
    \end{array}\right)\\\nonumber
    &\hspace{12cm}\times\Dbb\left(E_B\right)\Gb_{n-2}\\\nonumber
     &-\delta_{n2}\sqrt{3}\frac{4}{3}M_NH_2\Sigma_0(E_B)(\kappa_0-\tau_3\kappa_1)\oneb-\delta_{n2}\sqrt{3}\frac{1}{\Omega}(\widetilde{\kappa}_0(\Lambda)+\tau_3\widetilde{\kappa}_1(\Lambda))\oneb .
\end{align}
In the Wigner-SU(4) limit ($\delta = \delta_r = 0$) nearly every term drops out except terms with $\deltaLom{m}$ and $\deltaLtm{m}$, the three-body force $H_2$, and the three-nucleon magnetic moment counterterm.  Treating $\deltaLom{m}$ and $\deltaLtm{m}$ as perturbative corrections in the \eftnopi-Wigner-SU(4) dual expansion, we set them to zero and take the values of $L_1^{(m)}$ and $L_2^{(m)}$ in Eqs.~\eqref{eq:L1wig} and~\eqref{eq:L2wig}, respectively. (A more rigorous analysis of the power counting including $\deltaLom{m}$ and $\deltaLtm{m}$ in the dual expansion  requires more careful study.) 
In this case, the first non-vanishing contribution to the inhomogeneous term appears at \NNLO in \eftnopi and is from $H_2$ and the three-nucleon magnetic moment counterterm.
Thus, in the Wigner-SU(4) limit the M1 two-body triton photo-disintegration amplitude in the $J'=1/2$ channel is zero at LO and NLO in \eftnopi if the values of $L_1^{(0)}$ and $L_2^{(0)}$ that also make the M1 $np$ capture amplitude zero at  NLO in \eftnopi are used. This is also the case for the $J'=3/2$ channel, which is not shown here.

Combining Eq.~\eqref{3magmomW} with Eq.~\eqref{eq:inhomwig}, the inhomogeneous term ${\Bbt^{[n]}}{}_{L'S',LS}^{J,J'}(p,k)$ for an outgoing \DS channel in the Wigner-SU(4) limit with $\deltaLom{m}=\deltaLtm{m}=0$ is given by
\begin{align}
    \label{eq:inhomNNLOwig}
    &{\Bbt^{[n]}}{}_{0\frac{1}{2},0\frac{1}{2}}^{\frac{1}{2},\frac{1}{2}}(p,k)=-\delta_{n2}\sqrt{3}H_{\mathrm{LO}}\Sigma_0'(E_B)(\mu_t-\mu_p)\oneb.
\end{align}
Thus in the Wigner-SU(4) limit, along with $\deltaLom{m}=\deltaLtm{m}=0$, the first nonzero contribution to the inhomogenous ${\Bbt^{[n]}}{}_{L'S',LS}^{J,J'}(p,k)$ term for the two-body triton photo-disintegration amplitude appears at NNLO. As can be seen from Eq.~\eqref{eq:inhomwig}, the LO \eftnopi contribution is $\mathcal{O}(\delta)$ while the NNLO \eftnopi contribution (Eq.~\eqref{eq:inhomNNLOwig}) is $\mathcal{O}(\delta^0)$.  Using the combined Wigner-SU(4) and \EFT power counting suggests that both these terms are LO in the dual expansion and the LO inhomogeneous term in this modified counting is
\begin{align}
    \label{eq:inhomwigLO}
    &{\Bbt^{[0]}_W}{}_{0\frac{1}{2},0\frac{1}{2}}^{\frac{1}{2},\frac{1}{2}}(p,k)=\frac{\delta}{\sqrt{3}k_0}4\tau_3\kappa_1\left(\!\!\begin{array}{rr}
    0 & 1\\
    -1 & 0
    \end{array}\!\!\right)\Dbb\left(E_B\right)\Gb_{0}-\sqrt{3}H_{\mathrm{LO}}\Sigma_0'(E_B)(\mu_t-\mu_p)\oneb,
\end{align}
where the value for the three-nucleon magnetic moment counterterm is fixed by the difference between the triton magnetic moment and the proton magnetic moment. To be completely rigorous the vertex function and dibaryon propagator should also be expanded in powers of $\delta$.  This will become tedious at higher orders in \EFT because although the Wigner-SU(4) expansion is a good expansion for bound states it is expected to be a poor expansion for scattering states. Therefore, in two-body triton photo-disintegration only ${\Bbt^{[n]}}{}_{L'S',LS}^{J,J'}(p,k)$  should be expanded in powers of $\delta$, while the kernel and remaining inhomogeneous terms are  not expanded in powers of $\delta$.
If only the bound states are expanded in powers of $\delta$ and the same three-body force is used for scattering and bound states, the binding energy of the triton will change at each order in $\delta$ and associated corrections must be included in the two-body triton photo-disintegration amplitude.  Although feasible this is rather involved and we do not pursue this expansion further.

\section{\label{sec:obs}Observables}

The relationship between the two-body triton photo-disintegration amplitude in the spin basis and partial-wave basis is given by
\begin{align}
    &M_{m_1'm_2',\lambda m_2}(\vect{p},\vect{k})=\sum_{\alpha}\sqrt{4\pi}\sqrt{2L+1}\CG{1}{\frac{1}{2}}{S'}{m_1'}{m_2'}{m_S'}\CG{L}{S}{J}{0}{m_S}{M}\CG{L'}{S'}{J'}{m_{L'}}{m_S'}{M'}{Y_{L'}^{m_{L'}}}^{*}(\hat{\boldsymbol{p}})\\[-4mm]\nonumber
    &\hspace{8.5cm}\delta_{L0}\delta_{S\frac{1}{2}}\delta_{m_Sm_2}\Mc^{J'M',JM\lambda}_{L'S',LS}(p,k),
\end{align}
where $\Mc^{J'M',JM\lambda}_{L'S',LS}(p,k)$ is defined in Eq.~\eqref{eq:amprel}, $\lambda$ is the polarization of the incoming photon, $m_2$ is the spin of the triton, and $m_1'$ ($m_2'$) is the spin of the outgoing deuteron (neutron). $m_L$ ($m_L'$) and $m_S$ ($m_S'$) are the magnetic quantum numbers of $L$ ($L'$) and $S$ ($S'$) respectively.  $\alpha$ is a sum over all quantum numbers except $m_1'$, $m_2'$, $\lambda$, and $m_2$. $\vect{k}$ is chosen to be along the $z$-axis. Using Eq.~\eqref{eq:amprel} this can be written in terms of the amplitude calculated from the coupled integral equations yielding 
\begin{align}
    \label{eq:cross}
    &M_{m_1'm_2',\lambda m_2}(\vect{p},\vect{k})=\sum_{\alpha}\sqrt{4\pi}\sqrt{2L+1}\CG{1}{\frac{1}{2}}{S'}{m_1'}{m_2'}{m_S'}\CG{L}{S}{J}{0}{m_S}{M}\CG{L'}{S'}{J'}{m_{L'}}{m_S'}{M'}\\\nonumber    &\hspace{3cm}\delta_{L0}\delta_{S\frac{1}{2}}\delta_{m_Sm_2}{Y_{L'}^{m_{L'}}}^{*}(\hat{\boldsymbol{p}})T^{J,J'}_{L'S',LS}(p,k)\epsilon_{n\ell m}k_n\epsilon_{\gamma}^\ell(\lambda)\CG{J}{1}{J'}{M}{m}{M'}.
\end{align}
The unpolarized two-body triton photo-disintegration cross section in the zero-recoil limit is given by
 \begin{align}
 &\sigma=\frac{1}{4}\frac{1}{2k_0}\sum_{m_1',m_2',m_2,\lambda}\int\!\!\frac{d^3p_n}{(2\pi)^3}\int\!\!\frac{d^3p_d}{(2\pi)^3}\left|M_{m_1'm_2',\lambda m_2}(\vect{p},\vect{k})\right|^2\\\nonumber
&(2\pi)\delta\left(E_B+k_0-\frac{p_n^2}{2M_N}-\frac{p_d^2}{4M_N}+\frac{\gamma_d^2}{M_N}\right)(2\pi)^3\boldsymbol{\delta}^3\left(\vec{p}_n+\vec{p}_d\right).
\end{align}
Inserting Eq.~\eqref{eq:cross}, carrying out the integrals, and summing over the polarizations gives the expression
\begin{align}
&\sigma(\gamma t\to nd)=\frac{M_Np}{12\pi k_0}\sum_{\beta} k_0^2\frac{2}{3}(2J'+1)\left|T^{J,J'}_{L'S',LS}(p,k)\right|^2,
\end{align}
where $\beta$ is a sum over all quantum numbers.  

The cross-section of $nd\to \gamma t$ can be related to the cross-section for $\gamma t\to nd$ by detailed balance~\cite{bethe2006elementary}:
\begin{equation}
    \sigma(nd\to \gamma t)=\frac{2}{3}\frac{k_0^2}{p^2}\sigma(\gamma t\to nd),
\end{equation}
which finally gives the $nd$ capture cross section
\begin{align}
&\sigma(nd\to \gamma t)=\frac{M_Nk_0}{27\pi p}\sum_{\beta} k_0^2(2J'+1)\left|T^{J,J'}_{L'S',LS}(p,k)\right|^2
\end{align}
Expanding the amplitude perturbatively to \NNLO gives the \NNLO  $nd$ capture cross section
\begin{align}
&\sigma(nd\to \gamma t)=\frac{M_Nk_0}{27\pi p}\sum_{\beta} k_0^2(2J'+1)\\\nonumber
&\hspace{.5cm}\left\{\left|{T^{[0]}}{}^{J,J'}_{L'S',LS}(p,k)+{T^{[1]}}{}^{J,J'}_{L'S',LS}(p,k)\right|^2+2\mathrm{Re}\Bigg[{T^{[0]}}{}^{J,J'}_{L'S',LS}(p,k)\left({T^{[2]}}{}^{J,J'}_{L'S',LS}(p,k)\right)^*\Bigg]\right\}.
\end{align}
Although in principle all quantum numbers are necessary, at low energies it is sufficient to restrict ourselves to values $L=L'=0$ since terms with higher orbital angular momenta are suppressed.

\section{\label{sec:results}Results} 
\subsection{Predictions and fits}

\begin{figure}[hbt]
     \centering
     \includegraphics[width=125mm]{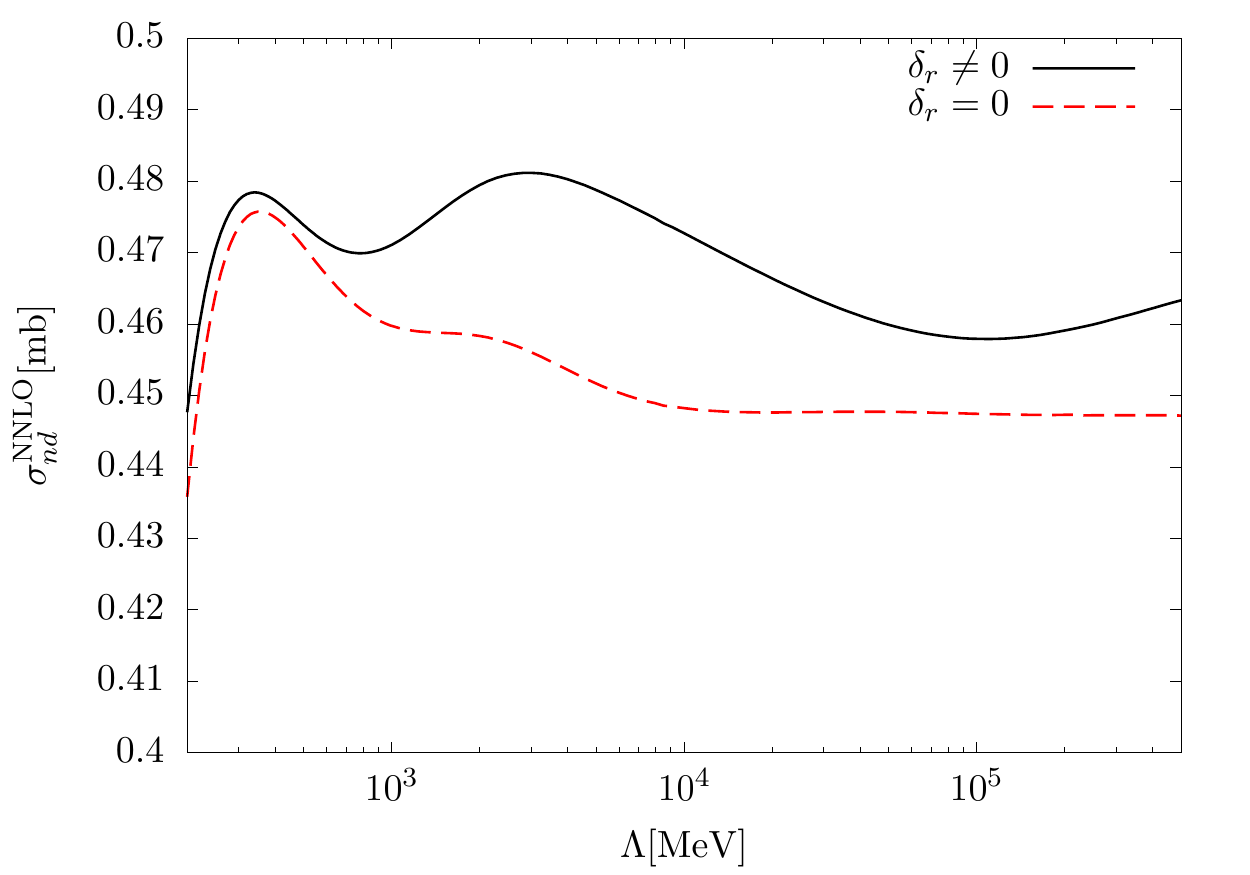}
     \caption{Plot of cutoff dependence of $\sigma_{nd}$ at \NNLO in \EFT for $\delta_r\neq 0$ and $\delta_r=0$. $L_1^{(0)}$ ($L_1^{(1)}$) is fit to $\mu_t$ ($\sigma_{np}$) at NLO (NNLO).  }
    \label{fig:cutoff-dep}
 \end{figure}

The dashed line in Fig.~\ref{fig:cutoff-dep} illustrates the cutoff convergence of $\sigma_{nd}$ with $\delta_r = 0$ at NNLO, while the solid line in Fig.~\ref{fig:cutoff-dep} shows an apparent convergence problem for $\sigma_{nd}$  with  $\delta_r \ne 0$ at NNLO.  This could be caused either by a slow convergence (that is, the current result will eventually converge) or by a divergence (that is, we are missing  a counterterm.). In principle using larger cutoffs could resolve this question, but at cutoffs $\Lambda> 10^6$~MeV numerical instabilities occur from cancellations of large numbers. Different numerical techniques will be needed in order to reach larger cutoffs and explore this question numerically.  A careful detailed asymptotic analysis could also address whether there is a slow convergence or divergence at \NNLO. This is left for future work.

In this section, we present results for $\sigma_{nd}$ up to NNLO when $\delta_r = 0$  (see Eq.~\eqref{eq:Wignervalues}). Choosing $\delta_r = 0$ is equivalent to taking the effective ranges in the $^3 S_1$ and  $^1 S_0$ channels to be equal.  Corrections to this limit are of the same size as N$^3$LO corrections. In taking $\delta_r=0$ we choose the value $\rho$ in Eq.~\eqref{eq:Wignervalues} for both the $^3S_1$ and $^1S_0$ channel.  For results using $\delta_r \ne 0$ see Appendix~\ref{app:finiteDr}. 

\begin{figure}[hbt]
    \centering
    \includegraphics{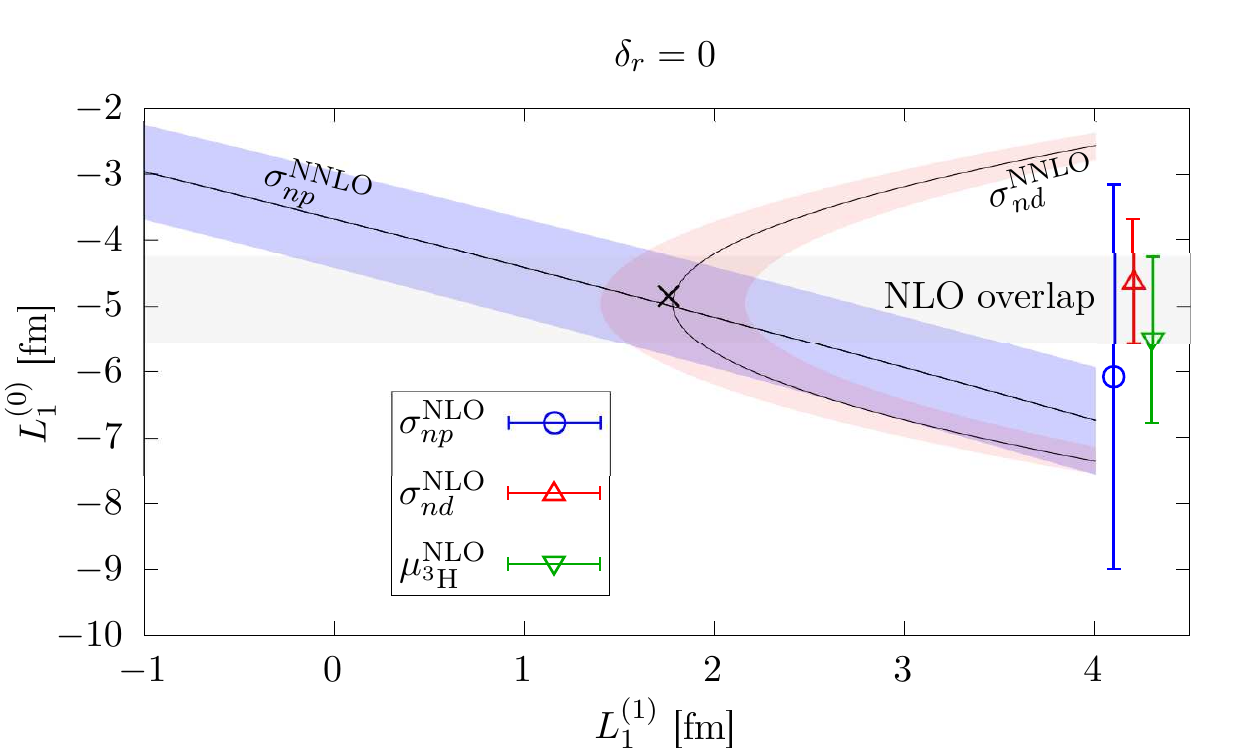}
    \caption{Plot of $L_1^{(0)}$ and $L_1^{(1)}$ dependence for  $\sigma_{np}$, $\sigma_{nd}$, and $\mu_{\jjvH}$ for $\delta_r=0$. Bands for $\sigma_{np}$ and $\sigma_{nd}$ give allowed values of $L_1^{(0)}$ and $L_1^{(1)}$ that reproduce the experimental values for observables within the naive theoretical uncertainty of \EFT at \NNLO. The symbols with error bars  are the range of values for $L_1^{(0)}$ that reproduce the experimental values for $\sigma_{np}$, $\sigma_{nd}$, and $\mu_t$ within naive theoretical uncertainty at NLO in \EFT (they are independent of $L_1^{(1)}$  and are shown on the right of the plot).  The values of $L_1^{(0)}$ that satisfy all three observables within naive theoretical uncertainty at NLO are shown as a gray band labeled ``NLO overlap."  Finally, the black ``$\times$" gives the value of $L_1^{(0)}$ and $L_1^{(1)}$ that is a best simultaneous fit to all three observables at NLO and $\sigma_{np}$ and $\sigma_{nd}$ at NNLO respectively.}
    \label{fig:L1plotWR}
\end{figure}

We fit $L_2^{(0)}$ and $L_2^{(1)}$ to the deuteron magnetic moment in the limit $\delta_r=0$ while keeping physical values for $\rho$. $\gamma$, and $\delta$.  Fig.~\ref{fig:L1plotWR} shows the $L_1^{(0)}$ and $L_1^{(1)}$ dependence of the cold $np$ capture cross section ($\sigma_{np}$) and the cold $nd$ capture cross section ($\sigma_{nd}$) at \NNLO for  $\delta_r=0$.  
The solid lines represent the values of $L_1^{(0)}$ and $L_1^{(1)}$ that reproduce the experimental value for the observable exactly, while the bands about the lines represent the values of $L_1^{(0)}$ and $L_1^{(1)}$ that reproduce the experimental value within the naive theoretical uncertainty of \EFT at \NNLO.  That naive theoretical error comes from the expansion parameter in \EFT, $Q \sim p/ m_\pi \sim$ $(Z_t-1)/2=0.3454$, or $Q^3\approx$ 4\% error at \NNLO.  The naive error on the observable is approximately the observable multiplied by $Q^n$ for an order $(n-1)$ result. Since this expansion is for the amplitude and the cross section comes from squaring the amplitude, this error is doubled for the cross section. While NNLO bands are shown for $\sigma_{np}$ and $\sigma_{nd}$, a \NNLO prediction of the triton magnetic moment ($\mu_{\jjvH}$) is not shown since the three-nucleon magnetic moment counterterm is fit to the experimental value for the triton magnetic moment at \NNLO.

In Fig.~\ref{fig:L1plotWR} we also indicate the range of values of $L_1^{(0)}$ that reproduce the experimental values for $\sigma_{np}$, $\sigma_{nd}$,  and $\mu_t$ at NLO. These are shown as symbols with error bars, placed at an arbitrary position on the horizontal axis because they are independent of  $L_1^{(1)}$.  The error bars are the naive NLO \eftnopi uncertainty.
The gray band, labeled ``NLO overlap," shows the range of values of $L_1^{(0)}$ that agree with all three experiments within naive theoretical errors at NLO. The \NNLO bands for $\sigma_{np}$ and $\sigma_{nd}$ have a region of overlap with the ``NLO overlap" band.  This means there are values of $L_1^{(0)}$ and $L_1^{(1)}$ that simultaneously satisfy $\sigma_{np}$, $\sigma_{nd}$, and $\mu_t$ within naive theoretical errors at NLO and \NNLO. The black cross in Fig.~\ref{fig:L1plotWR} is the location of the values of $L_1^{(0)}$ and $L_1^{(1)}$ that is obtained by a best simultaneous fit to to $\sigma_{np}$, $\sigma_{nd}$, and $\mu_t$ for $L_1^{(0)}$ at NLO and to $\sigma_{np}$ and $\sigma_{nd}$ for $L_1^{(1)}$ at \NNLO. All results have converged with respect to the cutoff.

Table~\ref{tab:L1depWr} shows the \EFT results\footnote{For benchmarking our results in Tables~\ref{tab:L1depWr}--\ref{tab:chandeltaneq0}, we provide more digits than the precision dictates.} for $\sigma_{np}$, $\sigma_{nd}$, and $\mu_{\jjvH}$ at LO, NLO, and \NNLO for several different choices of fitting $L_1^{(0)}$ and $L_1^{(1)}$. 
The  values for $L_1^{(m)}$ found from fitting to $\sigma_{np}$, $\sigma_{nd}$, or $\mu_t$ are given a superscript $p$, $d$, and $t$, respectively. Multiple superscripts indicate a simultaneous fit to those respective observables.
Errors given in Table~\ref{tab:L1depWr} are estimated in two ways. Those shown
in parentheses are the naive  uncertainty of \EFT for each observable arising from the $Q$ expansion. Errors shown in square brackets are the error on the observable that comes from propagating the error found in the LEC fit. Square bracket errors are only calculated for cases where $L_1^{(0)}$ $\big(L_1^{(1)}\big)$ is fit to a single observable at NLO (\NNLO). 
Details about this error propagation are given in App.~\ref{app:Errors}.  The errors obtained from these two different approaches are expected to be close to each other if the \eftnopi expansion is well behaved for the observables under consideration.\footnote{They could differ by a factor of two depending on if the observables are proportional to the amplitude or amplitude squared. See App.~\ref{app:Errors}.}

\begin{table}[hbt]
    \centering
    \begin{tabular}{|c|c|c|c|c|c|}
    \hline
        & $\sigma_{np}$ [mb] & $\sigma_{nd}$ [mb] &  $\mu_{\jjvH}$ & $L_1^{(0)}$~[fm]  & $L_1^{(1)}$~[fm] \\\hline
         LO & \hspace{1mm}325.2(224.6)\hspace{1mm} & 0.314(217) & \hspace{1.7mm}2.75(95) \hspace{1mm} &  - & - \\\hline
         \multirow{6}{*}{\shortstack[c]{LO\\[1mm]+\\[1mm]NLO}} & {334.2(79.7)} & 0.320(76)\phantom{4} & 2.82(34) & -6.08$^{p}$\phantom{[123]}& - \\
          & {334.2(79.7)} & 0.320[377] & 2.82[82] &   \hspace{1mm}-6.08[2.92]$^p$\hspace{1mm}& -  \\
          &349.6(83.4) & 0.393(94)\phantom{4} & {2.98(36)} &  -5.51$^t$\phantom{[123]}& - \\
          & 349.6[34.6] & 0.393[164] & {2.98(36)} &   -5.51[1.27]$^t$& - \\
          & 343.0(81.8)& 0.362(86)\phantom{4} & 2.91(35) &   -5.76$^{p,t}$\phantom{[12]}& - \\
          & 367.9(87.8)& 0.480(114) &3.17(38) &  -4.84$^{p,d,t}$\phantom{[1]}&  - \\\hline
         \multirow{6}{*}{\shortstack[c]{LO\\[1mm]+\\[1mm]NLO\\[1mm]+\\[1mm]\NNLO}} & {334.2(27.5)} & 0.408(34)\phantom{4} & 2.98(12) &  -6.08$^p$\phantom{[1.01]}& 3.17$^p$\phantom{[1.01]} \\
         & {334.2(27.5)} & 0.408[130] & {2.98[28]} & -6.08$^p$\phantom{[123]}& \hspace{1mm}3.17[{1.01}]$^p$\hspace{1mm} \\
         & {334.2(27.5)} & 0.447(37)\phantom{4} & {2.98(12)} & -5.51$^t$\phantom{[123]}& 2.45$^p\phantom{[1.01]}$ \\
         & {334.2(27.5)} & 0.447[130] & {2.98[28]} & -5.51$^t$\phantom{[123]}& 2.45[{1.01}]$^p$ \\
         & {334.2(27.5)} & 0.427(35)\phantom{4}   & {2.98(12)} & -5.76$^{p,t}$\phantom{[12]}& 2.76$^p$\phantom{[1.01]} \\
         & 339.3(28.0)  & 0.511(42)\phantom{4}  & {2.98(12)} & -4.84$^{p,d,t}$\phantom{[1]}& 1.76$^{p,d}$\phantom{[1.2]} \\\hline\hline
         Exp & 334.2$\pm$0.5 & 0.508$\pm$0.015 & 2.979 & - & - \\\hline
    \end{tabular}
    \caption{The values of $\sigma_{np}$ ,  $\sigma_{nd}$,  $\mu_{\jjvH}$, and  $L_1^{(0)}$ and $L_1^{(1)}$ where applicable, at different orders in \EFT with $\delta_r = 0$. Experimental values are also shown for $\sigma_{np}$~\cite{COX1965497}, $\sigma_{nd}$~\cite{Jurney:1982zz}, and $\mu_t$.  The superscripts $p$, $d$, and $t$ indicate the LEC is fit to $\sigma_{np}$, $\sigma_{nd}$, or $\mu_t$ respectively. Multiple superscripts indicate a simultaneous fit to those respective observables. The three-nucleon magnetic moment counterterm is fit to $\mu_t$ at \NNLO. Errors in the parenthesis are the naive theoretical uncertainty of \EFT for each observable. Errors in the square brackets are obtained by propagating the uncertainty of $L_1^{(0)}$ or $L_1^{(1)}$ shown in the square brackets and computed based on the theoretical error of the observable used to fit them. See App.~\ref{app:Errors} for details. The experimental error for $\mu_t$ is very small and thus not shown.}\label{tab:L1depWr}
\end{table}

At LO all predicted observables in Table~\ref{tab:L1depWr} are consistent with experiments within naive theoretical errors. At higher orders,  we use $L_2^{(0)}=-1.25$~fm  and $L_2^{(1)}=0.786$~fm, found by fitting to the deuteron magnetic moment in the limit $\delta_r=0$ while using physical values for $\rho$. $\gamma$, and $\delta$.  
Fitting $L_1^{(0)}$  at NLO to $\sigma_{np}$ alone gives $L_1^{(0)}$ a central value of $-6.08$~fm and yields (first row in LO + NLO block of Table~\ref{tab:L1depWr}) a NLO prediction  for $\sigma_{nd}$ that is well outside the experimental number within its naive theoretical uncertainty of 0.076~mb but a $\mu_{\jjvH}$  that agrees with the experimental number within its naive theoretical error.
In contrast, propagating the naive \eftnopi error of 79.7~mb for $\sigma_{np}$ at NLO gives an uncertainty of 2.92~fm for 
 $L_1^{(0)}$. Propagating this uncertainty for $L_1^{(0)}$ then gives  $\sigma_{nd}$ ($\mu_{\jjvH}$) an error of 0.377~mb (0.82 $\frac{e}{2M_N}$)
(second row in LO + NLO block of Table~\ref{tab:L1depWr}), and the prediction  for  $\sigma_{nd}$ now agrees  with experiment within these errors.
For  $\mu_{\jjvH}$, the ratio between the error of 0.82 $\frac{e}{2M_N}$ from the error propagation and its naive error of 0.34 $\frac{e}{2M_N}$ is close to a factor of two as expected (since $\mu_{\jjvH}$ is proportional to the amplitude whereas $\sigma_{nd}$ is proportional to the amplitude squared). For $\sigma_{nd}$, however, the ratio between the error of 0.377~mb from the error propagation and its naive error of 0.076~mb is much greater than one. In addition, we observe an increase of the errors for $\sigma_{nd}$ between the naive \eftnopi estimate at LO and the error propagation at NLO. This ratio between the NLO errors for $\sigma_{nd}$ obtained by the two methods and the increase of the error on $\sigma_{nd}$ between LO to NLO show that the ratio of the NLO  correction to $\sigma_{nd}$ over the LO result for $\sigma_{nd}$ is larger than the naive expectation from the \eftnopi expansion. This can be understood, as discussed in Sec.~\ref{sec:Wigner}, by the fact that the \eftnopi expansion alone is not the most appropriate expansion for computing $\sigma_{nd}$.
  
If $L_1^{(0)}$ at NLO is fit to  $\mu_t$ alone we find that $\sigma_{np}$ is reproduced within both the naive \eftnopi and propagated errors (third and fourth row of LO+NLO block of Table~\ref{tab:L1depWr}), while $\sigma_{nd}$ is only reproduced within the errors in the square brackets. Fitting $L_1^{(0)}$ to both $\sigma_{np}$ and $\mu_t$ we find (fifth row of LO+NLO block in Table~\ref{tab:L1depWr}) that both observables agree with experiment within naive theoretical error at NLO, but $\sigma_{nd}$ is underpredicted within naive theoretical error at NLO. A fit of $L_1^{(0)}$ to $\sigma_{np}$, $\sigma_{nd}$, and $\mu_t$ simultaneously seems to yield results for all three (last row of LO+NLO block in Table~\ref{tab:L1depWr}) that are consistent with experiment within naive theoretical errors. However, the results for $\sigma_{nd}$  for this simultaneous fit are misleading, since the LO and NLO values for the doublet channel (see Table~\ref{tab:sigmaJp-tritonfit}) are not consistent within naive errors.

To obtain the first five rows of the LO+NLO+NNLO block of Table~\ref{tab:L1depWr}, $L_1^{(1)}$ is fit to $\sigma_{np}$ at \NNLO.  The three-nucleon magnetic moment counter-term is then fit to $\mu_t$, and at \NNLO we can only predict $\sigma_{nd}$.  As can be seen from Table~\ref{tab:L1depWr}, fitting $L_1^{(0)}$ to $\sigma_{np}$ alone, or to $\sigma_{np}$ and $\mu_t$ simultaneously gives a \NNLO prediction for $\sigma_{nd}$ (first and fifth rows, respectively, of the LO+NLO+NNLO block of Table~\ref{tab:L1depWr}) that  underpredicts experiment within naive theoretical uncertainty, while using error propagation of $L_1^{(1)}$, for the fit to $\sigma_{np}$ (second row of LO+NLO+NNLO block of Table~\ref{tab:L1depWr}) yields consistency. Fitting $L_1^{(0)}$ to $\mu_t$ yields a \NNLO prediction for $\sigma_{nd}$ that agrees with experiment only if the propagated error is used (rows three and four of LO+NLO+NNLO block of Table~\ref{tab:L1depWr}). Performing a simultaneous fit of $L_1^{(0)}$  ($L_1^{(1)}$) to $\sigma_{np}$, $\sigma_{nd}$, and $\mu_t$ ($\sigma_{np}$ and $\sigma_{nd}$) at NLO (\NNLO) yields a value of $L_1^{(0)}$  ($L_1^{(1)}$) that gives predictions for all three observables that are consistent with experiment within naive theoretical uncertainty at NLO (\NNLO). However, the results for $\sigma_{nd}$ for this simultaneous fit are again misleading, since the NLO and \NNLO values for the quartet channel are not consistent within naive errors (see Table~\ref{tab:sigmaJp-tritonfit}).

\subsection{$\sigma_{nd}$ in the $J'=1/2$ and $J'=3/2$ channels}

Table \ref{tab:sigmaJp-tritonfit} shows $\sigma_{nd}$ in the incoming  doublet channel, $\sigma_{nd}(J'= \frac{1}{2})$, and  in the incoming quartet channel, $\sigma_{nd}(J'= \frac{3}{2})$, to each order in \eftnopi , using two different fits to obtain $L_1^{(0)}$  and $L_1^{(1)}$. In the top three rows of results in Table~\ref{tab:sigmaJp-tritonfit}, we show the LO results in row one, with naively propagated errors; the NLO results (row two) when  $L_1^{(0)}$ is fit to $\mu_t$; and the NNLO results (row three) when $L_1^{(1)}$ is fit to $\sigma_{np}$. Errors on both the NLO and NNLO results are found via error propagation through $L_1^{(m)}$. The errors for $\sigma_{nd}(tot)$ in the square brackets are given by the difference between the errors for $\sigma_{nd}$ in each channel, because a larger $L_1^{(0)}$ increases $\sigma_{nd}(J'=\frac{1}{2})$  while decreasing  $\sigma_{nd}(J'=\frac{3}{2})$; $\sigma_{nd}(J'=\frac{1}{2})$ is anti-correlated with $\sigma_{nd}(J'=\frac{3}{2})$ through $L_1^{(m)}$ (see App.~\ref{app:Errors}). For  this fit, $\sigma_{nd}(J'=\frac{1}{2})$ and $\sigma_{nd}(J'=\frac{3}{2})$ at LO (NLO)  agree with that at NLO (NNLO) within propagated errors.

\begin{table}[htb!]
\begin{tabular}{|l|c|c|c|}
    \hline
       &$\sigma_{nd}(J'=\frac{1}{2})$ [mb] & $\sigma_{nd}(J'=\frac{3}{2})$ [mb] &  $\sigma_{nd}(tot)$ [mb] \\\hline
    LO & 0.166(114)& 0.149(103)\phantom{4} & 0.314(217)\\
    +NLO$^t$ & 0.322[208] & 0.0704[445] & 0.393[164]\\
   +\NNLO \!$^p$ & 0.331[166]& 0.117[35]\phantom{44}\ & 0.447[130] \\\hline\hline
   LO & 0.166(114)& 0.149(103)\phantom{4} & 0.314(217)\\
    +NLO$^{p,d,t}$  & 0.433(103) & 0.0469(112)\phantom{4} & 0.480(114)\\
    +\NNLO \!$^{p,d}$   & 0.389(32)\phantom{4}& 0.122(10)\phantom{444} & 0.511(42)\phantom{4} \\\hline\hline
    Exp  & - & - & 0.508$\pm$0.015 \\\hline
\end{tabular}
    \caption{Individual contributions from the incoming $J'=1/2$ and $J'=3/2$ channels for $\sigma_{nd}$ at each order in \EFT in the limit $\delta_r=0$ with physical values for $\rho$, $\gamma$, and $\delta$. Two different fit procedures are shown, as is the experimental value for $\sigma_{nd}$~\cite{Jurney:1982zz}. In the top box of three rows, $L_1^{(0)}$ ($L_1^{(1)}$) is fit to $\mu_t$ ($\sigma_{np}$) at NLO (\NNLO). In the second box of three rows, $L_1^{(0)}$  ($L_1^{(1)}$) is fit to $\sigma_{np}$, $\sigma_{nd}$, and $\mu_t$ ($\sigma_{np}$ and $\sigma_{nd}$) simultaneously at NLO (NNLO). The error notation is the same as in Table~\ref{tab:L1depWr}. Details of error  propagation are discussed in  App.~\ref{app:Errors}. }
    \label{tab:sigmaJp-tritonfit}
\end{table}

In the lower box of three rows in Table~\ref{tab:sigmaJp-tritonfit}, we repeat the LO results in row one; the NLO results (row two) are now from a fit of $L_1^{(0)}$ to $\sigma_{np}$, $\sigma_{nd}$, and $\mu_t$ simultaneously; and the NNLO results (row three) are acquired using a fit of $L_1^{(1)}$ to $\sigma_{np}$ and  $\sigma_{nd}$ simultaneously. Errors shown in the parentheses are the naive \eftnopi errors.
The values for the total cross section ($\sigma_{nd}(tot)$) in the last column have already been presented in Table~\ref{tab:L1depWr}. 
Both fitting procedures indicate that while at LO $\sigma_{nd}(J'=\frac{1}{2})$ and $\sigma_{nd}(J'=\frac{3}{2})$ are of similar size, including NLO corrections yields a $\sigma_{nd}(J'=\frac{3}{2})$ that is significantly smaller than $\sigma_{nd}(J'=\frac{1}{2})$; there is a large cancellation between the LO contribution and the NLO correction to $\sigma_{nd}(J'=\frac{3}{2})$. After including NNLO corrections $\sigma_{nd}(J'=\frac{1}{2})$ remains dominant. The fact that the NLO corrections  to $\sigma_{nd}(J'=\frac{3}{2})$ are  comparable to the LO $\sigma_{nd}(J'=\frac{3}{2})$ suggests that the \EFT power counting may not be well-behaved.  In particular, at LO there are only contributions from one-nucleon currents that vanish in the Wigner-SU(4) limit while at NLO there are  currents that explicitly break Wigner-SU(4) symmetry and are not zero in the Wigner-SU(4) limit. In addition, although $\sigma_{nd}(tot)$ at each \eftnopi order agrees within naive errors with the next \EFT order, this is misleading, because the cross sections in each channel at LO and NLO (and also at NLO and NNLO in the $J'=3/2$ channel) do not agree within the naive theoretical uncertainties. A more rigorous error analysis propagating the errors from $L_1^{(m)}$ for this simultaneous fit should be carried out in the future. Extracting $\sigma_{nd}$ in each individual channel from experiment would also help clarify this situation.

\subsection{Comparison of $\sigma_{nd}$ with experiment and with other calculations}
\begin{figure}[hbt!]
    \centering
    \includegraphics[width=125mm]{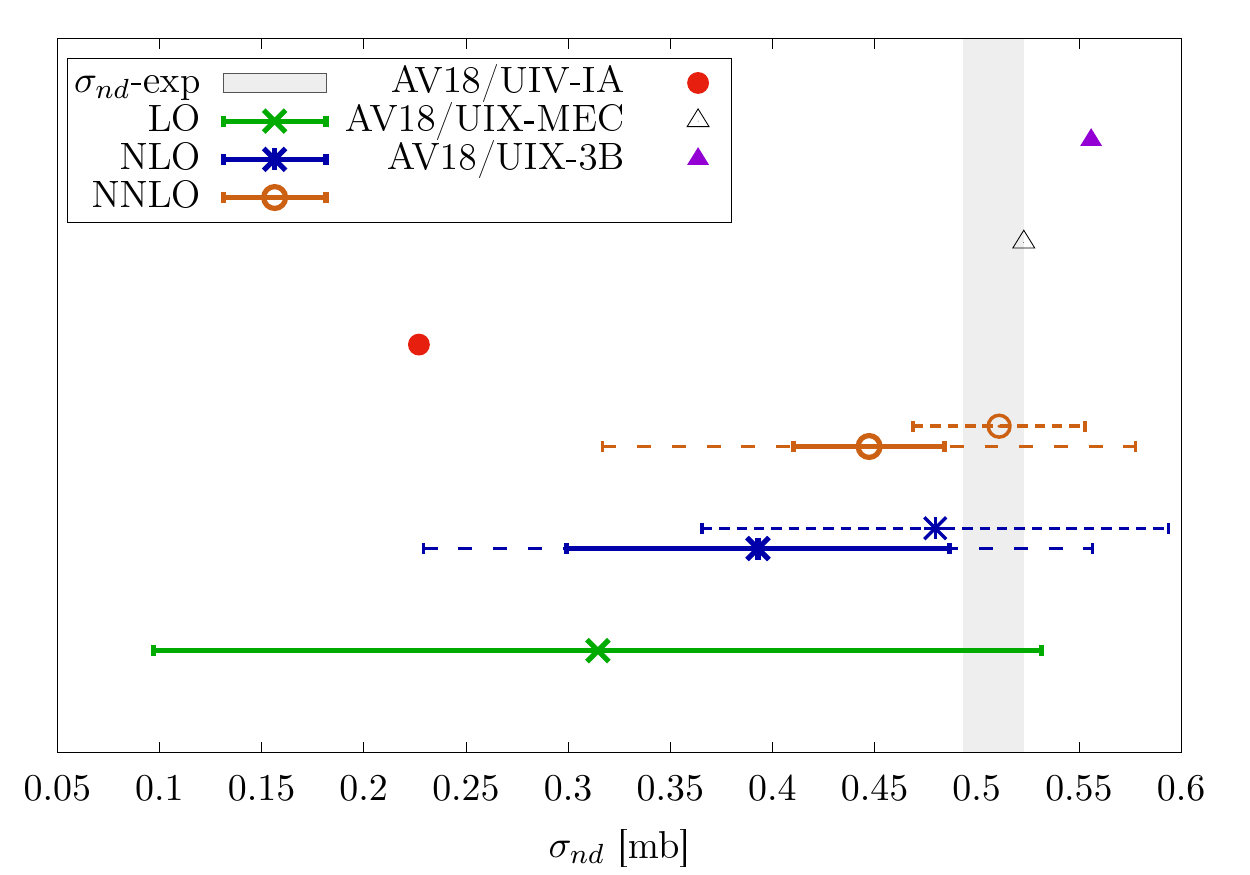} 
    \caption{Comparison of \EFT and potential model prediction for $\sigma_{nd}$ with experiment.  The shaded vertical band is the experimental value and its associated error.  The points without error bars come from the potential model calculations of Marcucci et al.~\cite{Marcucci:2005zc}.  \EFT calculations are shown for $\delta_r=0$.  The solid line shown with the LO \eftnopi result is the naive theoretical error estimate. For the NLO (NNLO) \eftnopi results where  $L_1^{(0)}$   ($L_1^{(1)}$) is fit to $\mu_{\jjvH}$ ($\sigma_{np}$), naive error bars appear as solid lines while the long dashed lines show the propagated error estimate. The NLO (NNLO) result with naive error bars where $L_1^{(0)}$   ($L_1^{(1)}$) is fit simultaneously to $\sigma_{np}$, $\sigma_{nd}$, and $\mu_t$ ($\sigma_{np}$ and $\sigma_{nd}$) is shown with short dashed lines. }
    \label{fig:capture-results}
\end{figure}

A comparison of our results for $\sigma_{nd}$ with calculations from Marcucci et al.~\cite{Marcucci:2005zc} is shown in Fig.~\ref{fig:capture-results}.
Ref.~\cite{Marcucci:2005zc} used the hyperspherical harmonic method with the Argonne $v_{18}$~\cite{Wiringa:1994wb} two-nucleon and Urbana IX~\cite{Pudliner:1995wk} three-nucleon potential to calculate the triton and $nd$ scattering wavefunction.  They calculated two- and three-nucleon currents by using meson exchange and minimal substitution. The solid red circle in Fig.~\ref{fig:capture-results} is the impulse approximation (IA) calculation of Ref.~\cite{Marcucci:2005zc}, which only includes one-nucleon currents, yielding 0.277~mb.  This value underpredicts the experimental value by roughly half.  Our LO in \eftnopi result is similar in that it only includes one-nucleon currents and similarly underpredicts the experimental value.  However, given the large error at LO our result is still consistent within theoretical \eftnopi error.  Including two-nucleon meson exchange currents (MEC) Ref.~\cite{Marcucci:2005zc} found 0.523~mb, and including three-body (3B) currents, they obtained 0.556~mb, with the former falling within the experimental error.  An earlier potential model calculation of $\sigma_{nd}$ can be found in Ref.~\cite{Viviani:1996ui}. A $\chi$EFT calculation of $\sigma_{nd}$ was also carried out ~\cite{Girlanda:2010vm}.
Using heavy-baryon chiral perturbation theory to N$^3$LO Ref.~\cite{Song:2008zf} found $\sigma_{nd}=0.490\pm0.008$~mb.  Their power counting did not include three-nucleon currents up to the order they were working.

The \EFT results shown in Fig~\ref{fig:capture-results} are for $\delta_r=0$ with physical values for $\rho$, $\gamma$, and $\delta$.  The solid NLO line shows the naive error when $L_1^{(0)}$ is fit to the triton magnetic moment, while the long dashed line shows the propagated error for the same fit of $L_1^{(0)}$. At NNLO the solid line again shows the naive error when $L_1^{(1)}$ is fit to $\sigma_{np}$, while the long dashed line shows the propagated error for the same fit.  Only the propagated errors at each order for these fits are consistent with experiment.  The short dashed NLO line shows the naive error when $L_1^{(0)}$ is fit simultaneously to $\sigma_{np}$, $\sigma_{nd}$, and $\mu_{\jjvH}$, while the short dashed line at \NNLO shows the naive error when $L_1^{(1)}$ is fit to $\sigma_{np}$ and $\sigma_{nd}$.  Both of these are consistent with experiment within naive theoretical uncertainties.  However, as pointed out in the previous subsection, this could be misleading since $\sigma_{nd}$ in the individual quartet and doublet channels changes  dramatically at each order, even though the total cross section $\sigma_{nd}(tot)$ at each \EFT order  agrees within naive theoretical uncertainties with the next order in \EFT.

\subsection{Correlation between $\sigma_{nd}$ and $a_{nd}$}

\begin{figure}[hbt]
     \centering
     \includegraphics[width=125mm]{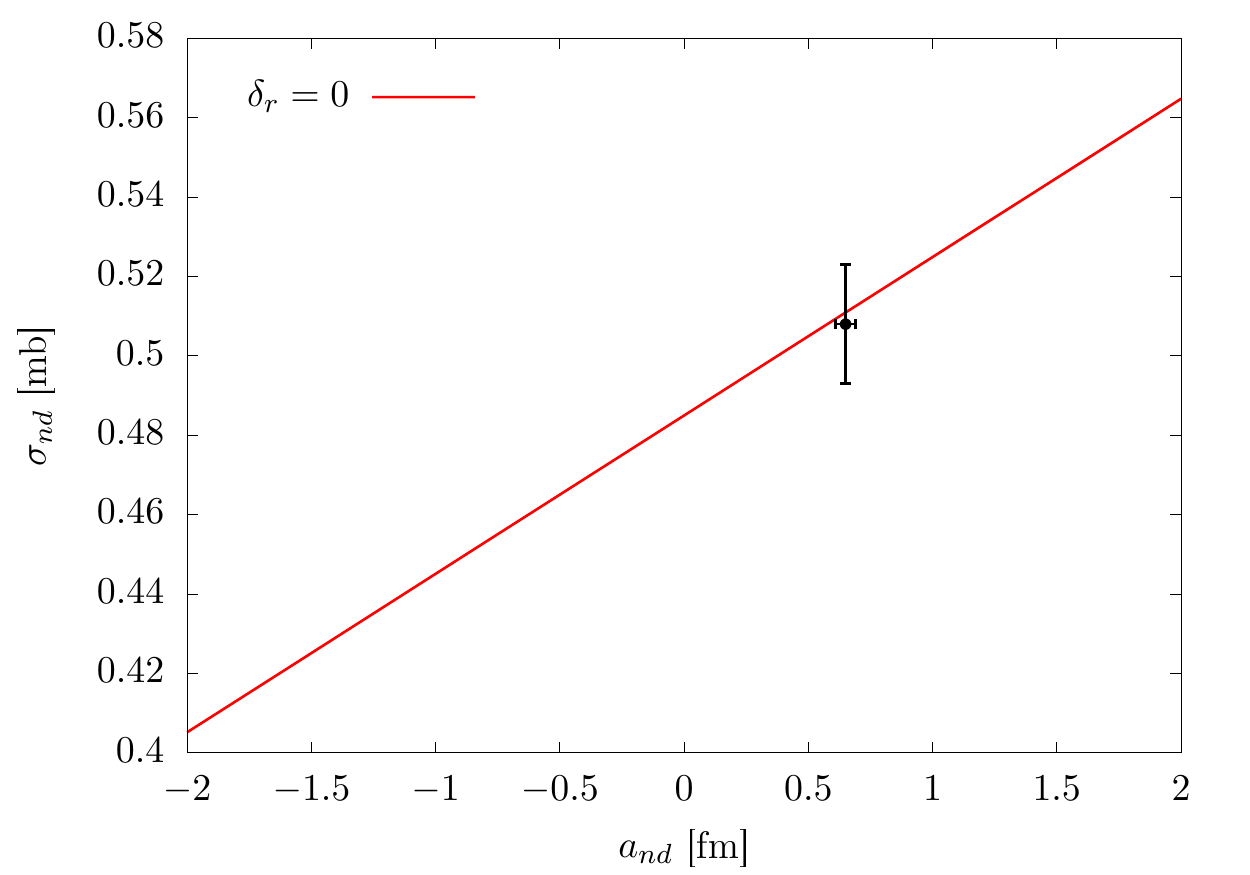}
     \caption{Plot of $a_{nd}$ vs. $\sigma_{nd}$ at \NNLO in \EFT for $\delta_r=0$.  The black dot (error bar) corresponds to the experimental value (uncertainty) for $a_{nd}$ and $\sigma_{nd}$.  \EFT results are for fitting $L_1^{(0)}$   ($L_1^{(1)}$) to $\sigma_{np}$, $\sigma_{nd}$, and $\mu_t$ simultaneously at NLO ($\sigma_{np}$ and $\sigma_{nd}$ simultaneously at NNLO). Theoretical errors are not shown here.
     }
    \label{fig:correlation}
 \end{figure}

The correlation between the doublet $S$-wave $nd$ scattering length $a_{nd}$ and $\sigma_{nd}$ is shown in Fig.~\ref{fig:correlation} for $\delta_r=0$ with physical values for $\rho$, $\gamma$, and $\delta$.
  That the correlation is linear comes as no surprise since at \NNLO both $a_{nd}$ and $\sigma_{nd}$ have a linear relationship to the \NNLO energy-dependent three-body force $H_2(\Lambda)$ and therefore have a linear relationship to each other.   The black dot with vertical and horizontal error bars in Fig.~\ref{fig:correlation} corresponds to the experimental values for $a_{nd}$ and $\sigma_{nd}$ and their respective errors. This correlation serves as an essential benchmark for any calculation of cold $nd$ capture.

\section{\label{sec:conclusion}Summary}

Using \EFT we calculated the cold $nd$ capture cross section to \NNLO.  We found that $\sigma_{nd}$ is sensitive to the isovector two-nucleon current in \EFT, similar to what is found using potential model calculations \cite{Marcucci:2005zc, Viviani:1996ui}. In addition, we found that a three-nucleon current is required at \NNLO for RG invariance of both $\sigma_{nd}$ and $\mu_{\jjvH}$ and we fit its value to reproduce the triton magnetic moment.  When using physical values for the two-nucleon effective ranges, we see a small residual cutoff dependence in the \NNLO calculation for $\sigma_{nd}$,  from either  a slow convergence or a possible divergence. This residual cutoff dependence goes away when the two-nucleon effective ranges are identical, i.e., $\delta_r=0$, which is a good approximation since the effective ranges in  the $^3 S_1$ and $^1 S_0$ channels are experimentally close to each other. Therefore, the $\delta_r=0$ results are presented in the body of this work.

Fitting the LEC of the two-nucleon isovector current $L_1^{(0)}$ ($L_1^{(1)}$) to $\mu_t$ at NLO ($\sigma_{np}$ at \NNLO) yields a NLO (\NNLO) prediction of $\sigma_{nd} = 0.393[164]$~mb ($\sigma_{nd} = 0.447[130]$~mb) when $\delta_r = 0$. The  errors in square brackets are obtained by error propagation of $L_1^{(m)}$. This agrees with the experimental value of 0.508(15)~mb~\cite{Jurney:1982zz} at a neutron laboratory velocity of 2200~m/s within propagated theoretical error at NLO and \NNLO.
Alternatively, fitting $L_1^{(0)}$ ($L_1^{(1)}$) simultaneously to $\sigma_{np}$, $\sigma_{nd}$, and $\mu_t$  ($\sigma_{np}$ and $\sigma_{nd}$) at NLO (\NNLO)  yields   $\sigma_{nd}=0.480(114)$~mb ($\sigma_{nd}=0.511(42)$~mb). For this choice of $L_1^{(0)}$ and $L_1^{(1)}$  we find that $\sigma_{nd}$, $\sigma_{np}$ and $\mu_{\jjvH}$ all agree with their experimental values using a naive estimate of theoretical uncertainty at LO, NLO, and \NNLO.

By repeated use of the integral equations for the triton vertex function and $nd$ scattering, as well as a shift in the definition of the integral equation for the two-body triton photo-disintegration amplitude, we were able to markedly simplify the expressions for the two-body triton photo-disintegration amplitude integral equations in the zero-recoil limit.  These simplified expressions exhibit the Wigner-SU(4) symmetry properties of the $nd$ capture amplitude, and make it readily apparent that in the Wigner-SU(4) limit ($\delta=\delta_r=0$) the one-nucleon current contributions to the $nd$ capture amplitude vanish.  Since the scale $\delta$ is smaller than the scale of three-body binding in the triton, the one-nucleon contributions are suppressed.
This suggests the utility of a dual power counting in powers of the Wigner-SU(4) breaking parameter $\delta$ and normal \EFT power counting as in Ref.~\cite{Vanasse:2016umz}.  At LO (first non-vanishing order) in the dual expansion (using Eq.~\eqref{eq:inhomwigLO})   $\sigma_{nd}=0.511$(353)~mb, surprisingly close to the experimental number (this error is naively estimated in the dual expansion using $Q\sigma_{nd}$).  However, we do not pursue this dual expansion  to higher orders here because while the  triton vertex function is perturbative in Wigner-SU(4) corrections,  low energy $nd$ scattering is not, which complicates the analysis. 

With the ability to calculate $\sigma_{nd}$ fully perturbatively to \NNLO we can now pursue the time-reversed process of two-body triton photo-disintegration.  However, at the higher energies generally considered for two-body triton photo-disintegration we will need to include the contribution from  electric dipole transitions arising from minimally coupled radiation photons.  We could also include the contribution to $\sigma_{nd}$ from the electric quadrupole moment arising from $SD$ mixing in the two-nucleon sector, which is strictly \NNLO, but potential model calculations~\cite{Marcucci:2005zc, Viviani:1996ui} indicate its impact is less than the \eftnopi theoretical error at \NNLO. Given that our  method yields expressions where the external nucleon and dibaryon legs are off-shell, we could also calculate Compton scattering and three-body photo-disintegration.  However, the energies relevant for three-body photo-disintegration may be too close to the breakdown scale of \EFT to make such calculations worthwhile. 

\appendix

\acknowledgments{We thank Daniel Phillips and Sebastian K{\"o}nig for useful discussions.  XL, HS, and RPS have received support from the U.S. Department of Energy, Office of Science, Office of Nuclear Physics, under Award Number DE-FG02-05ER41368.
HS is also funded 
in part by the DOE QuantISED program through the theory consortium ``Intersections of QIS and Theoretical Particle Physics'' at Fermilab with Fermilab Subcontract No. 666484,
in part by the Institute for Nuclear Theory with US Department of Energy Grant No. DE-FG02-00ER41132,
and in part by U.S. Department of Energy, Office of Science, Office of Nuclear Physics, InQubator for Quantum Simulation (IQuS) under Award No. DOE (NP) DE-SC0020970.
}

\section{\label{app:zerorecoil}The Zero-Recoil Limit}

Using the integral equations for the vertex functions and the definition of $\Sigma_n(E_B)$ the integral equation for the M1 moment of the two-body triton photo-disintegration amplitude can be simplified considerably.  With these simplifications summing all of the diagrams in the outgoing \DS channel without final state interactions in the zero-recoil limit and combining like terms gives
\begin{align}
    &{\Bb^{[n]}}_{0\frac{1}{2},0\frac{1}{2}}^{\frac{1}{2},\frac{1}{2}}(p,k)=-\frac{M_N}{\sqrt{3}}\left(\begin{array}{cc}
    4\kappa_0 & 2\tau_3\kappa_1 \\
    2\tau_3\kappa_1 & 0
    \end{array}\right)\Db\left(E_B-\frac{p^2}{2M_N},\vectn{p}\right)\Gb_n\left(p,E_B\right)\\\nonumber
    &\hspace{1cm}\frac{1}{\sqrt{\frac{3}{4}p^2-M_NE_B}+\sqrt{\frac{3}{4}p^2-M_NE-i\epsilon}}\\\nonumber
    &-\frac{1}{\sqrt{3}k_0}\left(\begin{array}{cc}
    4\kappa_0 & 2\tau_3\kappa_1\\
    2\tau_3\kappa_1 & 0
    \end{array}\right)\left\{\Gb_n(E_B,p)-\sum_{m=1}^n\mathbf{R}_m\left(E_B-\frac{p^2}{2M_N},\vectn{p}\right)\Gb_{n-m}(E_B,p)\right\}\\\nonumber
    &+\frac{1}{\sqrt{3}k_0}\left(\begin{array}{cc}
    -\frac{1}{2}\kappa_0 & -\frac{3}{2}\kappa_0+2\tau_3\kappa_1\\
    \frac{3}{2}\kappa_0+2\tau_3\kappa_1 & \frac{9}{2}\kappa_0
    \end{array}\right)\left\{\Gbt_n(p,E)-\sum_{m=1}^n\mathbf{R}_m\left(E-\frac{p^2}{2M_N},\vectn{p}\right)\Gbt_{n-m}(p,E)\right\}\\\nonumber
    &+\sum_{m=0}^n\frac{\pi H^{(m)}(\Lambda)}{3\sqrt{3}k_0}\left(\begin{array}{rr}
 0 & -12\kappa_0\\
0 & 12\kappa_0
 \end{array}
 \right)\Db\left(E_B-\frac{q^2}{2M_N},\vectn{q}\right)\otimes_q\Gb_{n-m}(E_B,q)\\\nonumber
 &+\delta_{n0}\frac{1}{\sqrt{3}k_0}\left(\begin{array}{cc}
    \frac{9}{2}\kappa_0 & \frac{3}{2}\kappa_0\\
    -\frac{3}{2}\kappa_0 & -\frac{1}{2}\kappa_0
    \end{array}\right)\oneb+\delta_{n2}\frac{4}{3}M_Nk_0H_2\Sigma_0(E_B)\frac{\kappa_0+\tau_3\kappa_1}{\sqrt{3}k_0}\oneb\\\nonumber
    &-\frac{M_N}{\sqrt{3}}\sum_{m=1}^n\left(\begin{array}{cc}
    2L_2^{(m-1)} & \tau_3 L_1^{(m-1)} \\
    \tau_3L_1^{(m-1)} & 0
    \end{array}\right)\Db\left(E_B-\frac{p^2}{2M_N},\vectn{p}\right)\Gb_{n-m}\left(p,E_B\right).
\end{align}
To simplify this further we note a c.c.~space matrix $\mathbf{M}$ times the boosted vertex functions can be written in terms of the c.m.~vertex function by using the integral equation for the boosted vertex function, yielding 
\begin{align}
    &\mathbf{M}\left\{\Gbt_n(p,E)-\sum_{m=1}^n\mathbf{R}_m\left(E-\frac{p^2}{2M_N},\vectn{p}\right)\Gbt_{n-m}(p,E)\right\}=\delta_{n0}\mathbf{M}\oneb\\\nonumber
    &-\frac{2\pi}{qp}Q_0\left(\frac{q^2+p^2-M_NE}{qp}\right)\mathbf{M}\left(\begin{array}{rr}
    1 & -3 \\
    -3 & 1
    \end{array}\right)\otimes_q\Db\left(E_B-\frac{q^2}{2M_N},\vectn{q}\right)\Gb_n(E_B,q).
\end{align}
Using the identity
\begin{equation}
    \mathbf{1}=\left(\begin{array}{rr}
    1 & -3\\
    -3 & 1
    \end{array}\right)\Db\left(E-\frac{q^2}{2M_N},\vectn{q}\right)\Db^{-1}\left(E-\frac{q^2}{2M_N},\vectn{q}\right)\left(-\frac{1}{8}\right)\left(\begin{array}{cc}
    1 & 3 \\
    3 & 1
    \end{array}\right) ,
\end{equation}
the expression can be rewritten as
\begin{align}
    &\mathbf{M}\left\{\Gbt_n(p,E)-\sum_{m=1}^n\mathbf{R}_m\left(E-\frac{p^2}{2M_N},\vectn{p}\right)\Gbt_{n-m}(p,E)\right\}=\delta_{n0}\mathbf{M}\oneb\\\nonumber
    &-\frac{2\pi}{qp}Q_0\left(\frac{q^2+p^2-M_NE}{qp}\right)\left(\begin{array}{rr}
    1 & -3\\
    -3 & 1
    \end{array}\right)\Db\left(E-\frac{q^2}{2M_N},\vectn{q}\right)\\\nonumber
    &\otimes_q\left[\Db^{-1}\left(E-\frac{q^2}{2M_N},\vectn{q}\right)\left(-\frac{1}{8}\right)\left(\begin{array}{cc}
    1 & 3 \\
    3 & 1
    \end{array}\right)\mathbf{M}\left(\begin{array}{rr}
    1 & -3 \\
    -3 & 1
    \end{array}\right)\Db\left(E_B-\frac{q^2}{2M_N},\vectn{q}\right)\Gb_n(E_B,q)\right].
\end{align}
The second line of this equation is nearly the kernel for the $nd$ scattering integral equation, except it is missing the three-body force.  Defining the function $\Jb_n(q,E)$
\begin{align}
    &\Jb_n(q,E)=\\\nonumber
    &\Db^{-1}\left(E-\frac{q^2}{2M_N},\vectn{q}\right)\left(-\frac{1}{8}\right)\left(\begin{array}{cc}
    1 & 3 \\
    3 & 1
    \end{array}\right)\mathbf{M}\left(\begin{array}{rr}
    1 & -3 \\
    -3 & 1
    \end{array}\right)\Db\left(E_B-\frac{q^2}{2M_N},\vectn{q}\right)\Gb_n(E_B,q),
\end{align}
the expression can be simplified to
\begin{align}
    &\mathbf{M}\left\{\Gbt_n(p,E)-\sum_{m=1}^n\mathbf{R}_m\left(E-\frac{p^2}{2M_N},\vectn{p}\right)\Gbt_{n-m}(p,E)\right\}=\delta_{n0}\mathbf{M}\oneb\\\nonumber
    &-\frac{2\pi}{qp}Q_0\left(\frac{q^2+p^2-M_NE}{qp}\right)\left(\begin{array}{rr}
    1 & -3\\
    -3 & 1
    \end{array}\right)\Db\left(E-\frac{q^2}{2M_N},\vectn{q}\right)\otimes_q\Jb_n(q,E).
\end{align}
To bring this into agreement with the kernel for $nd$ scattering we add and subtract a three-body force term:
\begin{align}
    &\mathbf{M}\left\{\Gbt_n(p,E)-\sum_{m=1}^n\mathbf{R}_m\left(E-\frac{p^2}{2M_N},\vectn{p}\right)\Gbt_{n-m}(p,E)\right\}=\delta_{n0}\mathbf{M}\oneb\\\nonumber
    &-\left[\frac{2\pi}{qp}Q_0\left(\frac{q^2+p^2-M_NE}{qp}\right)\left(\begin{array}{rr}
    1 & -3\\
    -3 & 1
    \end{array}\right)+\pi H_{\mathrm{LO}}\left(\begin{array}{rr}
    1 & -1\\
    -1 & 1
    \end{array}\right)\right]\Db\left(E-\frac{q^2}{2M_N},\vectn{q}\right)\otimes_q\Jb_n(q,E)\\\nonumber
    &+\pi H_{\mathrm{LO}}\left(\begin{array}{rr}
    1 & -1\\
    -1 & 1
    \end{array}\right)\Db\left(E-\frac{q^2}{2M_N},\vectn{q}\right)\otimes_q\Jb_n(q,E).
\end{align}
Finally, the second line can be written in terms of the $nd$ scattering integral equation kernel in the doublet channel, Eq.~\eqref{eq:scattkernel}, giving
\begin{align}
    \label{eq:scattid}
    &\mathbf{M}\left\{\Gbt_n(p,E)-\sum_{m=1}^n\mathbf{R}_m\left(E-\frac{p^2}{2M_N},\vectn{p}\right)\Gbt_{n-m}(p,E)\right\}=\delta_{n0}\mathbf{M}\oneb\\\nonumber
    &+\Kb^{\frac{1}{2}}_{0\frac{1}{2},0\frac{1}{2}}(q,p,E)\Db\left(E-\frac{q^2}{2M_N},\vectn{q}\right)\otimes_q\Jb_n(q,E)\\\nonumber
    &+\pi H_{\mathrm{LO}}\left(\begin{array}{rr}
    1 & -1\\
    -1 & 1
    \end{array}\right)\Db\left(E-\frac{q^2}{2M_N},\vectn{q}\right)\otimes_q\Jb_n(q,E).
\end{align}
Thus the contribution to the inhomogeneous term of the integral equation for the two-body triton photo-disintegration amplitude from the boosted vertex functions can be rewritten in terms of the function $\Jb_n(q,E)$ that only depends on the unboosted vertex function.  In addition, the term with the $nd$ scattering integral equation kernel can be absorbed into the definition of the kernel for the integral equation of the two-body triton photo-disintegration amplitude with outgoing \DS channel.  To do this we redefine the two-body triton photo-disintegration amplitude ${\Tb^{[n]}}{}^{J,J'}_{L'S',LS}(p,k)$ by
\begin{align}
    \label{eq:redef}
     &{\Tbt^{[n]}}{}^{J,J'}_{L'S',LS}(p,k)
    ={\Tb^{[n]}}{}^{J,J'}_{L'S',LS}(p,k)+\Jb_n(p,E).
 \end{align}
 
 This causes the second line of Eq.~\eqref{eq:scattid} to be absorbed into the integral equation, and the third line to be added to the inhomogeneous term for the integral equation of ${\Tbt^{[n]}}{}^{J,J'}_{L'S',LS}(p,k)$.  At LO $\Jb_0(p,E)$ must be added to the inhomogenous term for ${\Tbt^{[0]}}{}^{J,J'}_{L'S',LS}(p,k)$ and at higher orders $\Jb_n(p,E)$ multiplied by the functions $\mathbf{R}_m\left(E-\frac{p^2}{2M_N},\vectn{p}\right)$ is added to the inhomogenous term for ${\Tbt^{[n]}}{}^{J,J'}_{L'S',LS}(p,k)$ due to the redefinition.  The resulting inhomogenous term for the integral equation of ${\Tbt^{[n]}}{}^{J,J'}_{L'S',LS}(p,k)$ after combining like terms is given by
\begin{align}
    &{\Bbt^{[n]}}{}_{0\frac{1}{2},0\frac{1}{2}}^{\frac{1}{2},\frac{1}{2}}(p,k)=-\frac{M_N}{\sqrt{3}}\left(\begin{array}{cc}
    4\kappa_0 & 2\tau_3\kappa_1 \\
    2\tau_3\kappa_1 & 0
    \end{array}\right)\Db\left(E_B-\frac{p^2}{2M_N},\vectn{p}\right)\Gb_n\left(p,E_B\right)\\\nonumber
    &\hspace{1cm}\frac{1}{\sqrt{\frac{3}{4}p^2-M_NE_B}+\sqrt{\frac{3}{4}p^2-M_NE-i\epsilon}}\\\nonumber
    &-\frac{1}{\sqrt{3}k_0}\left(\begin{array}{cc}
    4\kappa_0 & 2\tau_3\kappa_1\\
    2\tau_3\kappa_1 & 0
    \end{array}\right)\left\{\Gb_n(E_B,p)-\sum_{m=1}^n\mathbf{R}_m\left(E_B-\frac{p^2}{2M_N},\vectn{p}\right)\Gb_{n-m}(E_B,p)\right\}\\\nonumber
    &+\frac{1}{\sqrt{3}k_0}\Db^{-1}\left(E-\frac{p^2}{2M_N},\vectn{p}\right)\left(\begin{array}{cc}
    4\kappa_0 & 2\tau_3\kappa_1\\
     2\tau_3\kappa_1 & 0
    \end{array}\right)\Db\left(E_B-\frac{p^2}{2M_N},\vectn{p}\right)\Gb_n(p,E_B)\\\nonumber
    &-\frac{1}{\sqrt{3}k_0}\sum_{m=0}^n\mathbf{R}_m\left(E-\frac{p^2}{2M_N},\vectn{p}\right)\Db^{-1}\left(E-\frac{p^2}{2M_N},\vectn{p}\right)\left(\begin{array}{cc}
    4\kappa_0 & 2\tau_3\kappa_1\\
     2\tau_3\kappa_1 & 0
    \end{array}\right)\\\nonumber
    &\hspace{9.5cm}\Db\left(E_B-\frac{p^2}{2M_N},\vectn{p}\right)\Gb_{n-m}(p,E_B)\\\nonumber
    &+\sum_{m=0}^n\frac{\pi H^{(m)}}{\sqrt{3}k_0}\left(\begin{array}{cc}
     4\kappa_0-2\tau_3\kappa_1&-4\kappa_0+2\tau_3\kappa_1 \\
     -4\kappa_0+2\tau_3\kappa_1&4\kappa_0-2\tau_3\kappa_1
    \end{array}\right)\Db\left(E_B-\frac{q^2}{2M_N},\vectn{q}\right)\otimes_q\Gb_{n-m}(E_B,q)\\\nonumber
 &+\delta_{n0}\frac{1}{\sqrt{3}k_0}\left(\begin{array}{cc}
    4\kappa_0 & 2\tau_3\kappa_1\\
    2\tau_3\kappa_1 & 4\kappa_0
    \end{array}\right)\oneb+\delta_{n2}\frac{4}{3}M_Nk_0H_2\Sigma_0(E_B)\frac{\kappa_0+\tau_3\kappa_1}{\sqrt{3}k_0}\oneb\\\nonumber
    &-\frac{M_N}{\sqrt{3}}\sum_{m=1}^n\left(\begin{array}{cc}
    2L_2^{(m-1)} & \tau_3 L_1^{(m-1)} \\
    \tau_3L_1^{(m-1)} & 0
    \end{array}\right)\Db\left(E_B-\frac{p^2}{2M_N},\vectn{p}\right)\Gb_{n-m}\left(p,E_B\right).
\end{align}
Using Eqs.~\eqref{eq:Sigmadef}, \eqref{eq:HLOdef}, and \eqref{eq:HNLOdef} the three-body force term can be rewritten, then combining like terms, and using the definition of $\oneb$, the result is
\begin{align}
    &{\Bbt^{[n]}}{}_{0\frac{1}{2},0\frac{1}{2}}^{\frac{1}{2},\frac{1}{2}}(p,k)=-\frac{M_N}{\sqrt{3}}\boldsymbol{\mathcal{M}}\Dbb\left(E_B\right)\Gb_n\frac{1}{\sqrt{\frac{3}{4}p^2-M_NE_B}+\sqrt{\frac{3}{4}p^2-M_NE-i\epsilon}}\\\nonumber
    &-\frac{1}{\sqrt{3}k_0}\boldsymbol{\mathcal{M}}\left\{\Gb_n-\sum_{m=1}^n\Bar{\mathbf{R}}_m\left(E_B\right)\Gb_{n-m}\right\}\\\nonumber
    &+\frac{1}{\sqrt{3}k_0}\Dbb^{-1}\left(E\right)\boldsymbol{\mathcal{M}}\Dbb\left(E_B\right)\Gb_n-\frac{1}{\sqrt{3}k_0}\sum_{m=1}^n\Bar{\mathbf{R}}_m\left(E\right)\Dbb^{-1}\left(E\right)\boldsymbol{\mathcal{M}}\Dbb\left(E_B\right)\Gb_{n-m}\\\nonumber
 &-\delta_{n2}\frac{4}{3}M_NH_2\Sigma_0(E_B)\frac{3(\kappa_0-\tau_3\kappa_1)}{\sqrt{3}}\oneb-\frac{M_N}{\sqrt{3}}\sum_{m=1}^n\left(\begin{array}{cc}
    2L_2^{(m-1)} & \tau_3 L_1^{(m-1)} \\
    \tau_3L_1^{(m-1)} & 0
    \end{array}\right)\Dbb\left(E_B\right)\Gb_{n-m} ,
\end{align}
where for brevity  we define
\begin{equation}
    \boldsymbol{\mathcal{M}}=\left(\begin{array}{cc}
    4\kappa_0 & 2\tau_3\kappa_1\\
     2\tau_3\kappa_1 & 0
    \end{array}\right),\Dbb(E)=\Db\left(E-\frac{p^2}{2M_N},\vectn{p}\right),\Bar{\mathbf{R}}_m(E)=\mathbf{R}_m\left(E-\frac{p^2}{2M_N},\vectn{p}\right) ,
\end{equation}
and drop the explicit energy and momentum dependence for the vertex functions since the arguments are the same for each term.  Noting the identity
\begin{align}
    \frac{M_Nk_0}{\sqrt{\frac{3}{4}p^2-M_NE_B}+\sqrt{\frac{3}{4}p^2-M_NE-i\epsilon}}=\Dbb^{-1}\left(E\right)-\Dbb^{-1}\left(E_B\right),
\end{align}
further terms in the inhomogeneous term cancel and after rearranging terms can be rewritten as
\begin{align}
    &{\Bbt^{[n]}}{}_{0\frac{1}{2},0\frac{1}{2}}^{\frac{1}{2},\frac{1}{2}}(p,k)=\frac{1}{\sqrt{3}k_0}\Dbb^{-1}\left(E_B\right)\left[\boldsymbol{\mathcal{M}},\Dbb\left(E_B\right)\right]\Gb_n\\\nonumber
    &+\frac{1}{\sqrt{3}k_0}\sum_{m=1}^n\left\{\boldsymbol{\mathcal{M}}\Bar{\mathbf{R}}_m(E_B)\Dbb^{-1}(E_B)-\Bar{\mathbf{R}}_m\left(E\right)\Dbb^{-1}\left(E\right)\boldsymbol{\mathcal{M}}\right\}\Dbb(E_B)\Gb_{n-m}\\\nonumber
 &-\delta_{n2}\frac{4}{3}M_Nk_0H_2\Sigma_0(E_B)\frac{3(\kappa_0-\tau_3\kappa_1)}{\sqrt{3}k_0}\oneb-\frac{M_N}{\sqrt{3}}\sum_{m=1}^n\left(\begin{array}{cc}
    2L_2^{(m-1)} & \tau_3 L_1^{(m-1)} \\
    \tau_3L_1^{(m-1)} & 0
    \end{array}\right)\Dbb\left(E_B\right)\Gb_{n-m}.
\end{align}

Computing the resulting matrices explicitly gives finally,
\begin{align}
    &{\Bbt^{[n]}}{}_{0\frac{1}{2},0\frac{1}{2}}^{\frac{1}{2},\frac{1}{2}}(p,k)=\frac{\gamma_t-\gamma_s}{\sqrt{3}k_0}2\tau_3\kappa_1\left(\!\!\begin{array}{rr}
    0 & 1\\
    -1 & 0
    \end{array}\!\!\right)\Dbb\left(E_B\right)\Gb_n\\\nonumber
    &-\frac{2\tau_3\kappa_1}{M_N\sqrt{3}k_0}\sum_{m=0}^{n-1}\left\{\left(c_{0t}^{(m)}-c_{0s}^{(m)}\right)(M_NE-\frac{3}{4}p^2)+c_{0t}^{(m)}\gamma_t^2-c_{0s}^{(m)}\gamma_s^2\right\}\left(\!\!\begin{array}{rr}
    0   & 1 \\
    -1     & 0
    \end{array}\!\!\right)\Dbb(E_B)\Gb_{n-1-m}\\\nonumber
 &+\sum_{m=0}^{n-1}\frac{1}{\sqrt{3}}\left(\begin{array}{cc}
    -4\kappa_0c_{0t}^{(m)}-2M_NL_2^{(m)} & -2\tau_3\kappa_1c_{0s}^{(m)}-\tau_3M_NL_1^{(m)}\\
    -2\tau_3\kappa_1c_{0t}^{(m)}-\tau_3M_NL_1^{(m)} & 0
    \end{array}\right)\Dbb\left(E_B\right)\Gb_{n-1-m}\\\nonumber
    &-\delta_{n2}\frac{4}{3}M_Nk_0H_2\Sigma_0(E_B)\frac{3(\kappa_0-\tau_3\kappa_1)}{\sqrt{3}k_0}\oneb.
\end{align}

 These are the inhomogenous terms for the integral equations of ${\Tbt^{[n]}}{}^{\frac{1}{2},\frac{1}{2}}_{0\frac{1}{2},0\frac{1}{2}}(p,k)$. The actual two-body triton photo-disintegration amplitude is related to this via Eq.~\eqref{eq:redef}.  Calculating $\Jb_n(p,E)$ explicitly,  the relationship between the actual triton photo-disintegration amplitude and ${\Tbt^{[n]}}{}^{J,J'}_{L'S',LS}(p,k)$ is given by
 \begin{align}
     {\Tbt^{[n]}}{}^{J,J'}_{L'S',LS}(p,k)
    ={\Tb^{[n]}}{}^{J,J'}_{L'S',LS}(p,k)+\frac{e}{2M_N}\frac{\gamma_t-\gamma_s}{\sqrt{3}k_0}2\tau_3\kappa_1\left(\!\!\begin{array}{cc}
    0 & 0\\[-2mm]
    1 & 0
    \end{array}\!\!\right)\Dbb\left(E_B\right)\Gb_n(p,E_B),
 \end{align}
 when $p$ is taken on-shell. The second term on the right hand side does not contribute for a deuteron state. 
 Therefore, the two-body triton photo-disintegration  amplitude  ${\Tbt^{[n]}}{}^{J,J'}_{L'S',LS}(p,k)$ is equivalent to the two-body  triton photo-disintegration amplitude ${\Tb^{[n]}}{}^{J,J'}_{L'S',LS}(p,k)$ for on-shell neutron momentum.  Similar arguments can be made for an outgoing \QS channel and the inhomogeneous term in this channel is given by
\begin{align}
&{\Bbt^{[n]}}{}_{0\frac{3}{2},0\frac{1}{2}}^{\frac{1}{2},\frac{3}{2}}(p,k)=\frac{\gamma_t-\gamma_s}{\sqrt{3}k_0}2\tau_3\kappa_1\left(\begin{array}{cc}
    0 & 1 \\
    0 & 0
    \end{array}\right)\Dbb\left(E_B\right)\Gb_n\\\nonumber
&-\frac{2\tau_3\kappa_1}{\sqrt{3}M_Nk_0}\sum_{m=0}^{n-1}\left\{\left(c_{0t}^{(m)}-c_{0s}^{(m)}\right)(M_NE-\frac{3}{4}p^2)+c_{0t}^{(m)}\gamma_t^2-c_{0s}^{(m)}\gamma_s^2\right\}\left(\!\!\begin{array}{cc}
    0 & 1\\
    0 & 0
    \end{array}\!\!\right)\Dbb\left(E_B\right)\Gb_{n-1-m}\\\nonumber
&-\frac{1}{\sqrt{3}}\sum_{m=0}^{n-1}\left(\begin{array}{cc}
    -M_NL_2^{(m)}-2\kappa_0c_{0t}^{(m)} & \tau_3 M_NL_1^{(m)}+2\tau_3\kappa_1c_{0s}^{(m)} \\
    0 & 0
    \end{array}\right)\Dbb\left(E_B\right)\Gb_{n-1-m}.
\end{align}

\section{\label{app:Wigner}Wigner-SU(4) Symmetry}

The single-nucleon current $nd$ capture M1 moment is given by the matrix element
\begin{equation}
    \frac{e}{2M_N}\langle^{2S+1}L_{J},\vect{p}|\sum_{i=1}^3(\kappa_0+\kappa_1\tau_3^{(i)})\sigma_3^{(i)}|\jjvH\rangle
\end{equation}
where $^{2S+1}L_J$ is either the $^2S_\frac{1}{2}$ or $^4S_{\frac{3}{2}}$ scattering state and $|\jjvH\rangle$ is the triton wavefunction.  Summing over nucleons the operators can be written as
\begin{equation}
    \sum_{i=1}^3\sigma_3^{(i)}=2S_z\quad,\quad\sum_{i=1}^3\tau_3^{(i)}\sigma_3^{(i)}=Y_{zz} ,
\end{equation}
where $S_z$ is the spin of the three nucleon system in the $z$ direction and $Y_{zz}$ is an SU(4) operator.  In the Wigner-SU(4) limit the triton wavefunction becomes Wigner-SU(4) symmetric and is an eigenstate of the $Y_{zz}$ operator.  Therefore, in the Wigner-SU(4) limit (also see Ref.~\cite{Mehen:1999qs})
\begin{equation}
    \frac{e}{2M_N}\langle^{2S+1}L_{J},\vect{p}|\kappa_0S_z+\kappa_1Y_{zz}|\jjvH\rangle=\frac{e}{2M_N}(\kappa_0+\kappa_1)\langle^{2S+1}L_{J},\vect{p}|\jjvH\rangle=0,
\end{equation}
where the second equality follows from the fact that the scattering states and bound states are orthogonal. Thus at LO in \EFT in the Wigner-SU(4) limit the M1 moment for $nd$ capture is zero.  Further, the triton wavefunction is always an eigenstate of $S_z$ and therefore by the same arguments the single-nucleon current contribution to the M1 moment  only depends on $\kappa_1$. 

\section{\label{app:dual-expansion}\eftnopi-Wigner-SU(4) dual expansion of cold $nd$ capture}
The two-nucleon currents in the nucleon (as opposed to auxiliary field) formalism of \eftnopi are given by \cite{Chen:1999tn}
\begin{equation}
    \mathcal{L}_{2,1}^{mag'}=
    e\,{}^{\cancel{\pi}}\!L_1 (\hat{N}^T P_i\hat{N})^\dagger(\hat{N}^T \bar{P}_3 \hat{N})\mathbf{B}_i\
    -e\,{}^{\cancel{\pi}}\!L_2 i\epsilon_{ijk} (\hat{N}^T P_i\hat{N})^\dagger(\hat{N}^T P_j \hat{N})\mathbf{B}_k + \mathrm{H.c.} 
    \label{eq:MEClagNLO_nuc}
\end{equation}
% %
The matching between the LEC in the auxiliary field formalism $L_1^{(m)}$ [$L_2^{(m)}$] and the LEC in the nucleon formalism ${}^{\cancel{\pi}}\!L_1$ [${}^{\cancel{\pi}}\!L_2$] can be obtained by comparing Eqs.~\eqref{eq:YNLO} and \eqref{eq:YNNLO} [Eq.~\eqref{eq:mud}] with $\sigma_{np}$ [$\mu_d$] in the nucleon formalism~\cite{Chen:1999tn,Rupak:1999rk}. For both the $^1S_0$ and $^3S_1$ channel, we used the $Z$ parameterization, which expands around the $^1S_0$ and $^3S_1$ dibaryon pole (see Eq.~\eqref{eq:dibprop}), respectively. In contrast, Refs.~\cite{Rupak:1999rk, Chen:1999tn} only expand the \NN amplitude around the deuteron pole in the $^3S_1$ channel; in the $^1S_0$ channel they expand the \NN amplitude around zero momentum. Here we give the matching of the LEC ${}^{\cancel{\pi}}\!L_1$ [${}^{\cancel{\pi}}\!L_2$] (when the \NN amplitude is expanded around the dibaryon pole in each channel) to $\deltaLoz$ in Eq.~\eqref{eq:specL1val} [$\deltaLtz$ in Eq.~\eqref{eq:specL2val}]. At NLO in \eftnopi, we find 
\begin{align}
   \frac{\Ltwonopi(\mu - \gamma_t)^2 M_N}{\pi}
    &=\deltaLtz,
    \label{eq:matching-for-L20}
\end{align}
and
\begin{align}
    \frac{L_{np} M_N}{2\pi} = L_{1}^{(0)}
\end{align}
where the $\mu$-independent parameter $L_{np}$ is defined similarly to Ref.~\cite{Rupak:1999rk}
\begin{align}
    L_{np} &\equiv \left((\mu - \gamma)^2 - \delta^2\right) \Lonenopi - \frac{2\kappa_1\pi}{M_N}\left(\frac{\rho(\mu-\gamma)^2+\rho\delta^2 + 2\delta_r\delta(\mu-\gamma)}{\left((\mu-\gamma)^2-\delta^2\right)}\right)
    \label{eq: L-np}.
\end{align}
In the Wigner-SU(4) limit, the second term in Eq.~\eqref{eq: L-np} becomes $\mu$-independent and the matching for $\Lonenopi$ becomes
\begin{align}
    \frac{\Lonenopi(\mu - \gamma)^2M_N}{2\pi}
    &= \deltaLoz.
\end{align}
At NNLO in \eftnopi, we define $\Lonenopi^{(1)}$ ($\Ltwonopi^{(1)}$) as the perturbative correction to $\Lonenopi$ ($\Ltwonopi$) in the nucleon formalism. Their matching to the NNLO LECs in the auxiliary field formalism is
\begin{align}
   \frac{\Ltwonopi^{(1)}(\mu - \gamma_t)^2 M_N}{\pi}    &=\deltaLto,
    \label{eq:matching-for-L21}
\end{align}
and
\begin{align}
    \frac{\widetilde{L}_{np} M_N}{2\pi} = L_{1}^{(1)} ,
\end{align}
where the $\mu$-independent parameter $\widetilde{L}_{np}$ is again defined similarly to Ref.~\cite{Rupak:1999rk}
\begin{align}
    \widetilde{L}_{np} &\equiv \left((\mu - \gamma)^2 - \delta^2\right) \Lonenopi^{(1)} + \frac{2\kappa_1\pi}{M_N}\left(\frac{\gamma\left[(\rho^2 + \delta_r^2)\left((\mu-\gamma)^2+\delta^2\right)+4\delta\delta_r\rho(\mu-\gamma)\right]}{{\left((\mu-\gamma)^2-\delta^2\right)}}
    \right.\nonumber\\
    &\hspace{5.5cm} \left.+ \frac{2\delta\left[\rho\delta_r\left((\mu-\gamma)^2+\delta^2\right)+\delta(\rho^2 + \delta_r^2)(\mu-\gamma)\right]}{{\left((\mu-\gamma)^2-\delta^2\right)}}\right).
    \label{eq: Ltilde-np}
\end{align}
In the Wigner-SU(4) limit, the matching for $\Lonenopi^{(1)}$ becomes
\begin{align}
    \frac{\Lonenopi^{(1)}(\mu - \gamma)^2M_N}{2\pi} 
    &= \deltaLoo.
    \label{eq:matching for L1}
\end{align}

We can write the $nd$ capture amplitude as  
\begin{align}
    A_{nd} = A^{(1)}_{\deltanopi^0} +  \sum_{n\in\{1,2\}}A^{(n)}_{\deltanopi^1} + \sum_{n\in\{1,2,3\}}A^{(n)}_{\deltanopi^2} ,
\end{align}
where the subscript indicates the order in $\deltanopi$ counting in the \eftnopi expansion. The $``(n)"$ superscripts on $A$ indicate that the contribution is only from $n$-nucleon currents at the given order denoted by the subscript.

In the Wigner-SU(4) expansion of the $nd$ capture amplitude, we have shown in Sec.~\ref{sec:Wigner} that amplitudes with $\kappa_1$ are suppressed by $\deltawig = \frac{\delta}{\kappa_*}$, where ${\kappa_*}$ is a three-nucleon scale that is greater than $\delta$, except for the $H_2$ term. In contrast, amplitudes with $\deltaLom{m}$, $\deltaLtm{m}$, or $(\widetilde{\kappa}_0(\Lambda)-\widetilde{\kappa}_1(\Lambda))$ are not suppressed. However, as argued in Sec.~\ref{sec:Wigner}, corrections from $\deltaLom{m}$ and  $\deltaLtm{m}$ seem to be higher order. In fact, unlike $\Lonenopi$ in $\sigma_{np}$, $\Ltwonopi$ is not needed  to  cancel any $\mu$ dependence from other contributions in $\mu_{d}$ at NLO, and Ref.~\cite{De-Leon:2020glu} suggests $\Ltwonopi$ can be treated as a  higher order term. In addition, Ref.~\cite{Richardson:2020iqi} shows that $\Ltwonopi$ is suppressed in the large-$N_c$ expansion compared to $\Lonenopi$, where $N_c$ is the number of colors in quantum chromodynamics. As we do not intend to pursue a full rigorous dual expansion here, for simplicity we assume that the suppression of contributions from $\deltaLom{m}$ and $\deltaLtm{m}$ can be counted as suppressions on the order of $\deltawig$. The dual expansion for the $nd$ capture amplitude, supplemented by this assumption, reads 
\begin{align}
    A_{nd} = 
    \begin{cases}
     \vspace{0.5cm}
        \underbrace{
         0 
            \vphantom{\left(A^{(1,~H_2)}_{\deltanopi^2\deltawig^0}\right)}
         }_{\mathcal{O}(\deltanopi^0)}
         + 
         \underbrace{
         A^{(1)}_{\deltanopi^0\deltawig^1} 
            \vphantom{\left(A^{(1,~H_2)}_{\deltanopi^2\deltawig^0}\right)}
         }_{\mathcal{O}(\deltanopi^1)}
         + 
         \underbrace{\left(
         A^{(1)}_{Q^1 \deltawig^1} +
         A^{(1,~H_2)}_{\deltanopi^2\deltawig^0} +
         A^{(2)}_{\deltanopi^1\deltawig^1} +
         A^{(3)}_{\deltanopi^2\deltawig^0}
         \right)}_{\mathcal{O}(\deltanopi^2)}, & \textrm{if}~\deltawig \sim Q\\
         \underbrace{
         0 
            \vphantom{\left(A^{(1,~H_2)}_{\deltanopi^2\deltawig^0}\right)}
         }_{\mathcal{O}(\deltanopi^0)}
         +\underbrace{
         0 
            \vphantom{\left(A^{(1,~H_2)}_{\deltanopi^2\deltawig^0}\right)}
         }_{\mathcal{O}(\deltanopi^1)}+
         \underbrace{\left(
         A^{(1)}_{\deltanopi^0\deltawig^1}+
         A^{(1,~H_2)}_{\deltanopi^2\deltawig^0}+
         A^{(3)}_{\deltanopi^2\deltawig^0}
         \right)}_{\mathcal{O}(\deltanopi^2)}, & \textrm{if}~\deltawig \sim Q^2
    \end{cases}
    \label{eq:dual-expansion-app}
\end{align}
where the subscript now indicates both the $\deltanopi$ and $\deltawig$ counting, and the dual counting in terms of $\deltanopi$ is indicated below each term. $A^{(1,~H_2)}_{\deltanopi^2}$ is the contribution from the single-nucleon current only associated with the energy-dependent three-nucleon force with LEC $H_2$.
In Eq.~\eqref{eq:dual-expansion-app} we only consider terms in the dual counting up to $\deltanopi^2$.

\section{\label{app:magwigner} Triton Magnetic Moment in Wigner-SU(4) basis}

The LO triton magnetic moment in the Wigner-SU(4) basis is given by
\begin{align}
&\mu_0^{\jjvH}=(\kappa_0+\kappa_1)+2\pi M_{N}\frac{2}{3}\kappa_1\left(\widetilde{\boldsymbol{\Gamma}}_{W,0}(q)\right)^{T}\otimes_qM(q,\ell)\left(\!\!\begin{array}{cc}
0 & 0\\[-2mm]
0 & 1
\end{array}\!\!\right)\otimes_\ell\widetilde{\boldsymbol{\Gamma}}_{W,0}(\ell),
\end{align}
where the Wigner-SU(4) basis function is defined by
\begin{equation}
    \widetilde{\boldsymbol{\Gamma}}_{W,n}(q)=\left(\begin{array}{rr}
    1 & -1\\
    1 & 1
    \end{array}\right)\widetilde{\boldsymbol{\Gamma}}_{n}(q)=\left(\begin{array}{c}
    \Gamma_{\Ws,n}(q)\\
    \Gamma_{\Was,n}(q)
    \end{array}\right),
\end{equation}
with $\Gamma_{\Ws,n}(q)$ being the Wigner-SU(4)-symmetric part and $\Gamma_{\Was,n}(q)$ the Wigner-SU(4) anti-symmetric part.  In the Wigner-SU(4) limit the Wigner-SU(4) anti-symmetric piece vanishes.  Thus the LO triton magnetic moment in the Wigner-SU(4) limit is $\kappa_0+\kappa_1$.  The NLO correction to the triton magnetic moment in the Wigner-SU(4) basis is given by
\begin{align}
&\mu_1^{\jjvH}=4\pi M_{N}\frac{2}{3}\kappa_1\left(\widetilde{\boldsymbol{\Gamma}}_{W,1}(q)\right)^{T}\otimes_qM(q,\ell)\left(\!\!\begin{array}{cc}
0 & 0\\[-2mm]
0 & 1
\end{array}\!\!\right)\otimes_\ell\widetilde{\boldsymbol{\Gamma}}_{W,0}(\ell)\\\nonumber
&-\pi M_{N}\left(\widetilde{\boldsymbol{\Gamma}}_{W,0}(q)\right)^{T}\otimes_q\left\{\frac{\pi}{2}\frac{\delta(q-\ell)}{q^{2}}
\left(\begin{array}{cc}
-\frac{2}{3}\deltaLtz-\frac{2}{3}\deltaLoz &-\frac{2}{3}\kappa_1\delta_r -\frac{2}{3}\deltaLtz \\
-\frac{2}{3}\kappa_1\delta_r-\frac{2}{3}\deltaLtz & -\frac{4}{3}\kappa_1\rho +\frac{2}{3}\deltaLoz-\frac{2}{3}\deltaLtz
\end{array}\right)\right\}\otimes_\ell\widetilde{\boldsymbol{\Gamma}}_{W,0}(\ell).
\end{align}
In this form it is apparent that the only non-zero contribution of the NLO correction to the triton magnetic moment in the Wigner-SU(4) limit is from $\deltaLoz$ and $\deltaLtz$. The \NNLO correction to the magnetic moment in the Wigner-SU(4) basis is
\begin{align}
&\mu_2^{\jjvH}=4\pi M_{N}\frac{2}{3}\kappa_1\left(\widetilde{\boldsymbol{\Gamma}}_{W,2}(q)\right)^{T}\otimes_qM(q,\ell)\left(\!\!\begin{array}{cc}
0 & 0\\[-2mm]
0 & 1
\end{array}\!\!\right)\otimes_\ell\widetilde{\boldsymbol{\Gamma}}_{W,0}(\ell)\\\nonumber
&+2\pi M_{N}\frac{2}{3}\kappa_1\left(\widetilde{\boldsymbol{\Gamma}}_{W,1}(q)\right)^{T}\otimes_qM(q,\ell)\left(\!\!\begin{array}{cc}
0 & 0\\[-2mm]
0 & 1
\end{array}\!\!\right)\otimes_\ell\widetilde{\boldsymbol{\Gamma}}_{W,1}(\ell)\\\nonumber
&-2\pi M_{N}\left(\widetilde{\boldsymbol{\Gamma}}_{W,0}(q)\right)^{T}\otimes_q\left\{\frac{\pi}{2}\frac{\delta(q-\ell)}{q^{2}}
\left(\begin{array}{cc}
-\frac{2}{3}\deltaLtz-\frac{2}{3}\deltaLoz &-\frac{2}{3}\kappa_1\delta_r -\frac{2}{3}\deltaLtz \\
-\frac{2}{3}\kappa_1\delta_r-\frac{2}{3}\deltaLtz & -\frac{4}{3}\kappa_1\rho +\frac{2}{3}\deltaLoz-\frac{2}{3}\deltaLtz
\end{array}\right)\right\}\otimes_\ell\widetilde{\boldsymbol{\Gamma}}_{W,1}(\ell)\\\nonumber
&-\pi M_{N}\left(\widetilde{\boldsymbol{\Gamma}}_{W,0}(q)\right)^{T}\otimes_q\left\{\vphantom{\left(\begin{array}{cc}
-\frac{2}{3}\deltaLtz-\frac{2}{3}\deltaLoz &-\frac{2}{3}\frac{\kappa_1}{M_N}(c_{0t}^{(1)}-c_{0s}^{(1)}) -\frac{2}{3}\deltaLtz \\
-\frac{2}{3}\kappa_1(c_{0t}^{(1)}-c_{0s}^{(1)})-\frac{2}{3}\deltaLtz & -\frac{4}{3}\kappa_1(c_{0t}^{(1)}+c_{0s}^{(1)}) +\frac{2}{3}\deltaLoz-\frac{2}{3}\deltaLtz
\end{array}\right)}\frac{\pi}{2}\frac{\delta(q-\ell)}{q^{2}}\right.
\\\nonumber
&\hspace{1cm}\left.\left(\begin{array}{cc}
-\frac{2}{3}\deltaLto-\frac{2}{3}\deltaLoo &-\frac{2}{3}\frac{\kappa_1}{M_N}(c_{0t}^{(1)}-c_{0s}^{(1)}) -\frac{2}{3}\deltaLto \\
-\frac{2}{3}\kappa_1(c_{0t}^{(1)}-c_{0s}^{(1)})-\frac{2}{3}\deltaLto & -\frac{4}{3}\kappa_1(c_{0t}^{(1)}+c_{0s}^{(1)}) +\frac{2}{3}\deltaLoo-\frac{2}{3}\deltaLto
\end{array}\right)\right\}\otimes_\ell\widetilde{\boldsymbol{\Gamma}}_{W,0}(\ell)\\\nonumber
&-\frac{4}{3}M_NH_2\frac{\Sigma_0^2(E_B)}{\Sigma_0'(E_B)}(\kappa_0+\kappa_1)-\frac{1}{\Omega H_{\mathrm{LO}}\Sigma_0'(E_B)}(\widetilde{\kappa}_0(\Lambda)-\widetilde{\kappa}_1(\Lambda)).
\end{align}
The only non-zero contribution in the Wigner-SU(4) limit comes from $\deltaLom{n}$, $\deltaLtm{n} $, the $H_2$ term, and from $\widetilde{\kappa}_0(\Lambda)-\widetilde{\kappa}_1(\Lambda)$ .

\section{\label{app:finiteDr} Results for $\delta_r \neq 0 $}

In Sec.~\ref{sec:results} we presented  the results for $\sigma_{nd}$  using $\delta_r=0$ (that is, where the effective ranges in the singlet and triplet channels are identical). Those results are valid up to corrections at the same order as N$^3$LO corrections. In this appendix we present the results using $\delta_r \ne 0$. 
Although the LO and NLO results can be taken as predictions the NNLO results can only be considered preliminary until we can establish whether convergence with respect to the cutoff at NNLO has occurred for $\sigma_{nd}$ using $\delta_r \ne 0$. Table~\ref{tab:L1dep} is the equivalent of Table~\ref{tab:L1depWr} except with $\delta_r \ne 0$. 
Table~\ref{tab:chandeltaneq0} is equivalent to the results shown in Table~\ref{tab:sigmaJp-tritonfit} except that now $\delta_r \ne 0$. 

\begin{table}[hbt]
    \centering
    \begin{tabular}{|c|c|c|c|c|c|}
    \hline
        & $\sigma_{np}$ [mb] & $\sigma_{nd}$ [mb] &  $\mu_{\jjvH}$ & $L_1^{(0)}$~[fm]  & $L_1^{(1)}$~[fm] \\\hline
         LO & \hspace{1mm}325.2(224.6)\hspace{1mm} & 0.314(217) & \hspace{1.7mm}2.75(95) \hspace{1mm} &  - & - \\\hline
         \multirow{6}{*}{\shortstack[c]{LO\\[1mm]+\\[1mm]NLO}} & {334.2(79.7)} & 0.180(43)\phantom{4} & 2.62(31) & -6.90$^{p}$\phantom{[123]}& - \\
          & {334.2(79.7)} & 0.180[377] & 2.62[82] &   \hspace{1mm}-6.90[2.92]$^p$\hspace{1mm}& -  \\
          &369.0(88.0) & 0.345(82)\phantom{4} & {2.98(36)} &  -5.62$^t$\phantom{[123]}& - \\
          & 369.0[34.6] & 0.345[164] & {2.98(36)} &   -5.62[1.27]$^t$& - \\
         & 354.1(84.5)& 0.274(65)\phantom{4} & 2.83(34) &   -6.17$^{p,t}$\phantom{[12]}& - \\
         & 393.9(94.0)& 0.463(110) & 3.24(39) &  -4.71$^{p,d,t}$\phantom{[1]}&  - \\\hline
         \multirow{6}{*}{\shortstack[c]{LO\\[1mm]+\\[1mm]NLO\\[1mm]+\\[1mm]\NNLO}} & {334.2(27.5)} & 0.408(34)\phantom{4} & {2.98(12)} &  -6.90$^p$\phantom{[123]}& 3.85$^p$\phantom{[1.01]} \\
         & {334.2(27.5)} & 0.408[130]& {2.98[28]} & -6.90$^p$\phantom{[123]}& \hspace{1mm}3.85[{1.01}]$^p$\hspace{1mm} \\
         & {334.2(27.5)} & 0.463(38)\phantom{4} & {2.98(12)} & -5.62$^t$\phantom{[123]}& 2.15$^p\phantom{[1.01]}$ \\
         & {334.2(27.5)} & 0.463[130] & 2.98[28] & -5.62$^t$\phantom{[123]}& 2.15[{1.01}]$^p$ \\
         & {334.2(27.5)} & 0.421(35)\phantom{4} & {2.98(12)} & -6.17$^{p,t}$\phantom{[12]}& 2.89$^p$\phantom{[1.01]} \\
         & 352.8(29.1) & 0.519(43)\phantom{4}  & {2.98(12)} & -4.71$^{p,d,t}$\phantom{[1]}& 1.57$^{p,d}$\phantom{[1.2]} \\\hline\hline
         Exp & 334.2$\pm$0.5 & 0.508$\pm$0.015 & 2.979 & - & - \\\hline
    \end{tabular}
    \caption{ Same as Table~\ref{tab:L1depWr} except with $\delta_r \ne0$.  Results are evaluated at a cutoff of $\Lambda=500,000$~MeV. (NNLO results are preliminary.) \label{tab:L1dep}}
\end{table}

\begin{table}[htb!]
\begin{tabular}{|l|c|c|c|}
    \hline
     &$\sigma_{nd}(J'=\frac{1}{2})$ [mb] & $\sigma_{nd}(J'=\frac{3}{2})$ [mb] &  $\sigma_{nd}(tot)$ [mb] \\\hline
    LO & 0.166(114)& 0.149(103)\phantom{4} & 0.314(217)\\
    +NLO$^{t}$& 0.305[208] & 0.0401[445] & 0.345[164]\\
    +\NNLO\!$^{p}$ & 0.336[166]& 0.127[35]\phantom{44} & 0.463[130] \\\hline\hline
    LO & 0.166(114)& 0.149(103)\phantom{4} & 0.314(217)\\
    +NLO$^{p,d,t}$ & 0.455(108) & 0.00809(193)& 0.463(110)\\
    +\NNLO\!$^{p,d}$ & 0.392(32)\phantom{4} & 0.127(10)\phantom{444} & 0.519(43)\phantom{4}\\\hline\hline
    Exp & - & - & 0.508$\pm$0.015 \\\hline
\end{tabular}
    \caption{Same is in Table~\ref{tab:sigmaJp-tritonfit} only for $\delta_r \ne 0$. Results are evaluated at a cutoff of $\Lambda=500,000$~MeV.  (NNLO results are preliminary.)}
    \label{tab:chandeltaneq0}
\end{table}

\newpage

\section{\label{app:Errors} Error Analysis}

Consider an observable $\mathcal{O}$ whose experimental value is $\mathcal{O}^{\textrm{exp}}$ and \eftnopi prediction at LO, NLO, or NNLO is $\mathcal{O}_{m}$, where $m=$ 0, 1, or 2, respectively. Naively, the error of $\mathcal{O}_{m}$ can be estimated by
\begin{align}
\label{eq:naiveError}
    \DeltaN (\mathcal{O}_{m}) = \left|{\beta Q^{m+1}\mathcal{O}_{m}}\right|
\end{align}
where $\beta = 1$ ($\beta=2$) is for observables proportional to an amplitude (amplitude squared). The subscript ``$\textrm{N}$" on $\DeltaN$ indicates that this is the naive \eftnopi error estimate. If  $\mathcal{O}_{m}$ depends on a LEC $C_m$ that first appears at $m$-th order, we can fit $C_m$ to $\mathcal{O}^{\textrm{exp}}$ and find the uncertainty of $C_m$, defined as $\Delta C_m$, from $ \DeltaN (\mathcal{O}_{m}) $:
\begin{align}
\label{eq:errorLEC}
    \Delta C_m &= \left|\frac{ \DeltaN (\mathcal{O}_{m}) }{\partial \mathcal{O}_{m}/ \partial C_m }\right|\\ \nonumber
    &=\left|\frac{\beta Q^{m+1} \mathcal{O}^{\textrm{exp}}}{\partial \mathcal{O}_{m}/ \partial C_m }\right|,
\end{align}
where we have used Eq.~\eqref{eq:naiveError} and replaced $\mathcal{O}_{m}$ with $\mathcal{O}^{\textrm{exp}}$ since it is used to fit $C_m$. In this example, only $C_m$ is allowed to flow and all other LECs are fixed. At first order in $C_m$, $\mathcal{O}_{m}$ depends linearly on $C_m$, and  $\partial \mathcal{O}_{m}/ \partial C_m$ is used to extract the relevant prefactor. This is how the error bars for $L_1^{(0)}$ in Fig.~\ref{fig:L1plotWR} are obtained. $\Delta C_m$ can then be used to propagate the error from $\DeltaN (\mathcal{O}_{m})$ to another observable, $\mathcal{O}'$:
\begin{align}
\label{eq:errorProp}
    \DeltaP (\mathcal{O}'_m) = \left|\frac{\partial \mathcal{O}'_{m}}{\partial C_m}\right|  \Delta C_m
\end{align}
where $\mathcal{O}'_m$ is the \eftnopi prediction at $m$-th order for $\mathcal{O}'$ and the subscript ``$\textrm{P}$" of $\DeltaP$ indicates that $\DeltaP (\mathcal{O}'_{m})$ is obtained from the error propagation through $\Delta C_m$. To compare $\DeltaP (\mathcal{O}'_{m})$ to $\DeltaN (\mathcal{O}'_{m})$, consider the ratio
\begin{align}
\label{eq:ratioErrors}
    \frac{\DeltaN(\mathcal{O}'_m)}{\DeltaP(\mathcal{O}'_m) } &= \frac{\left|{\beta'Q^{m+1}\mathcal{O}'_{m}}\right|}{\left|\frac{\partial \mathcal{O}'_{m}}{\partial C_m}\cdot\frac{\beta Q^{m+1} \mathcal{O}^{\textrm{exp}}}{\partial \mathcal{O}_{m}/ \partial C_m }\right|}\\\nonumber
    &=\frac{\beta'}{\beta}\left|\frac{\frac{C_m}{\mathcal{O}^{\textrm{exp}}}\frac{\partial \mathcal{O}_m}{\partial C_m}}{\frac{C_m}{\mathcal{O}'_m}\frac{\partial \mathcal{O}'_m}{\partial C_m}}\right|\\\nonumber
    &=\frac{\beta'}{\beta}\left| \frac{\mathcal{O}^{C_m}_{m\textrm{, corr}}/\mathcal{O}^{\textrm{exp}}}{\mathcal{O}'^{C_m}_{m\textrm{, corr}}/\mathcal{O}'_m}\right|,
\end{align}
where $\mathcal{O}^{C_m}_{m\textrm{, corr}}$ ($\mathcal{O}'^{C_m}_{m\textrm{, corr}}$) represent the $m$-th order correction for $\mathcal{O}$ ($\mathcal{O}'$) from $C_m$ at the first order in $C_m$. $\beta'$ plays the same role as $\beta$ but for $\mathcal{O}'$. This analysis does not require knowledge of the experimental value for $\mathcal{O}'$. 
If the \eftnopi expansion is well behaved for both $\mathcal{O}$  and $\mathcal{O}'$, then $C_m$ should give a similar relative correction to each observable, which suggests the ratio given in Eq.~\eqref{eq:ratioErrors} is roughly $\beta'/\beta$. As shown in  Table.~\ref{tab:L1dep}, propagating the naive \eftnopi errors of $\mu_{\jjvH}$ through $L_1^{(0)}$, for example, the ratio between $\DeltaN(\sigma_{np})$ in parenthesis and $\DeltaP(\sigma_{np})$ in square brackets is roughly two, whereas the ratio between $\DeltaN(\sigma_{nd})$ and $\DeltaP(\sigma_{nd})$ is roughly nine. The former ratio is in line with the \eftnopi expansion, but the latter ratio is not. This lack of agreement with the \eftnopi expansion for $\sigma_{nd}$ can be understood in the context of Wigner-SU(4) symmetry as discussed in Sec.~\ref{sec:Wigner}. At NNLO in \eftnopi, we only allow $L_1^{(1)}$ to flow and propagate the naive \eftnopi error of $\sigma_{np}$ given by $2Q^2\sigma_{np}$. The errors from three-body LECs, such as the three-nucleon magnetic moment counterterm and the energy-dependent three-body force, are neglected in the current treatment.  

$\Delta C_m$  can be used to propagate the error of one LEC to  multiple observables, in which case the errors of those observables are correlated through $\Delta C_m$. An important example is $\sigma_{nd}(J'=1/2)$ and $\sigma_{nd}(J'=3/2)$. At NLO in \eftnopi and using the physical value for $\delta_r$, we obtain 
\begin{align}
    \sigma_{nd}^{\textrm{NLO}}(J'=1/2) &= \left(0.176 + 0.164\frac{\deltaLoz}{\textrm{fm}}\right)\textrm{mb} \\\nonumber
    &= \left(0.176 + 0.164\left(\frac{L_1^{(0)}}{\textrm{fm}}+6.41\right)\right)\textrm{mb}
\end{align}
and
\begin{align}
    \sigma_{nd}^{\textrm{NLO}}(J'=3/2) &=  \left(0.0676 - 0.0351\frac{\deltaLoz}{\textrm{fm}}\right)\textrm{mb} \\\nonumber
    &= \left(0.0676 - 0.0351\left(\frac{L_1^{(0)}}{\textrm{fm}}+6.41\right)\right)\textrm{mb}
\end{align}
where we have used Eq.~\eqref{eq:specL1val}. Note that the coefficients of $L_1^{(0)}$ for $ \sigma_{nd}^{\textrm{NLO}}(J'=1/2) $ and $ \sigma_{nd}^{\textrm{NLO}}(J'=3/2)$ have opposite signs, i.e., $ \sigma_{nd}^{\textrm{NLO}}(J'=1/2) $ and $ \sigma_{nd}^{\textrm{NLO}}(J'=3/2)$ are anti-correlated through $L_1^{(0)}$. The total cross section is given by their sum
\begin{align}
 \label{eq:sigmandNLO}
    \sigma_{nd}^{\textrm{NLO}}(tot) &=  \left(0.243 + 0.129\frac{\deltaLoz}{\textrm{fm}}\right)\textrm{mb} \\\nonumber
    &= \left(0.243 + 0.129\left(\frac{L_1^{(0)}}{\textrm{fm}}+6.41\right)\right)\textrm{mb}
\end{align}
Their errors (from the error propagation of $\Delta L_1^{(0)}$) are given by
\begin{align}
    \DeltaP \left(\sigma_{nd}^{\textrm{NLO}}(J'=1/2)\right) &= 0.164\frac{\Delta L_1^{(0)}}{\textrm{fm}}~\textrm{mb}\\\nonumber \DeltaP \left(\sigma_{nd}^{\textrm{NLO}}(J'=3/2)\right)  &= 0.0351\frac{\Delta L_1^{(0)}}{\textrm{fm}}~\textrm{mb}\\\nonumber \DeltaP \left(\sigma_{nd}^{\textrm{NLO}}(tot)\right)  &=  0.129\frac{\Delta L_1^{(0)}}{\textrm{fm}}~\textrm{mb}.
\end{align}
The uncertainty of the total cross section is given by the difference between the uncertainty in each individual channel due to the anti-correlation between $ \sigma_{nd}^{\textrm{NLO}}(J'=1/2) $ and $ \sigma_{nd}^{\textrm{NLO}}(J'=3/2)$ through $L_1^{(0)}$. The sensitivity of $\sigma_{nd}$ to $L_1^{(0)}$ can be observed   by comparing Eq.~\eqref{eq:sigmandNLO} with $\sigma_{np}$ at NLO 
 \begin{align}
 \label{eq:signpNLO}
     \sigma_{np}^{\textrm{NLO}} = \left(347.6 + 27.3\left(\frac{L_1^{(0)}}{\textrm{fm}}+6.41\right)\right)\textrm{mb} .
 \end{align}
 Considering the ratio of the number multiplying $\left(\frac{L_1^{(0)}}{\textrm{fm}}+6.41\right)$ over the first term in each expression,
 $0.129/ 0.243$ in Eq.~\eqref{eq:sigmandNLO} is much larger than $27.3/347.6$ in Eq.~\eqref{eq:signpNLO}.

\bibliography{ref}

\end{document}

%% file: preamble.tex
\usepackage{multirow}
\usepackage{siunitx}
\usepackage{tikz}
\usetikzlibrary{patterns}
\usetikzlibrary{decorations.pathmorphing}

\usepackage{graphicx,times}
\usepackage{latexsym}
\usepackage{mathtools}

\usepackage{amsmath,amssymb,amsbsy,amsfonts}
\usepackage{array}
\usepackage{graphics}
\usepackage{xcolor}
\usepackage{cancel}
\usepackage{xspace}
\usepackage[normalem]{ulem}

\usepackage[unicode, pdfprintscaling=None]{hyperref}
\hypersetup{
    colorlinks,
    linkcolor={red!50!black},
    citecolor={blue!50!black},
    urlcolor={blue!80!black}
}

\usepackage[capitalise]{cleveref}

\usepackage{xcolor}
\usepackage{makecell}

\newcommand{\EFT}{EFT($\cancel{\pi}$)\xspace}
\newcommand{\vect}[1]{\vec{\boldsymbol{#1}}}
\newcommand{\vectn}[1]{#1}

\newcommand{\Gb}{\boldsymbol{\mathcal{G}}}
\newcommand{\Jb}{\boldsymbol{\mathcal{J}}}
\newcommand{\Gbt}{\widetilde{\boldsymbol{\mathcal{G}}}}
\newcommand{\Kb}{\mathbf{K}}
\newcommand{\Tb}{\mathbf{T}}
\newcommand{\Tbt}{\widetilde{\mathbf{T}}}
\newcommand{\Bb}{\mathbf{B}}
\newcommand{\Bbt}{{\widetilde{\mathbf{B}}}}
\newcommand{\Db}{\mathbf{D}}
\newcommand{\oneb}{\tilde{\mathbf{1}}}
\newcommand{\DS}{${}^2\!S_{\frac{1}{2}}$\xspace}
\newcommand{\QS}{${}^4\!S_{\frac{3}{2}}$\xspace}
\newcommand{\CG}[6]{C_{#1,#2,#3}^{#4,#5,#6}}

\newcommand{\eftnopi}{EFT($\cancel{\pi}$)\xspace}
\newcommand{\lambdanopi}{\Lambda_{\cancel{\pi}}}
\newcommand{\Ws}{W\!s}
\newcommand{\Was}{W\!as}
\newcommand{\NNLO}{NNLO\xspace}
\newcommand{\jjvH}{{}^3\mathrm{H}}
\newcommand{\jjvHe}{{}^3\mathrm{He}}
\newcommand{\Dbb}{\Bar{\mathbf{D}}}
\newcommand{\NN}{$N\!N$\xspace}
\newcommand{\Mc}{\mathcal{M}}

\newcommand{\Lonenopi}{{}^{\cancel{\pi}}\!{L}_1}
\newcommand{\Ltwonopi}{{}^{\cancel{\pi}}\!{L}_2}
\newcommand{\deltaLoz}{\delta_{L_1^{(0)}}}
\newcommand{\deltaLoo}{\delta_{L_1^{(1)}}}
\newcommand{\deltaLtz}{\delta_{L_2^{(0)}}}
\newcommand{\deltaLto}{\delta_{L_2^{(1)}}}
\newcommand{\deltaLom}[1]{\delta_{L_1^{(#1)}}}
\newcommand{\deltaLtm}[1]{\delta_{L_2^{(#1)}}}
\newcommand{\DeltaN}{\Delta_{\textrm{N}}}
\newcommand{\DeltaP}{\Delta_{\textrm{P}}}
\usepackage{orcidlink}

\usepackage{relsize}

\usepackage{etoolbox}
\makeatletter
\newcommand*{\rom}[1]{\expandafter\@slowromancap\romannumeral #1@}
\newcommand{\deltanopi}{Q}
\newcommand{\deltawig}{\delta_W}
\newcolumntype{H}{>{\setbox0=\hbox\bgroup}c<{\egroup}@{}}